\documentclass[twocolumn,astrosymb,tighten,times,trackchanges]{aastex701}
\usepackage{adjustbox}

\begin{document}

\title{Small and Complex I: The Three Component Structure of $z \sim 0$ Massive Compact Quiescent Galaxies}

\author[orcid=0000-0002-7967-3376,sname='Slodkowski Clerici']{Katia Slodkowski Clerici}
%\altaffiliation{Universidade Federal do Rio Grande do Sul}
\affiliation{Universidade Federal do Rio Grande do Sul – Departamento de Astronomia – 91501-970, Porto Alegre-RS, Brazil}
\email[show]{clericikatia@gmail.com}  

\author[orcid=0000-0002-0953-8224,sname='A. Schnorr-Müller']{Allan Schnorr-Müller}
%\altaffiliation{Universidade Federal do Rio Grande do Sul}
\affiliation{Universidade Federal do Rio Grande do Sul – Departamento de Astronomia – 91501-970, Porto Alegre-RS, Brazil}
\email{allan.schnorr@ufrgs.br}

\author{Ana Carolina Santiago-Menezes}
%\altaffiliation{Universidade Federal do Rio Grande do Sul}
\affiliation{Universidade Federal do Rio Grande do Sul – Departamento de Astronomia – 91501-970, Porto Alegre-RS, Brazil}
\affiliation{European Southern Observatory, Alonso de Córdova 3107, Vitacura, Región Metropolitana, Chile}
\email{carolsantiago020@gmail.com}

\author{Marina Trevisan}
%\altaffiliation{Universidade Federal do Rio Grande do Sul}
\affiliation{Universidade Federal do Rio Grande do Sul – Departamento de Astronomia – 91501-970, Porto Alegre-RS, Brazil}
\email{marina.trevisan@ufrgs.br}

\author{Tiago Vecchi Ricci}
%\altaffiliation{Universidade Federal da Fronteira Sul}
\affiliation{Universidade Federal da Fronteira Sul – Campus Cerro Largo – 97900-000, Cerro Largo-RS, Brazil}
\email{tiago.ricci@uffs.edu.br}

\author{Rafael Merib-Dias}
%\altaffiliation{Universidade Federal do Rio Grande do Sul}
\affiliation{Universidade Federal do Rio Grande do Sul – Departamento de Astronomia – 91501-970, Porto Alegre-RS, Brazil}
\email{rafameribdias@hotmail.com}

\author{Felícia Palacios}
%\altaffiliation{Universidade Federal do Rio Grande do Sul}
\affiliation{Universidade Federal do Rio Grande do Sul – Departamento de Astronomia – 91501-970, Porto Alegre-RS, Brazil}
\email{felicia.palacios@gmail.com}

\author{Weslley Linck Becker}
%\altaffiliation{Universidade Federal do Rio Grande do Sul}
\affiliation{Universidade Federal do Rio Grande do Sul – Departamento de Astronomia – 91501-970, Porto Alegre-RS, Brazil}
\email{weslleylinck0@gmail.com}

\author{Fabricio Ferrari}
%\altaffiliation{Universidade Federal do Rio Grande}
\affiliation{Universidade Federal do Rio Grande - Instituto de Matemática, Estatística e Física - 96203-900, Rio Grande-RS, Brazil}
\email{fabricio.ferrari@gmail.com}

\begin{abstract}
We investigate the morphology and structural properties of 246 massive compact quiescent galaxies (MCGs; $\log M_{\star} \sim 10$--$11$, $\sigma_{\mathrm{e}} \sim 150$--$350\,$km\,s$^{-1}$, $R_{\mathrm{e}} \sim 0.7$--$2.5\,$kpc) at $z \sim 0$, selected as outliers in the stellar mass--velocity dispersion and velocity dispersion--size relations, using $g$-, $r$-, and $i$-band Hyper Suprime-Cam images. We compare them to a control sample of average-sized quiescent galaxies (CSGs) matched in stellar mass, star formation rate, redshift, and $g-i$ color. Both samples are dominated by S0 galaxies, comprising $93\%$ of MCGs and $71\%$ of CSGs, while ellipticals account for $4\%$ and $11\%$, respectively. The fraction of interacting or morphologically disturbed systems is low in both samples ($13\%$ for MCGs and $16\%$ for CSGs). Multi-component decompositions of the $g$- and $r$-band images show that $75\%$ of MCGs require a three-component model (bulge, disk, and envelope), while $21\%$ are best fit by two components and $4\%$ by a single Sérsic profile. Two-component MCGs are preferentially low-inclination systems, suggesting that the three-component fraction represents a lower limit. In contrast, only $7\%$ of CSGs exhibit a comparable three-component structure. Bars are present in $29\%$ of CSGs but are absent in MCGs. For three-component systems, MCGs and CSGs have similar bulge ($R_\mathrm{e}=0.39$ vs.\ $0.45$\,kpc) and envelope ($R_\mathrm{e}=6.4$ vs.\ $5.8$\,kpc) sizes, while MCG disks are significantly more compact ($R_\mathrm{e}=1.9$ vs.\ $3.3$\,kpc). The envelope component shows a broad ellipticity distribution ($\epsilon_\mathrm{Envelope} \sim 0.0$--$0.6$), which we interpret as corresponding to either a stellar halo or a thick disk.

\end{abstract}


\section{Introduction}
Galaxies are complex systems whose structural components encode key information about their formation and evolutionary histories. The most influential attempt to classify galaxy morphology was introduced by \citet{hubble26}, forming the foundation of classification schemes still in use today. In this framework, known as the Hubble sequence and commonly represented by the tuning-fork diagram, galaxies are organized according to their apparent morphology. Elliptical (E) galaxies occupy the left-hand side of the diagram and are characterized by smooth, featureless light distributions, with no prominent large-scale disks or spiral structure. Spiral galaxies populate the right-hand side and exhibit prominent stellar disks with well-defined spiral arms; this branch bifurcates into unbarred (S) and barred (SB) systems. Located between ellipticals and spirals are the lenticular (S0) galaxies, which combine features of both classes: they possess large-scale stellar disks similar to those of spirals, but lack prominent spiral arms and little or no ongoing star formation, resulting in overall red colors akin to those of elliptical galaxies. Numerous authors have since proposed revisions and extensions to this scheme. Among the most influential is the extension by \citet{deVaucouleurs59}, later formalized in \citet{vaucouleurs91}, which introduced a more continuous classification framework, added finer subdivisions along the spiral sequence, incorporated an intermediate bar class (SAB), and included morphological features such as rings and lenses. A more detailed and systematic classification of ring structures was later developed by \citet{buta95}.

With the availability of large samples of galaxies at intermediate and high redshifts ($z \gtrsim 1$) observed with high-resolution imaging, numerous studies have investigated the evolution of the Hubble sequence at intermediate and high stellar masses. Early work concluded that irregular and peculiar galaxies become increasingly common with redshift, dominating the galaxy population by $z \gtrsim 1.5$ (see \citealt{conselice14} and references therein). This picture has been substantially revised with the advent of the \emph{James Webb Space Telescope} (JWST). Studies based on JWST observations show that disk galaxies remain the most common morphological type out to at least $z \sim 5$ \citep{ferreira22,huertas24,lee24}. Nevertheless, both early- and late-type galaxies at $z \gtrsim 1$ differ markedly from their local counterparts. In particular, early-type galaxies are more compact, with effective radii smaller by factors of $\sim 2$--$3$ at fixed stellar mass \citep{vanderwel.etal.2014}. Moreover, while the high-mass end of the local galaxy population is dominated by giant ellipticals, many massive early-type galaxies at cosmic noon exhibit prominent disks, more closely resembling S0s \citep{bruce12,vanderwel12,huertas16,davari17,Hill.etal.2019}.

The small effective radii of compact early-type galaxies at $z \gtrsim 2$ (typically $\sim 0.5$--$2$~kpc) make spatially resolved studies of their spectroscopic and photometric properties extremely challenging with current facilities. Consequently, most spectroscopic analyses rely on integrated measurements, while morphological studies are often limited to global properties inferred from bulge--disk decompositions. These observational limitations have motivated the search for local analogues of high-redshift compact galaxies \citep{ferre-mateu12,trujillo14,ferre-mateu17,yildirim17,spiniello21,Schnorr.et.al.2021}. Work on these nearby systems has focused primarily on stellar population properties and kinematics. Detailed morphological analyses remain limited to a couple of case studies \citep{Yildirm.etal.2015} which have hinted at complex stellar structures that are not well captured by simple bulge--disk decompositions, requiring additional structural components.

Over the past two decades, multi-component photometric decomposition studies have substantially advanced our understanding of the structure and assembly history of local early-type galaxies. Elliptical galaxies were long regarded as simple objects, whose surface brightness profiles could be adequately described by a single Sérsic function with $2.5 \lesssim n \lesssim 10$. However, deeper imaging and higher-resolution observations have challenged this view, revealing the presence of multiple photometric components occupying distinct spatial scales \citep{Huang.etal.2013,Oh.etal.2017}, interpreted as the result of either in-situ star formation in the early universe or the later accretion of smaller systems \citep{huang13b}. Originally interpreted as a transitional class between spirals and ellipticals, lenticular galaxies are now understood to constitute a heterogeneous population. Many S0s exhibit structural complexity comparable to that of spiral galaxies, including stellar bars, faint spiral arms, lenses, rings, and thick disks \citep{laurikainen09,laurikainen11}. In addition, a significant fraction of S0s show low bulge-to-total flux ratios similar to those of late-type galaxies, supporting a scenario in which S0s originate from faded spirals. These findings motivated the proposal of an S0 sequence running parallel to the spiral sequence in the Hubble diagram \citep{vanderberg76,Cappellari.etal.2011,kormendy12}.  Nevertheless, this picture is incomplete. Some S0s host bulges that are brighter than those typically found in spirals \citep{burstein05}, and the bulge properties of the S0 and spiral populations differ in systematic ways \citep{gao18}. A comprehensive framework for S0 formation must therefore account for origins both as faded spirals and as true transitional objects between late-type galaxies and ellipticals.

In \citet{Clerici.etal.2024}, we analyzed the stellar population properties of a large sample of 1\,858 massive compact galaxies (hereafter MCGs) at \(z \approx 0\) drawn from the Sloan Digital Sky Survey. We found that MCGs host predominantly old ($\sim 10$\,Gyr) and $\alpha$-enhanced ($[\alpha/\mathrm{Fe}] \sim 0.2$) stellar populations, with solar to super-solar metallicities. In this work, the first in a series dedicated to the morphology of massive compact quiescent galaxies at $z \sim 0$, we present the results of a multi-component decomposition of the two-dimensional surface brightness distribution of deep, high-resolution $r$-band images for a representative subset of 246 MCGs selected from the \citet{Clerici.etal.2024} sample. The main goals of this study are: (i) to assess whether MCGs are predominantly S0s, ellipticals, or a mixture of both classes; (ii) to characterize the structural properties of their bulges and disks; (iii) to evaluate the prevalence of additional structural components beyond the bulge and disk, and to characterize them when present; and (iv) to compare our results with those obtained for a control sample of non-compact quiescent galaxies matched in stellar mass, star formation rate, redshift, and color, in order to identify the ways in which MCGs differ from their non-compact counterparts beyond compactness alone. 

This paper is organized as follows. In Sec.\,\ref{sec:data}, we describe the data and sample selection criteria. In Sec.\,\ref{sec:methods}, we present our multi-component decomposition strategy. Our results are presented in Sec.\,\ref{sec:results}, and discussed and contextualized in Sec.\,\ref{sec:discuss}. Finally, we summarize our findings in Sec.\,\ref{sec:conclusions}. Throughout this work, we adopt a simplified $\Lambda$CDM cosmology with $\Omega_{\rm M} = 0.3$, $\Omega_\Lambda = 0.7$, and $H_0 = 70\,\mathrm{km\,s^{-1}\,Mpc^{-1}}$.

\section{Data and sample selection} \label{sec:data}

\begin{figure}
\centering
 \includegraphics[width=\linewidth, trim=40 15 30 40, clip]{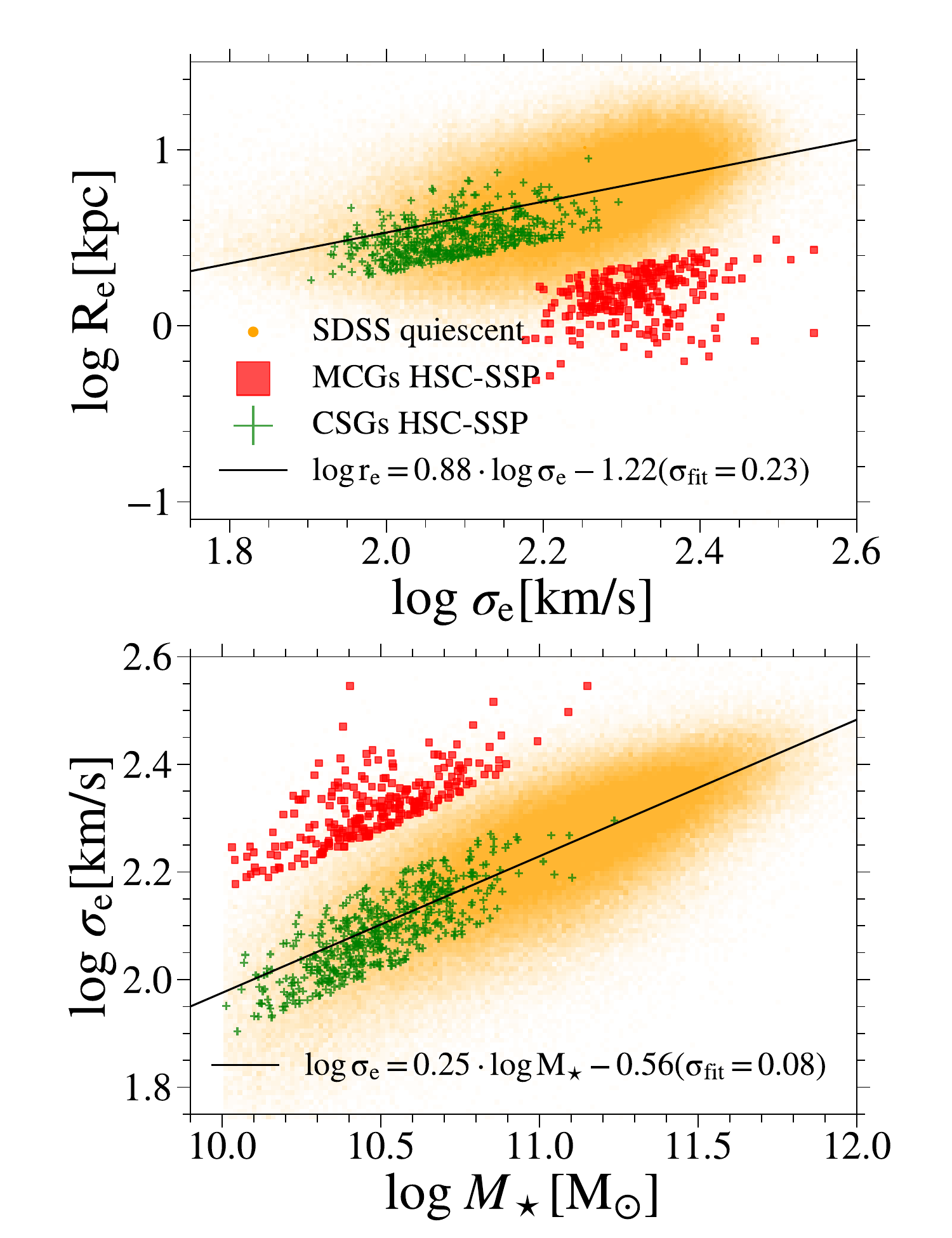}
    \caption{Sample selection: SDSS quiescent galaxies are shown in orange, MCGs in red, and CSGs in green. Top panel:  $\mathrm{log}\, R_{\mathrm{e}}$ versus  $\mathrm{log}\, \sigma_{\mathrm{e}}$ with a linear fit of  $\mathrm{log}\, r_{\mathrm{e}} = 0.8784 \cdot  \mathrm{log}\, \sigma_{\mathrm{e}} -1.2265$ ($\sigma_{\mathrm{fit}} = 0.2287$). Bottom panel: $\mathrm{log}\, \sigma_{\mathrm{e}}$ versus $\mathrm{log}\, M_{\star}$ and a linear fit of $\mathrm{log}\, \sigma_{\mathrm{e}} = 0.2538 \cdot \mathrm{log}\, M_{\star} -0.5622$ ($\sigma_{\mathrm{fit}} = 0.0865$).}
    \label{fig:final_sample}
\end{figure}

\subsection{SDSS Data}

To obtain global galaxy properties, we make use of several catalogs based on Sloan Digital Sky Survey (SDSS) data. Stellar masses ($M_{\star}$) and star formation rates (SFRs) are taken from the GALEX--SDSS--WISE Legacy Catalog (GSWLC; \citealt{salim18}). Effective radii ($R_\mathrm{e}$), defined as the semi-major axis of the half-light ellipse, are extracted from the catalog of \citet{simard11}. We adopt $R_\mathrm{e}$ values derived from Sérsic+exponential fits to the two-dimensional surface-brightness profiles of SDSS--DR7 \textit{r}-band images. Stellar velocity dispersions are retrieved from the SDSS spectroscopic catalog. These measurements are converted from the fiber-aperture velocity dispersion ($\sigma_\mathrm{ap}$) to an effective velocity dispersion ($\sigma_\mathrm{e}$) using the relation $\sigma_{\mathrm{ap}} = \sigma_\mathrm{e} [R_{\mathrm{ap}}/R_\mathrm{e}]^{-0.066}$ \citep{cappellari06}, where $R_{\mathrm{ap}}$ is the aperture radius. Morphological classifications are taken from the catalog of \citet{dominguez-sanchez18}. We classify galaxies using both the T-type and $\mathrm{P_{S0}}$ (which quantifies the probability that a galaxy with T-type $< 0$ is an S0) parameters. These were obtained with deep learning algorithms trained on the visual classification catalog of \citet{nair10}. Lastly, dark matter halo masses and central/satellite classifications are obtained from the group catalog of \citet{lim17}. This catalog assigns halo masses to galaxy groups using proxies based on either stellar mass or luminosity. In this work, we adopt halo masses and group classifications derived using luminosity as the proxy.

\subsection{HSC Data}
Given the small angular sizes of massive compact galaxies, SDSS imaging, with a typical angular resolution of $\sim1.2\arcsec$, is insufficient for a detailed analysis of their structural properties. We therefore make use of data from the Hyper Suprime-Cam Subaru Strategic Program (HSC--SSP), a deep, multi-band (\textit{grizy}) imaging survey carried out with the Hyper Suprime-Cam on the 8.2-m Subaru Telescope \citep{aihara18}. The survey comprises three layers: Wide ($\sim1400\,\mathrm{deg}^2$, $r \sim 26$\,mag), Deep ($\sim27\,\mathrm{deg}^2$, $r \sim 27$\,mag), and UltraDeep ($\sim3.5\,\mathrm{deg}^2$, $r \sim 28$\,mag). The HSC--SSP data products include reduced science images, variance maps, and point-spread-function (PSF) models for each galaxy and filter.

In this work, we use \textit{g}- and \textit{r}-band images from the Wide layer, publicly available as part of the third data release \citep{Aihara.etal.2022}. The median seeing is $0.79\arcsec$ in the \textit{g} band and $0.75\arcsec$ in the \textit{r} band. Although the \textit{i} band provides the best median seeing ($0.61\arcsec$), its lower saturation limit (18.3\,mag, compared to 18.1\,mag in the \textit{r} band and 17.4\,mag in the \textit{g} band) results in a large fraction of galaxies with saturated central regions. For this reason, \textit{i}-band data are not used in the multi-component photometric decompositions, and are instead employed solely for visual inspection. 

\subsection{Massive Compact Galaxy Sample}

In \citet{Clerici.etal.2024}, we presented a sample of $1\,858$ massive compact galaxies extracted from SDSS~DR14. MCGs were selected as outliers lying $2\sigma_{\mathrm{fit}}$ below the best-fit linear relation in the $\log \sigma_{\mathrm{e}}$--$\log R_\mathrm{e}$ plane and $2\sigma_{\mathrm{fit}}$ above the best-fit linear relation in the $\log M_{\star}$--$\log \sigma_{\mathrm{e}}$ plane. The MCG sample analyzed in this work was obtained by cross-matching the MCG sample of \citet{Clerici.etal.2024} with the {\scshape Wide} layer of HSC--SSP. A total of $246$ MCGs have available $r$-band imaging, and $225$ have $g$-band imaging. 

In Fig.\,\ref{fig:final_sample}, we show the distribution of MCGs (red squares) in the  $\log \sigma_{\mathrm{e}}$--$\log R_\mathrm{e}$ (top panel) and $\log M_{\star}$--$\log \sigma_{\mathrm{e}}$ (bottom panel) planes. For comparison, SDSS quiescent galaxies are shown as orange dots. Best fitting relations are shown as solid black lines.

\subsection{Control Sample}

\begin{figure*}
\centering
	\includegraphics[width=\textwidth, trim=18 20 20 20, clip]{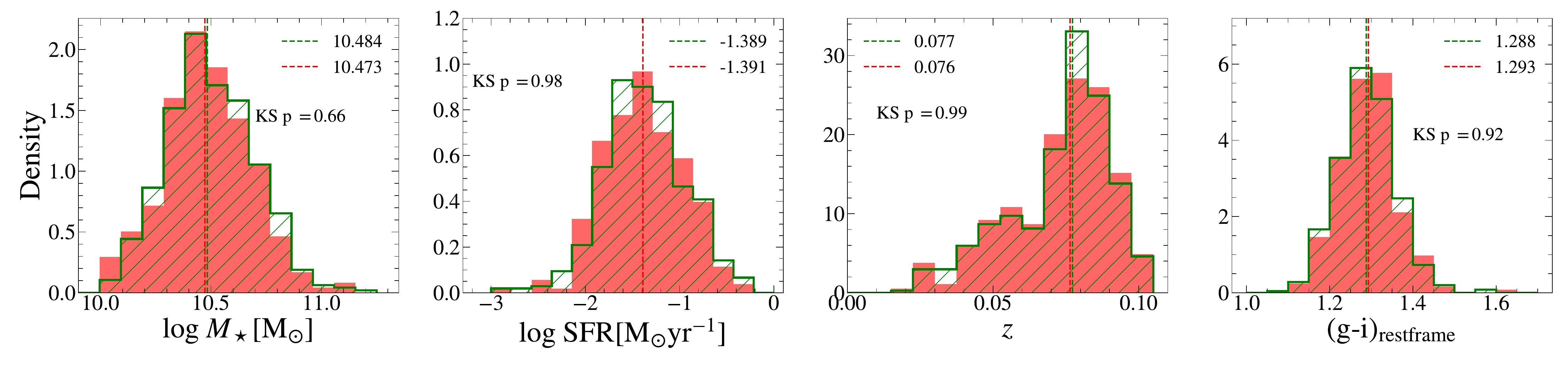}
    \caption{Comparison of stellar mass ($M_{\star}$), star formation rate (SFR), redshift ($z$), and $g-i$ color for MCGs from HSC-SSP (red filled histograms) and CSGs from HSC-SSP (green step histograms). The dashed lines indicate the median values of each sample. P-values from Kolmogorov–Smirnov tests are shown in each panel.}
    \label{fig:properties_sample}
\end{figure*}

In order to compare the morphological and structural properties of MCGs with those of typical quiescent galaxies at $z \approx 0$, we built a control sample from the quiescent galaxy population in the HSC-SSP survey that lies within $\pm1\sigma_{\mathrm{fit}}$ of the linear best fits to the $\log \sigma_\mathrm{e}$ vs.\ $\log R_\mathrm{e}$ and $\log M_{\star}$ vs.\ $\log \sigma_{\mathrm{e}}$ relations from \citet{Clerici.etal.2024}, with twice the size of the MCG sample. Galaxies in the MCG and control samples were matched based on four properties: stellar mass, star formation rate (SFR), $g-i$ color, and redshift ($z$). We chose to match the samples in SFR and $g-i$ color to avoid selecting recently quenched galaxies. The matching procedure was performed using the {\scshape MatchIt} R package \citep{Ho.et.al.2011} together with the Propensity Score Matching (PSM) technique \citep{Rosenbaum.and.Rubin.1983}. Control sample galaxies (hereafter CSGs) are shown as green crosses in Fig.\,\ref{fig:final_sample}. In Fig.\,\ref{fig:properties_sample} we show the distributions of $\log M_{\star}/M_\odot$, SFR, $z$, and $g-i$ color for the MCG and CSG samples.

\section{Methodology} \label{sec:methods}

\subsection{Multi-component Decomposition Strategy}

\begin{figure*}
    \centering
    \includegraphics[width=\textwidth,trim=10 10 10 10, clip]{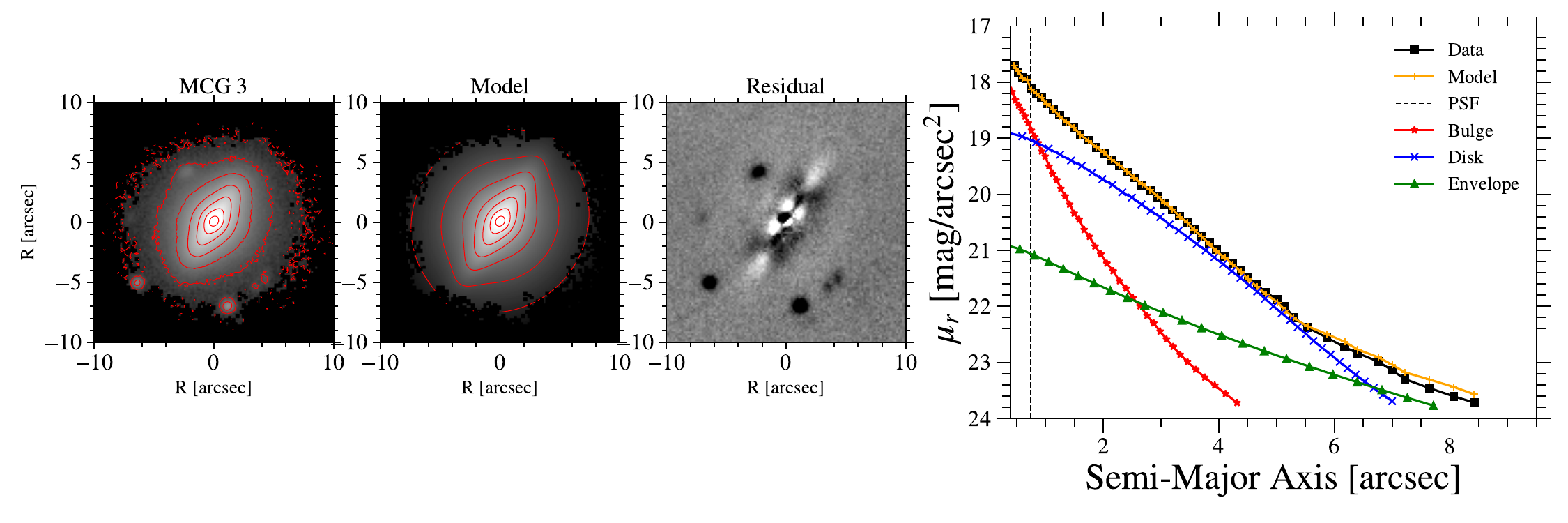}
    \caption{Example of an MCG fitted with three structural components (bulge, disk, and envelope). Displayed from left to right are the r-band image, the PSF convolved model, the residuals and the surface brightness profile of the galaxy and the individual model components. All panels are masked using the corresponding galaxy mask.}
    \label{fig:exemp_components}
\end{figure*}

We performed two-dimensional surface brightness modeling of the HSC images using the software {\scshape imfit} \citep{Erwin.2015}, which allows galaxy images to be fitted with combinations of analytic functions, including Sérsic, exponential, Ferrer, among other profiles. The required inputs for {\scshape imfit} are a galaxy image, an associated noise map, a point-spread function (PSF), a mask image, and a configuration file specifying the model components, as well as which parameters are free or fixed, their bounds, and their initial values.

For each galaxy, we extracted image cutouts with sizes scaled to the SDSS \textit{r}-band Petrosian radius (PetroRad$_r$) from the SDSS photometric catalog. Specifically, each cutout covered an area of $2.5 \times \mathrm{PetroRad}_r$ on a side, in arcseconds, and the same geometry was adopted for the corresponding noise maps in both the \textit{g} and \textit{r} bands. This choice ensures that the full surface brightness profile of each galaxy is included while minimizing contamination from unrelated sources that would otherwise require extensive masking.

Mask images were constructed using segmentation maps generated with Source Extractor version~2.28.0 ({\scshape SExtractor}; \citealp{Bertin.and.Arnouts.1996}). All detected sources other than the target galaxy were masked. In addition, pixels with a signal-to-noise ratio below 3 were excluded from the fits.
 
The two-dimensional surface brightness distributions of the galaxies were modeled using combinations of 2D Sérsic \citep{Sersic.1968} functions. The Sérsic profile is defined as

\begin{equation}
I(x,y) = I_e \, \exp \left\{ -b_n \left[ \left( \frac{r(x,y)}{r_e} \right)^{1/n} - 1 \right] \right\},
\end{equation}

\noindent where $I_e$ is the surface brightness at the effective radius $r_e$, $n$ is the Sérsic index, and $b_n$ is a normalization constant that depends on $n$ and is chosen such that $r_e$ encloses half of the total luminosity.

The elliptical radius is defined as
\begin{equation}
r(x,y) = \sqrt{x^2 + \left( \frac{y}{q} \right)^2},
\end{equation}
where $q = b/a$ is the axis ratio, with $a$ and $b$ denoting the semi-major and semi-minor axes, respectively. The coordinates $(x,y)$ are defined in a reference frame aligned with the major axis of the component.

In {\scshape imfit}, each 2D Sérsic component is characterized by the Sérsic index $n$, the effective surface brightness $I_e$, the effective radius $r_e$, the position angle (PA), the ellipticity $e = 1 - q$, and the central position $(x_0, y_0)$ of the ellipse.

We ran \textsc{imfit} using the differential evolution algorithm as the $\chi^2$ minimization method. Differential evolution is less prone to becoming trapped in local minima than the alternative optimizers available in \textsc{imfit}, and has been shown to yield robust results with minimal user intervention (see \citealt{gadotti26} for a detailed discussion). This algorithm does not require initial parameter guesses, but instead relies on user-defined lower and upper bounds for all free parameters. Below we describe how these bounds were defined.

We assumed that all structural components within a given galaxy share a common center. The image center was adopted as the initial reference position, and the central coordinates were allowed to vary by $\pm 10$ pixels during the fitting process. Limits on the position angle (PA) of each component were determined on a case-by-case basis using estimates from isophotal fitting. When the variation of the isophotal PA across the galaxy was smaller than $10^\circ$, the same bounds (set to $\pm 10^\circ$ around the median PA) were adopted for all components. If the PA variation exceeded $10^\circ$, the PA bounds were defined independently for each component.

Bulges were modeled using a Sérsic component with ellipticity bounds of $0.0 \leq \epsilon_\mathrm{Bulge} \leq 0.6$ and Sérsic index bounds of $0.5 \leq n_\mathrm{Bulge} \leq 8$. The effective radius was constrained to be larger than one pixel in all cases, while the upper bound was defined on a case-by-case basis. If the best-fitting effective radius approached or reached the upper limit, this bound was increased and the fitting procedure was repeated.

Disks were modeled with a Sérsic component with fixed Sérsic index $n = 1$ and ellipticity bounds of $0.0 \leq \epsilon_\mathrm{Disk} \leq 0.85$. The effective radius was initially constrained to $10 \leq R_{\mathrm{e,Disk}} \leq 20$ pixels. When the best-fitting value approached either boundary, the limits were adjusted accordingly and the fit was repeated.

In some galaxies, a third component was required to achieve a satisfactory representation of the surface brightness distribution. This component was modeled with a Sérsic function with ellipticity bounds of $0.0 \leq \epsilon \leq 0.85$, Sérsic index bounds of $0.5 \leq n \leq 8$, and effective radius bounds of $20 \leq R_{\mathrm{e}} \leq 100$ pixels.

To determine the optimal number of structural components for the MCG and CSG samples, we fitted each galaxy using one-, two-, and three-component Sérsic models. The preferred model was defined as the one with the minimum number of components required to reproduce both the observed surface brightness distribution and the geometric structure of the galaxy. Model selection was based on the following criteria:

\begin{enumerate}
    \item Reproduction of ellipticity and position angle profiles:  The best-fitting model was required to reproduce the observed radial variations of ellipticity and position angle derived from isophotal fitting. Typically, galaxies exhibit lower ellipticity in their central regions and higher ellipticity at larger radii. In several galaxies from both samples, however, $\epsilon$ increased up to an intermediate radius and subsequently decreased toward the outskirts. This non-monotonic behavior could not be consistently reproduced by one- or two-component models and required the inclusion of a third Sérsic component;
    \item Absence of significant residuals: The preferred model was required to show no coherent, large-scale residual structures after model subtraction. In several cases, although two-component models reproduced the observed $\epsilon$ and PA profiles, significant residuals persisted at large radii, indicating the presence of an additional extended component;
    \item Significant improvement of the fit quality: We required a relative decrease of at least $10\%$ in both the reduced chi-square and the Bayesian Information Criterion (BIC) compared to the model with fewer components;
    \item Physical plausibility of the solution: Even in cases where residuals were visually small, additional components were included when simpler models produced non-physical solutions, such as Sérsic indices reaching imposed boundaries, unrealistically small or large effective radii, or disk components compensating for missing central or extended light.
\end{enumerate}

In Fig. \ref{fig:exemp_components}, we show an example of an MCG fitted with three structural components (bulge, disk, and envelope), including the r-band image, the PSF-convolved model, the residuals, and the surface brightness profile with its individual components.

\subsection{Visual Inspection of Images}

As this work aims to compare the fractions of galaxies hosting bars and exhibiting morphological disturbances between the two samples, rather than to provide a precise census of such features, we rely on visual inspection of the galaxy images instead of more sophisticated automated methods. Visual classifications were performed using $50 \times 50$\,kpc image cutouts. We primarily inspected \textit{i}-band images because of their superior spatial resolution and depth, and used \textit{r}-band images when \textit{i}-band data were unavailable. In addition, residual maps from the multi-component photometric decompositions in the \textit{r} band were examined to aid in the identification of subtle structures, especially bars. During the inspection process, the contrast and brightness of the images were adjusted interactively to facilitate the detection of faint structures.

\section{Results} \label{sec:results}

\begin{figure}
\centering
    \includegraphics[width=0.8\linewidth, trim=30 30 30 30, clip]{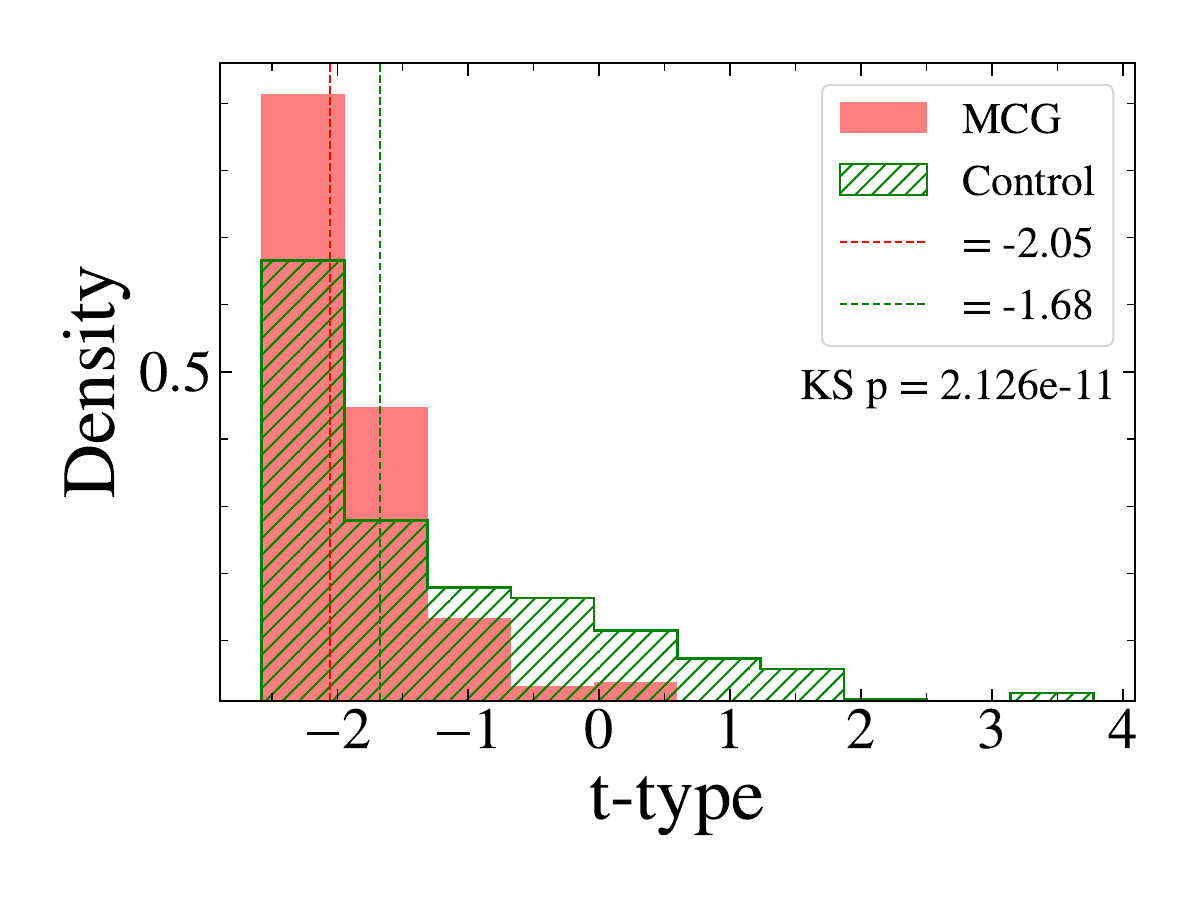}
    \includegraphics[width=0.8\linewidth, trim=10 30 30 20, clip]{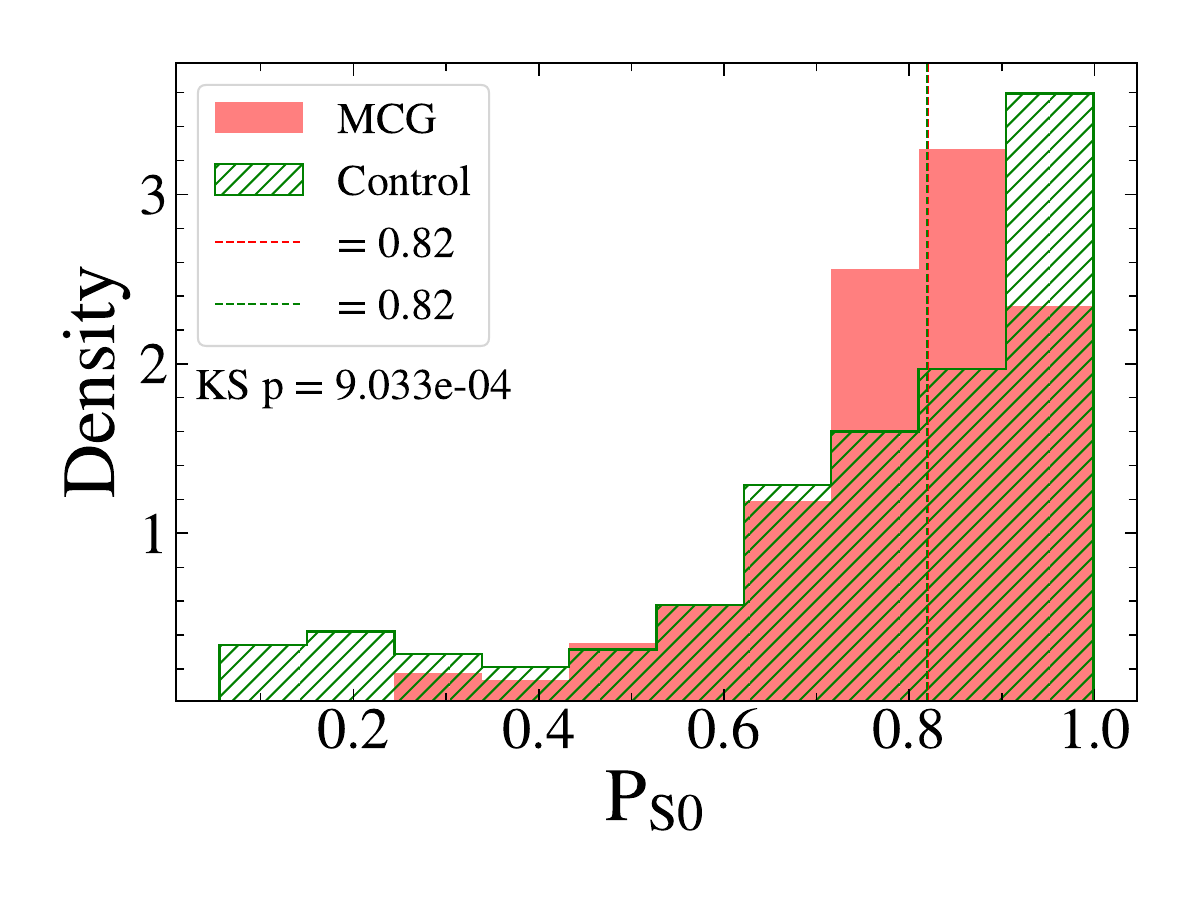}
    \caption{Morphological classification of MCGs and CSGs. In the top panel we show the numerical t-type, in the bottom panel we show $\mathrm{P_{S0}}$, which gives the probability of an early type galaxy (t-type $ < 0$) being classified as an S0. The parameters were extracted from the morphological catalog of \citet{dominguez-sanchez18}. }
    \label{fig:ttype}
\end{figure}

\begin{figure*}
    \centering
    \includegraphics[width=0.9\textwidth]{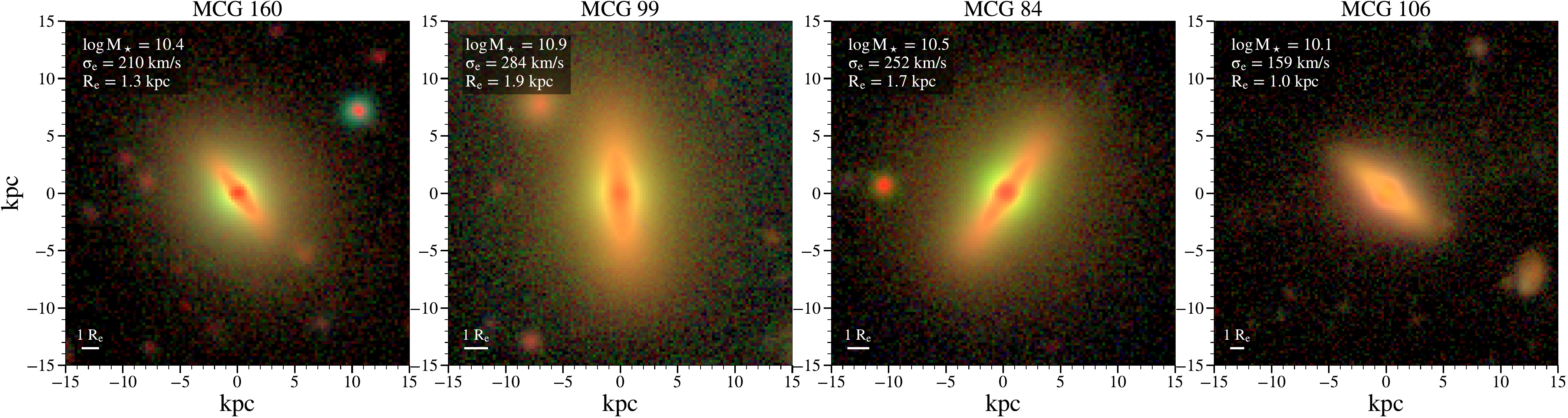}
    \vspace{0.1cm}
    
    \includegraphics[width=0.9\textwidth]{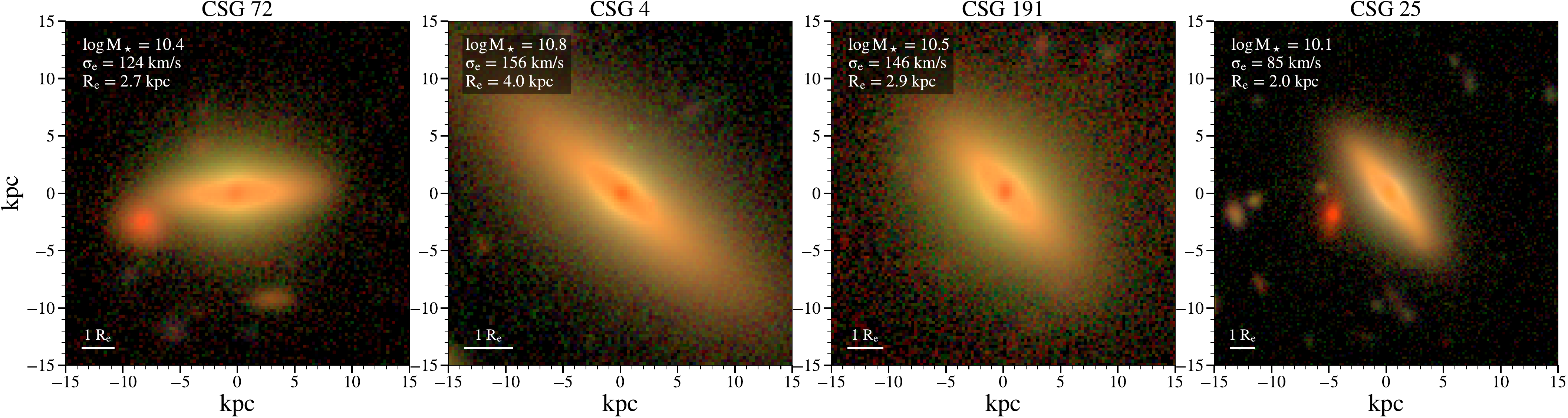}
    \caption{Example RGB composite images of representative three-component MCGs (top row) and CSGs (bottom row). Galaxies in both samples exhibit similar structures, with stellar disks embedded within a rounder, low--surface-brightness envelope.}
    \label{fig:3comp}
\end{figure*}

\begin{figure}
\centering
    \includegraphics[width=0.8\linewidth, trim=30 30 30 30, clip]{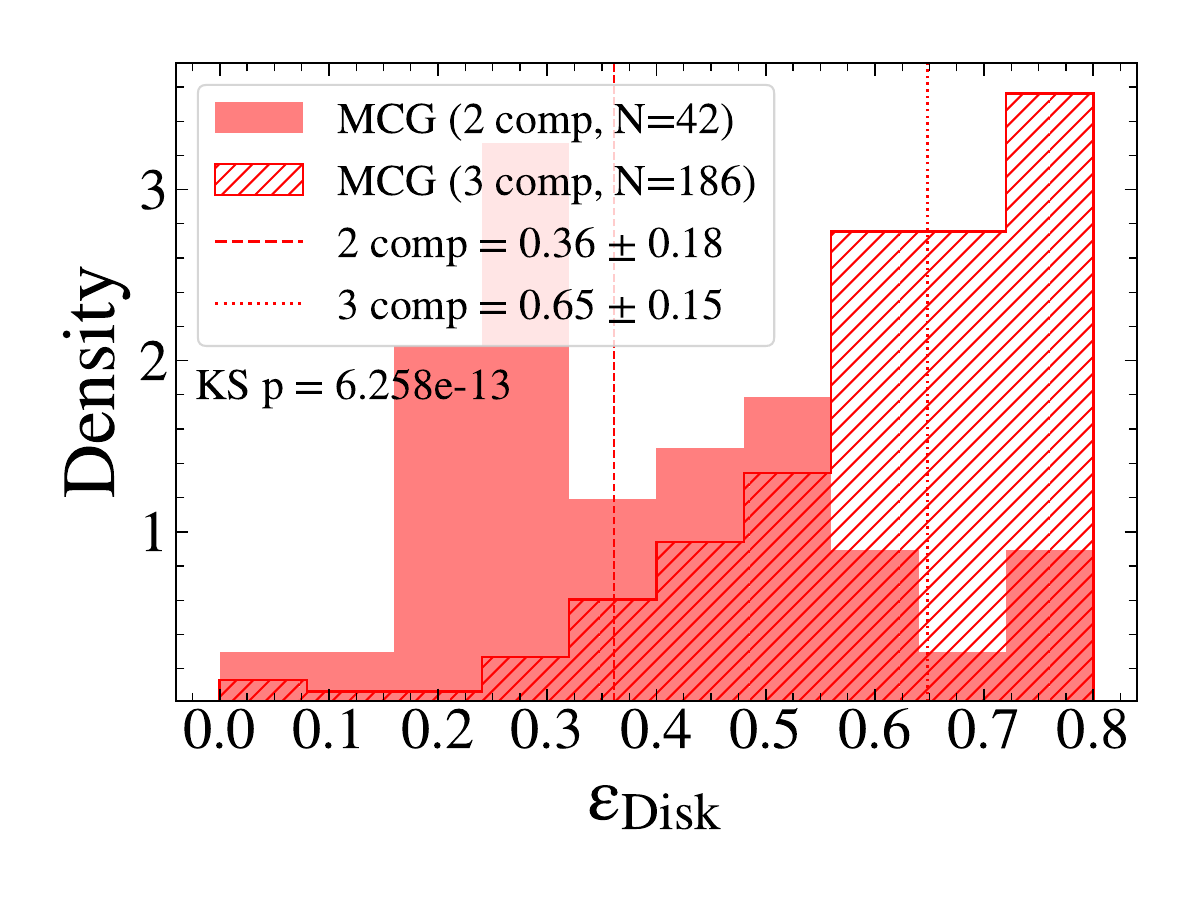}
    \includegraphics[width=0.8\linewidth, trim=10 30 30 20, clip]{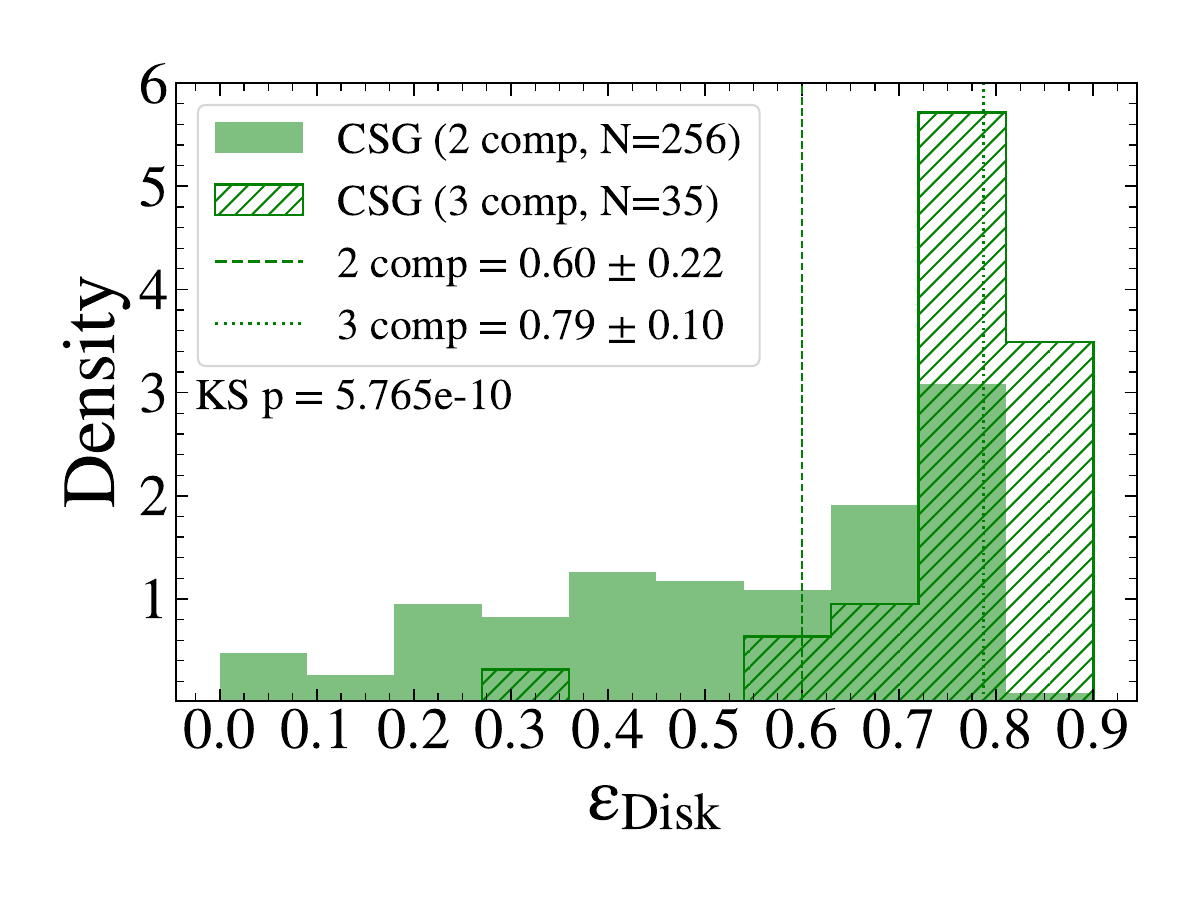}
    \caption{Disk ellipticity ($\epsilon_\mathrm{Disk}$) distributions for two- and three-component MCGs and CSGs. Severely disturbed and barred galaxies were excluded. Three-component systems tend to show higher $\epsilon_\mathrm{Disk}$, consistent with the detectability of the envelope being affected by inclination.}
    \label{fig:ell}
\end{figure}

\begin{figure*}
\centering
% --- Row 1 title ---
\vspace{0.1cm}
\makebox[\textwidth][c]{\textbf{Bulge Properties}}
\vspace{0.1cm}
% --- Row 1 ---
\begin{minipage}[t]{0.265\textwidth}
\includegraphics[width=\linewidth, trim=30 30 30 30,clip]{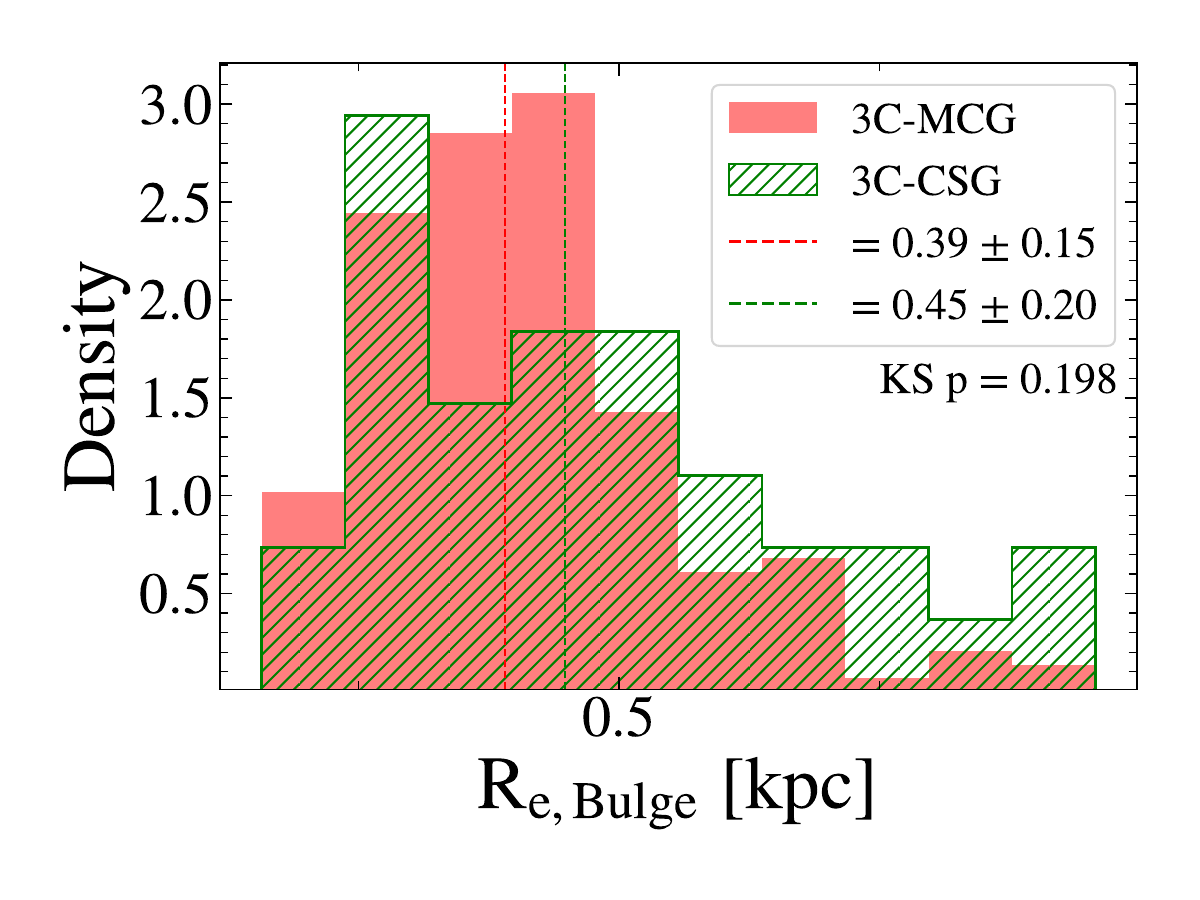}
\end{minipage}
\begin{minipage}[t]{0.265\textwidth}
\includegraphics[width=\linewidth, trim=30 30 30 30,clip]{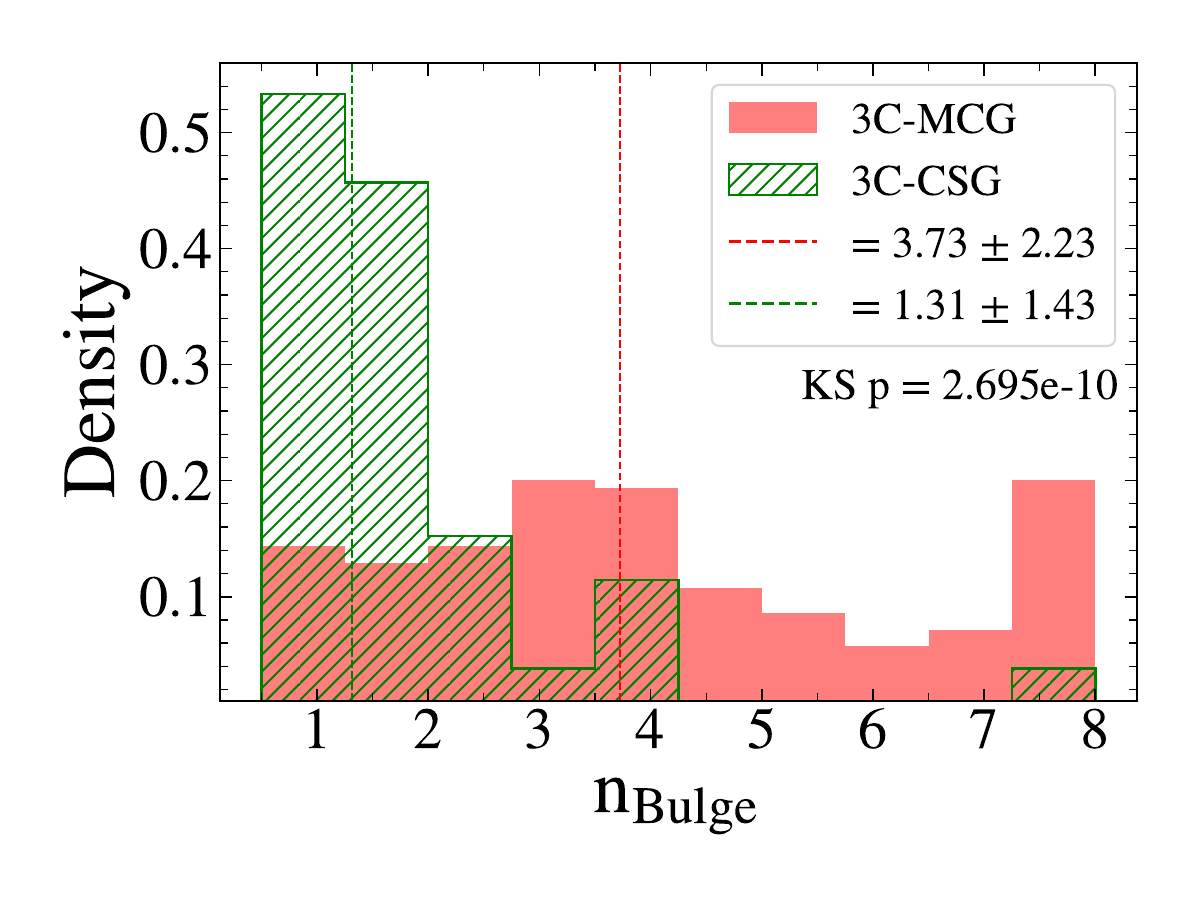}
\end{minipage}
\begin{minipage}[t]{0.265\textwidth}
\includegraphics[width=\linewidth, trim=30 30 30 30,clip]{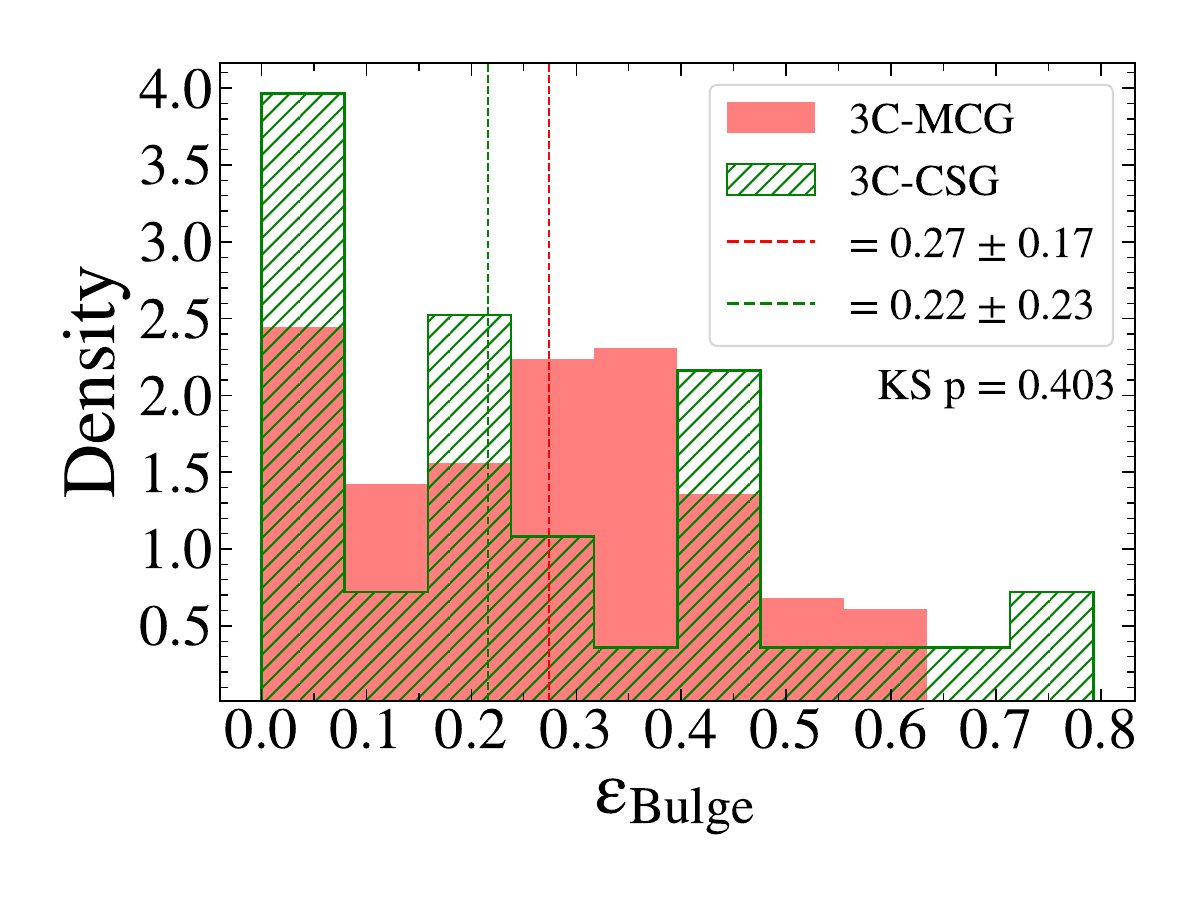}
\end{minipage}
\vspace{0.1cm}
% --- Row 2 title ---
\makebox[\textwidth][c]{\textbf{Disk Properties}}
\vspace{0.1cm}
% --- Row 2 ---
\begin{minipage}[t]{0.265\textwidth}
\includegraphics[width=\linewidth, trim=30 30 30 30,clip]{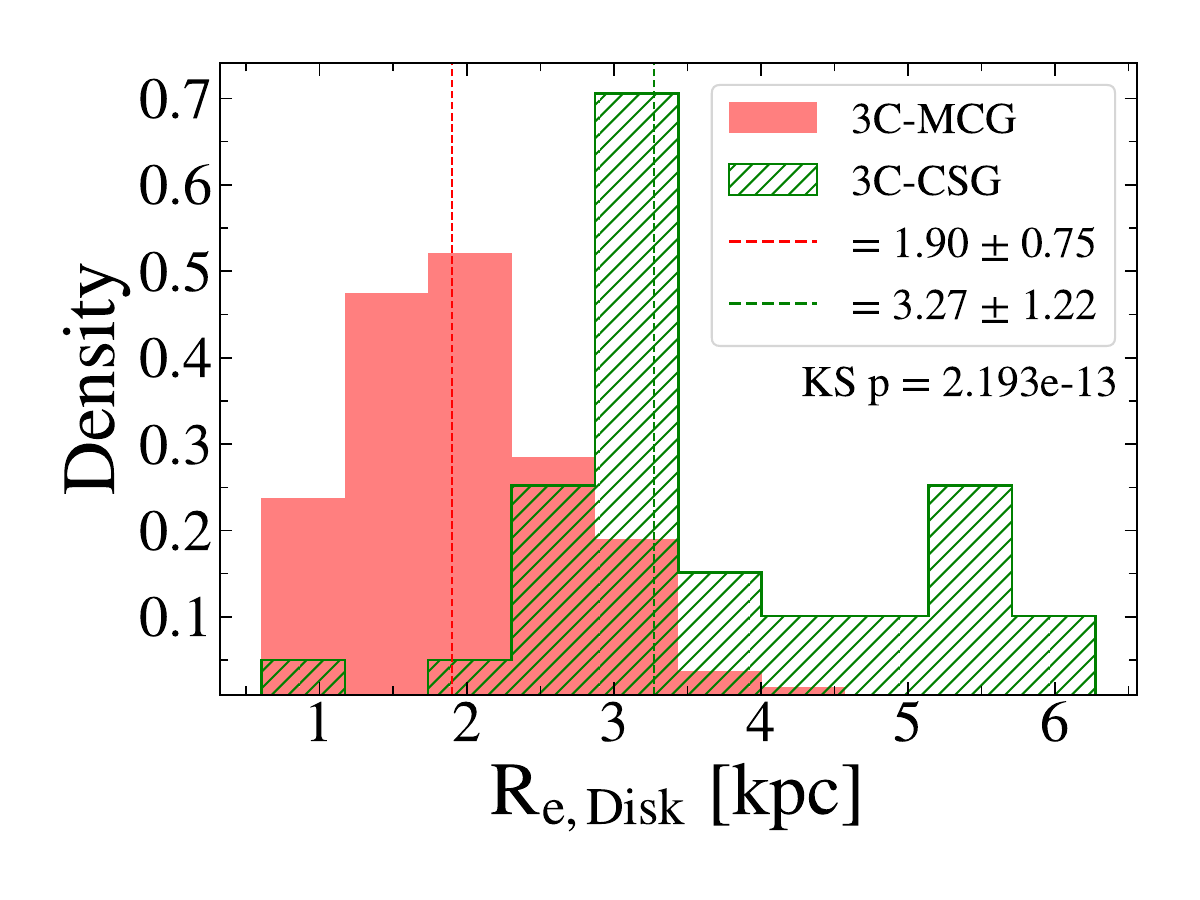}
\end{minipage}
\begin{minipage}[t]{0.265\textwidth}
\includegraphics[width=\linewidth, trim=30 30 30 30,clip]{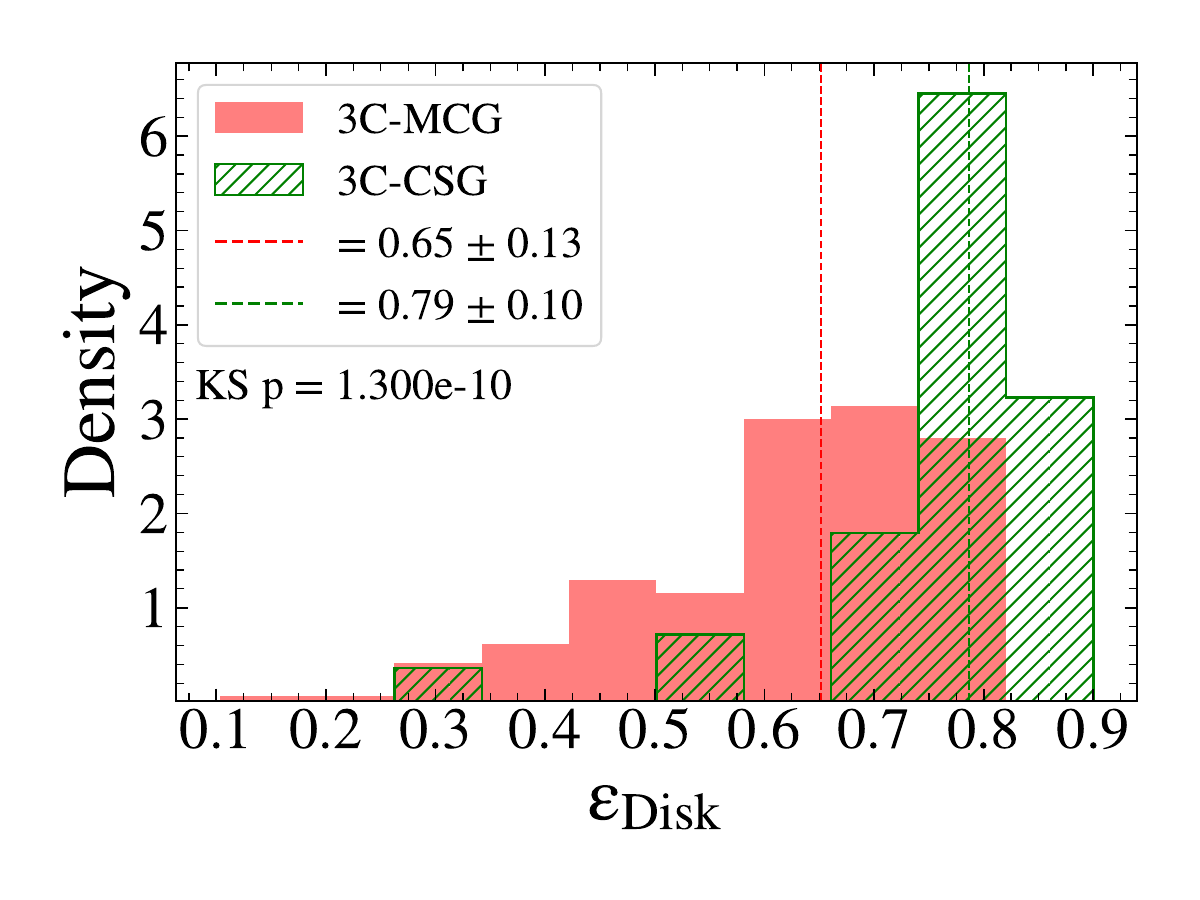}
\end{minipage}
\vspace{0.1cm}
% --- Row 3 title ---
\makebox[\textwidth][c]{\textbf{Envelope Properties}}
\vspace{0.2cm}
% --- Row 3 (center two figures together) ---
\makebox[\textwidth][c]{%
  \begin{minipage}[t]{0.265\textwidth}
  \includegraphics[width=\linewidth, trim=30 30 30 30,clip]{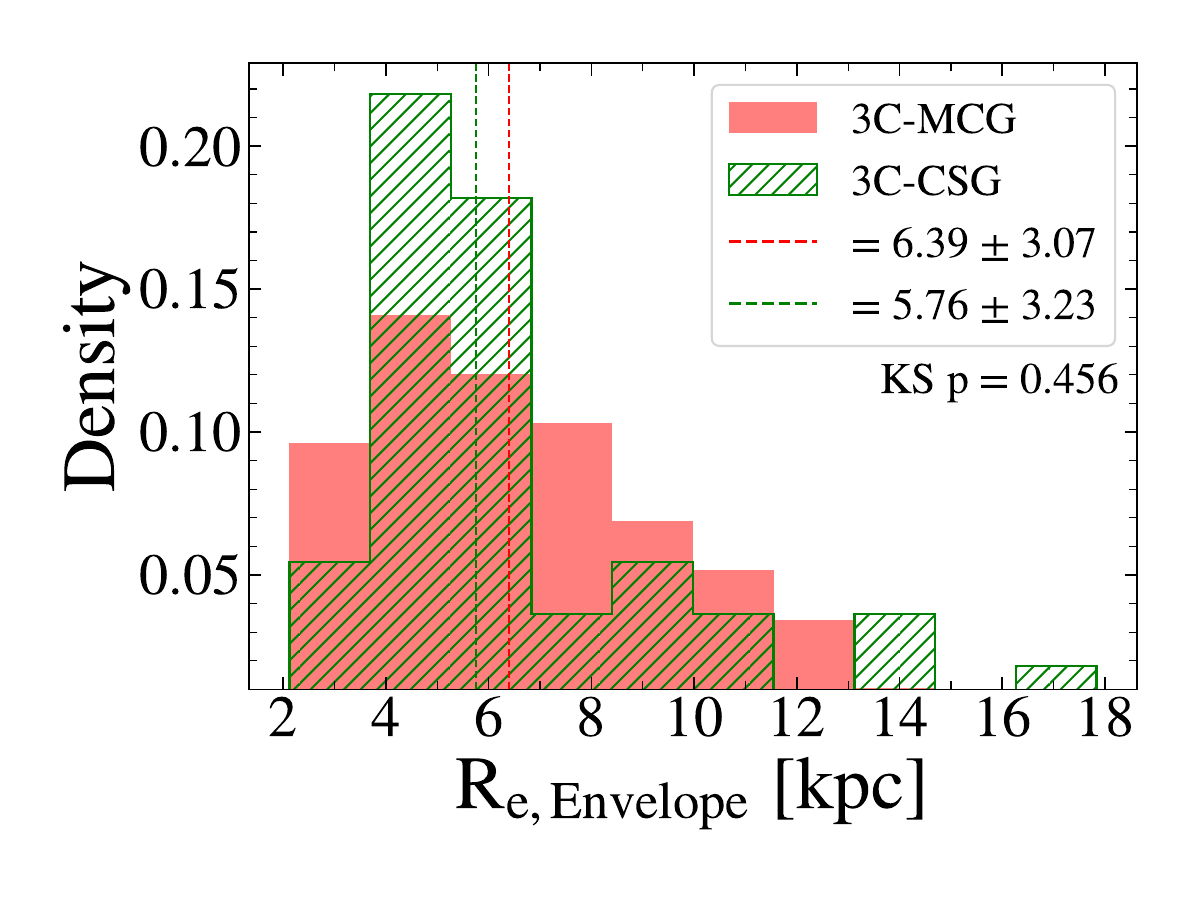}
  \end{minipage}%
  \hspace{0.01\textwidth}%
  \begin{minipage}[t]{0.265\textwidth}
  \includegraphics[width=\linewidth, trim=30 30 30 30,clip]{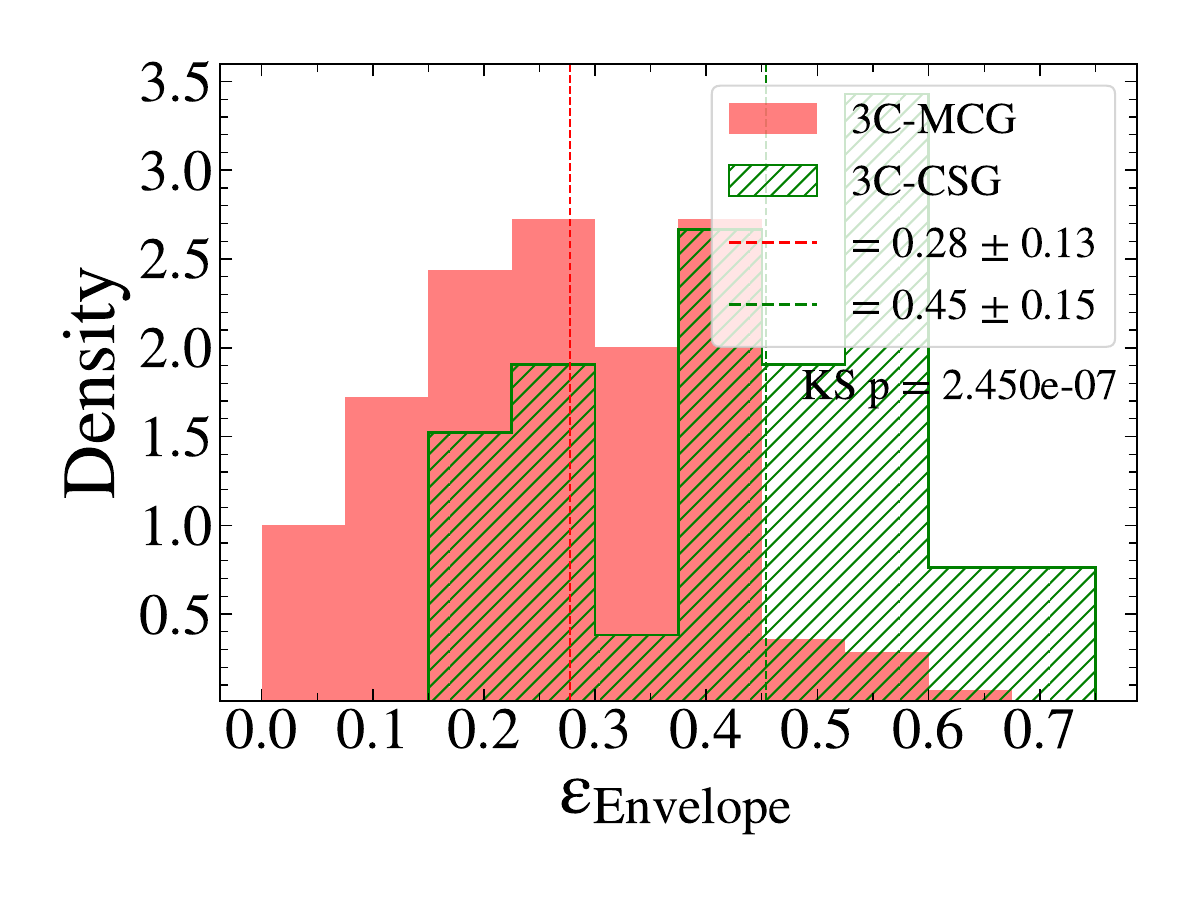}
  \end{minipage}%
  \begin{minipage}[t]{0.265\textwidth}
  \includegraphics[width=\linewidth, trim=30 30 30 30,clip]{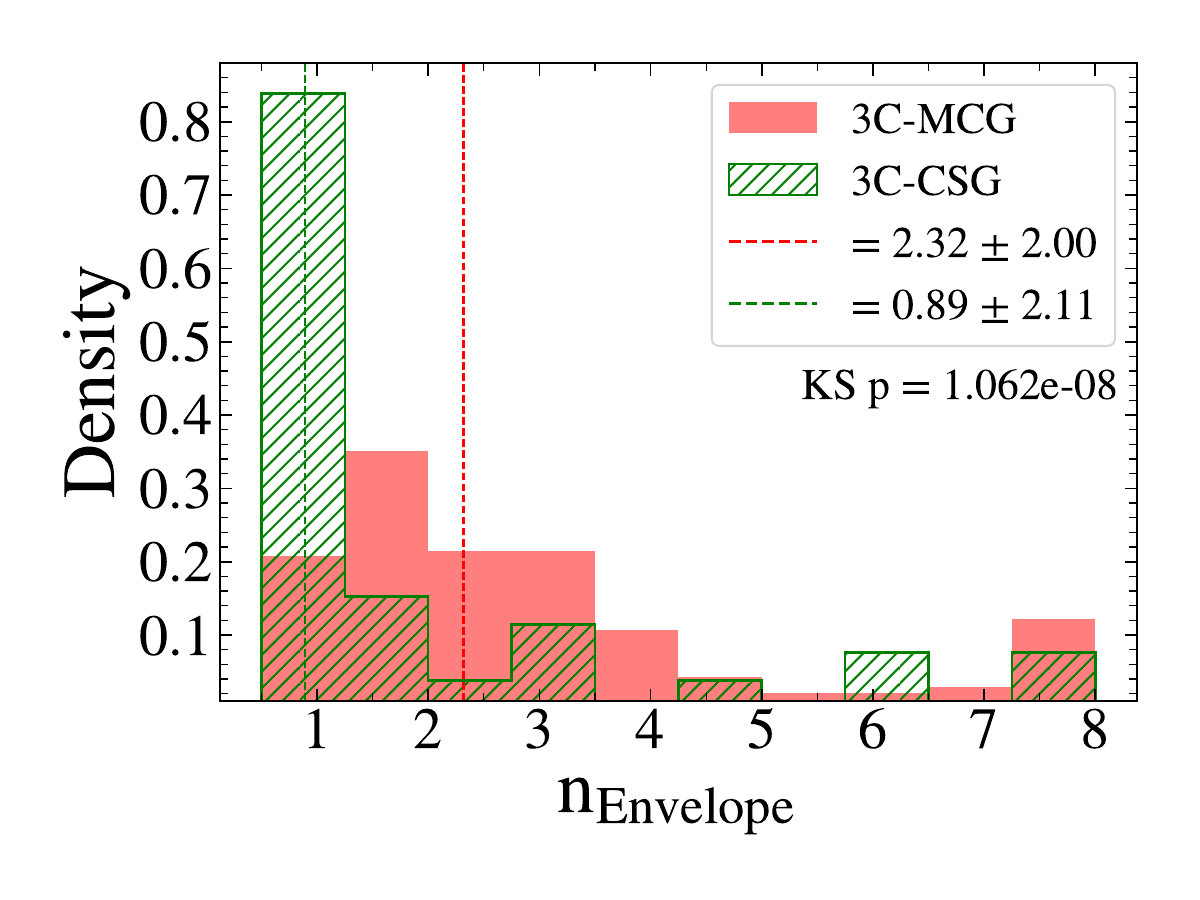}
  \end{minipage}

}
\vspace{0.1cm}
% --- Row 4 title ---
\makebox[\textwidth][c]{\textbf{Flux-to-total ratio}}
\vspace{0.1cm}
% --- Row 4 ---
\begin{minipage}[t]{0.265\textwidth}
\includegraphics[width=\linewidth, trim=30 30 30 30,clip]{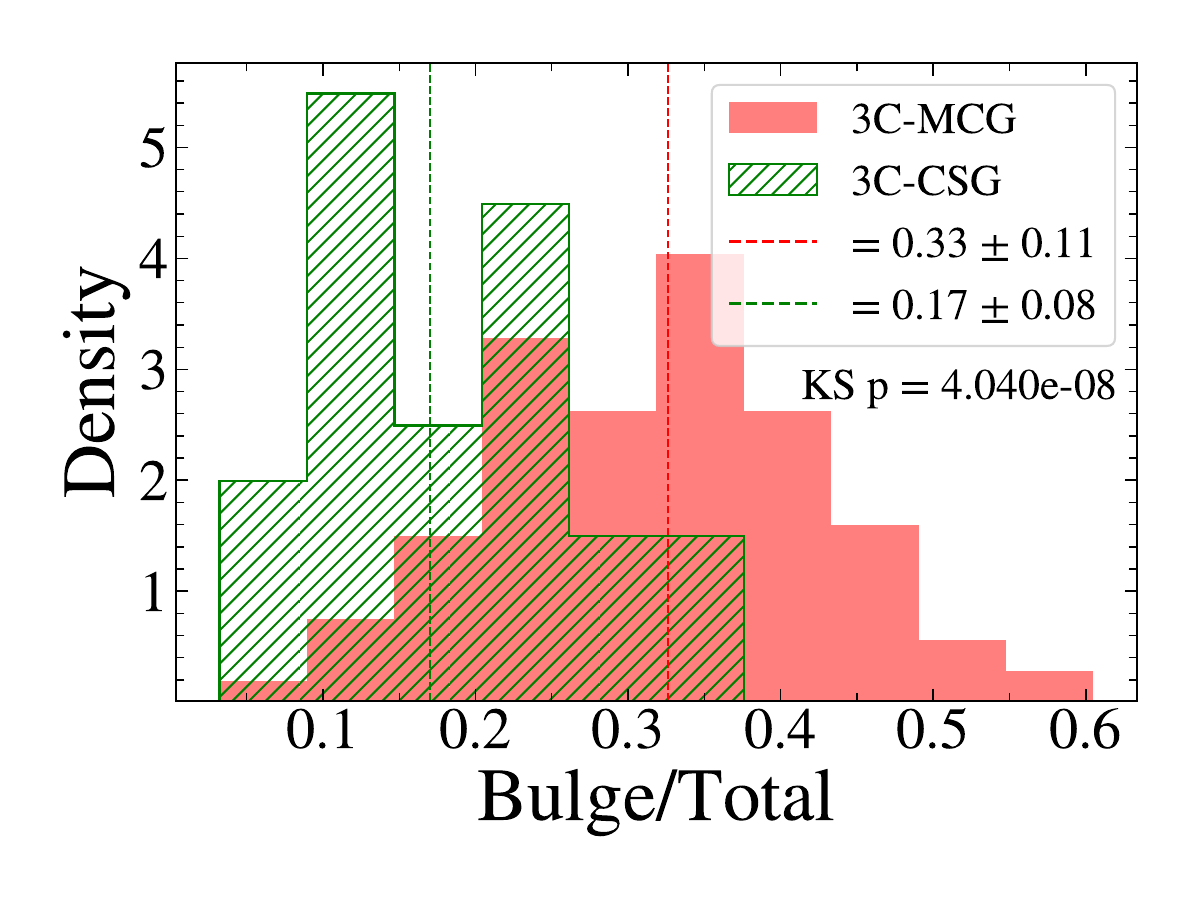}
\end{minipage}
\begin{minipage}[t]{0.265\textwidth}
\includegraphics[width=\linewidth, trim=30 30 30 30,clip]{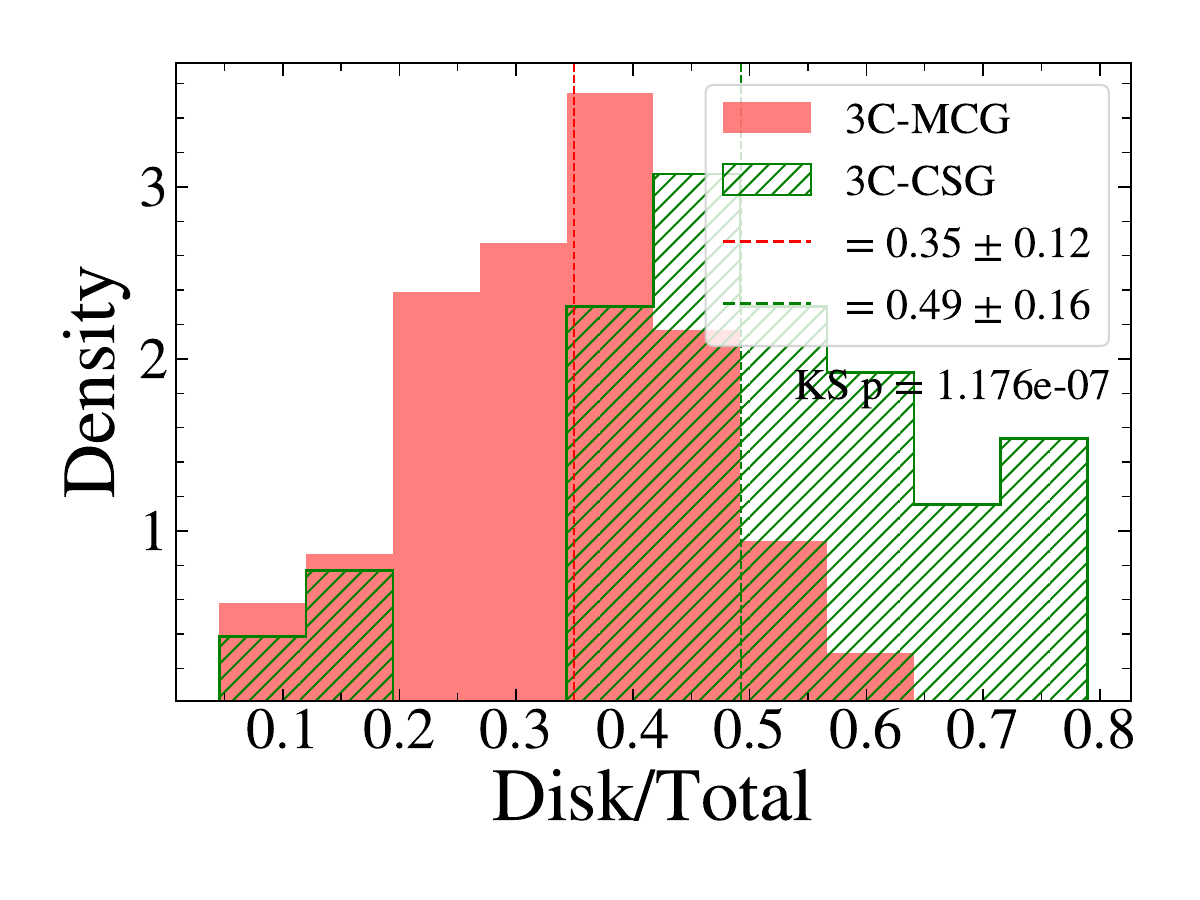}
\end{minipage}
\begin{minipage}[t]{0.265\textwidth}
\includegraphics[width=\linewidth, trim=30 30 30 30,clip]{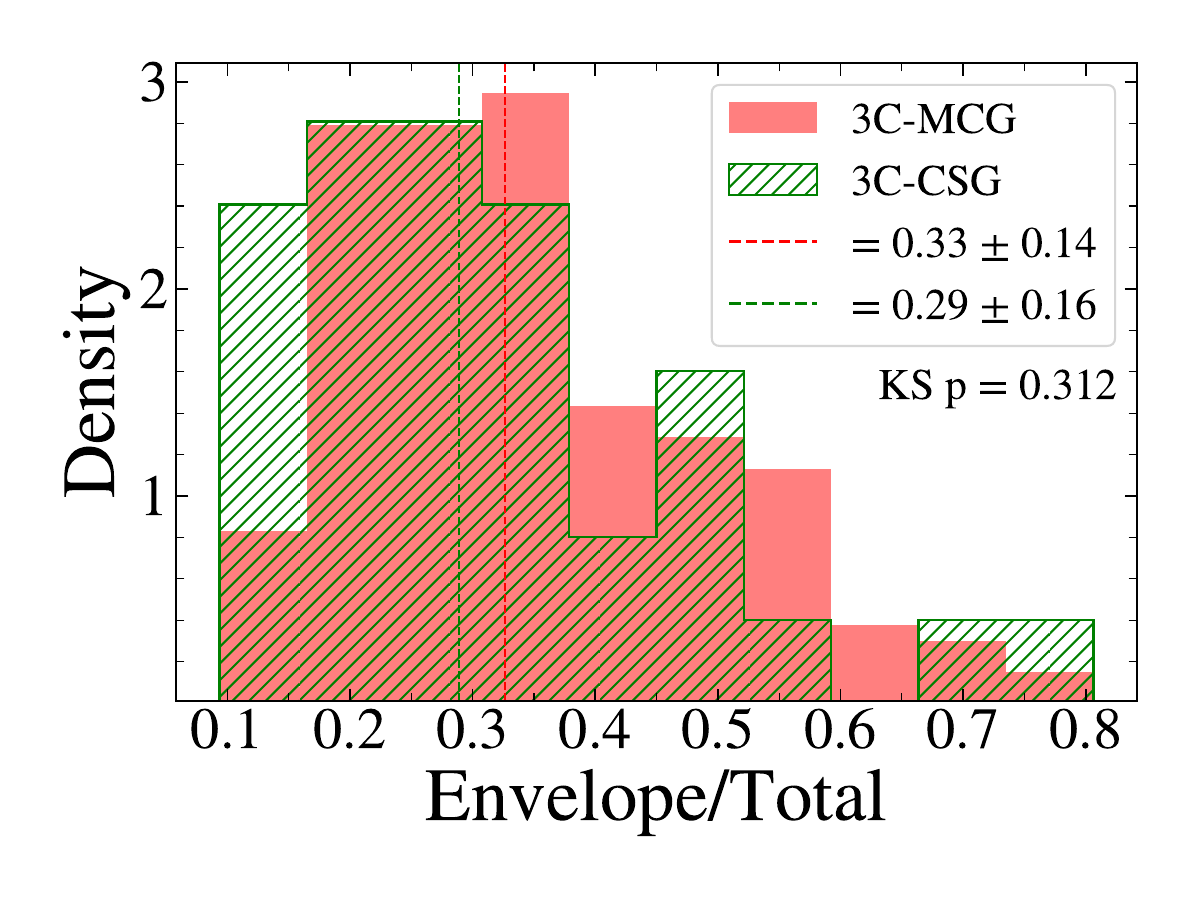}
\end{minipage}

\caption{Structural properties of 3C-MCGs (186 galaxies) and 3C-CSGs (35 galaxies). The first row shows the Sérsic index, effective radius, and ellipticity of the bulge component. The second row presents the disk scale length, scale height, and the ratio between them. In the third row, we show the effective radius and ellipticity of the stellar envelope, and the fourth row displays the flux-to-total ratio of each component. Dashed lines indicate the median values for each sample.}
\label{fig:morph}
\end{figure*}

\subsection{Morphological classification}

In Fig.\,\ref{fig:ttype} we compare the morphological classification of MCGs and CSGs. The top panel shows the distribution of T-types for both samples. MCGs and CSGs are predominantly early-type galaxies (T-type $ \leq 0$), although a significant fraction of CSGs have T-type $ > 0$. As the T-type alone does not provide a clean separation between lenticular and elliptical systems, the bottom panel shows the distribution of $\mathrm{P_{S0}}$, the probability of being classified as an S0, for early-type galaxies only. Lenticular galaxies dominate both samples, but the distributions differ significantly at the low and  high-$P_{\mathrm{S0}}$ ends. The fraction of ellipticals ($P_{\mathrm{S0}} < 0.5$) is slightly higher among CSGs ($11\%$) than MCGs ($4\%$). At the high end, the MCG distribution peaks at $P_{\mathrm{S0}} \sim 0.8$ while the CSG distribution peaks near $P_{\mathrm{S0}} \sim 1$, indicating the presence of morphological differences among the lenticular galaxies in the two samples.

\subsection{Morphologically Disturbed and Interacting Galaxies}

We visually inspected the $r$- and $i$-band images, as well as the residual maps, for morphological disturbances and signatures of interactions. We find comparable fractions in the two samples: $13\%$ of MCGs (33 galaxies) show signs of interaction, compared to $16\%$ (78 galaxies) of CSGs. Most galaxies exhibit only weak features, such as mild warps in the outer disk and faint tidal tails, shells, or streams. As these features do not introduce significant uncertainties in the derived structural parameters, such galaxies are retained in the sample. In contrast, galaxies displaying strong asymmetries or severe disturbances—8 MCGs and 24 CSGs—were excluded from the subsequent morphological analysis.

\subsection{Multi-component Decomposition: The Number of Structural Components} \label{sec:3comp}

Morphological decompositions of large galaxy samples are typically performed using either one (pure disk or spheroid) or two (bulge and disk) structural components, represented by a Sérsic function and by a combination of a Sérsic and an exponential function, respectively. Accordingly, we carried out one- and two-component fits for the galaxies in our sample. However, visual inspection of the images and residual maps reveals that a substantial fraction of galaxies are not well described by such simple models, instead requiring three structural components to achieve satisfactory fits. In most cases, this added complexity is associated with isophotal ellipticities that vary with radius, generally decreasing toward larger radii. The origin of this behavior differs between the two samples. In MCGs, the ellipticity gradient is caused by inclined disks embedded within a low–surface-brightness stellar envelope of lower ellipticity (hereafter referred to as envelope). This configuration is observed in 185 MCGs ($75\%$ of the sample), but in only 35 CSGs ($7\%$). In Fig.\,\ref{fig:3comp} we present examples of MCGs (top panels) and CSGs (bottom panels) exhibiting this structure.

In contrast, for CSGs the ellipticity variation is primarily driven by stellar bars. Inspection of the residual maps and the $r$- and $i$-band images shows that $29\%$ of CSGs host bars, which are frequently associated with rings connected to the ends of the bars. In addition, a small number of CSGs display inner lenses embedded within rings. No bars are detected in MCGs, and only three MCGs show rings. These rings differ from those observed in CSGs, as they are outer, bluer structures with a likely accreted origin. Accurate estimates of bulge and disk structural parameters in barred galaxies require explicit modelling of the bar (see \citealt{gao18} for a detailed discussion). Since bars are absent in MCGs, such modelling is beyond the scope of this work; we therefore exclude barred CSGs from the subsequent analysis.

Could the MCGs well fitted by two components simply be lower-inclination systems, for which the disk and envelope cannot be reliably separated? To test this hypothesis, we compare in Fig.\,\ref{fig:ell} the disk ellipticities of two- and three-component MCGs (top panel) and CSGs (bottom panel). For MCGs, the two-component distribution peaks at $\epsilon_\mathrm{Disk} \sim 0.4$, whereas three-component systems peak at $\epsilon_\mathrm{Disk} \sim 0.7$, consistent with the interpretation that inclination affects the detectability of the envelope. We note, however, that 13 two-component MCGs have $\epsilon_\mathrm{Disk} > 0.5$, indicating that MCGs lacking a detectable third component do exist, albeit rarely. For CSGs, three-component galaxies exhibit high $\epsilon_\mathrm{Disk}$ values, as expected for highly inclined systems, while two-component galaxies span a broad range of disk ellipticities, including many with $\epsilon_\mathrm{Disk} > 0.6$. Indeed, the number of two-component CSGs with $\epsilon_\mathrm{Disk} > 0.6$ (140) greatly exceeds the number of three-component CSGs, implying that prominent envelopes are uncommon among CSGs.

It is worth noting that the $\epsilon_\mathrm{Disk}$ distribution of MCGs is inconsistent with that expected for randomly oriented systems, instead showing a clear bias toward higher inclinations. The origin of this bias is likely related to our selection criteria, in particular the location of MCGs in the $\log M_\star$--$\log \sigma_e$ diagram. Because the velocity dispersion is measured from SDSS fiber spectra (with a fiber diameter of $3\arcsec$), unresolved rotation can contribute significantly to the measured $\sigma_e$. As a result, galaxies that are positive outliers from the central relation are preferentially expected to be highly inclined systems. This bias complicates a direct comparison with CSGs, as some structural parameters depend on inclination. Moreover, it cannot be straightforwardly corrected, since a reliable estimate of the inclination would require either a dynamical analysis or knowledge of the intrinsic disk thickness, which is currently unconstrained. To mitigate its impact, we therefore restrict the structural analysis that follows to three-component MCGs (186 galaxies) and CSGs (35 galaxies) only. From here on, when discussing these subsamples we refer to them as 3C-MCG and 3C-CSG to avoid confusion. We note that, despite their smaller sizes, these subsamples still have similar stellar mass, star formation rate, $(g-i)$ color, and redshift distributions.

\subsection{Multi-component Decomposition: Structural Parameters}

In Fig.\,\ref{fig:morph} we present the structural properties of the bulge, disk, and envelope, together with their flux-to-total ratios, for the 3C-MCG and 3C-CSG samples. A comparison of the bulge structural parameters is presented in the first row of Fig.\,\ref{fig:morph}. Bulge effective radii are similar, with median values of $R_\mathrm{e,Bulge} \sim 0.4$\,kpc. We note, however, that in several galaxies $R_\mathrm{e,Bulge}$ is comparable to the full width at half maximum of the point-spread function, rendering the bulge Sérsic index and ellipticity poorly constrained. We therefore refrain from discussing these parameters further.

The second row of Fig.\,\ref{fig:morph} shows the distributions of disk effective radii and ellipticities. In contrast to the bulges, the disks of 3C-MCGs are significantly more compact, with a median $R_{\mathrm{e,Disk}} \simeq 1.9$~kpc, compared to $3.3$~kpc for 3C-CSGs. The $\epsilon_{\mathrm{Disk}}$ distributions also differ markedly, with 3C-MCGs exhibiting systematically lower disk ellipticities. We emphasize, however, that these values correspond to projected rather than intrinsic ellipticities. Given the inclination bias discussed above for the MCG sample, strong conclusions should therefore be avoided. Nonetheless, the absence of 3C-MCGs with $\epsilon_{\mathrm{Disk}} > 0.8$ raises the possibility that 3C-MCG disks are intrinsically thicker than those of 3C-CSGs.

Envelope structural parameters are shown in the third row of Fig.\,\ref{fig:morph}. While the two samples have comparable envelope sizes (median $R_\mathrm{e,Envelope} = 6.4$\,kpc for 3C-MCGs and $5.8$\,kpc for 3C-CSGs), their ellipticities and Sérsic indices differ. 3C-MCG envelopes tend to be rounder (median $\epsilon = 0.28$ versus $0.45$) and have higher Sérsic indices than those of 3C-CSGs. In particular, 3C-CSGs typically show $n_\mathrm{Envelope} \sim 0.5$--$1.0$, whereas 3C-MCGs span a broader range, $n_\mathrm{Envelope} \sim 0.5$--$3.0$. We caution, however, that $n_\mathrm{Envelope}$ is only weakly constrained. For instance, fixing $n_\mathrm{Envelope}=1.0$ leads to only small increases in the reduced $\chi^2$ and changes of a few percent in the other structural parameters. Deeper observations will be required to place tighter constraints on $n_\mathrm{Envelope}$.

The fourth row of Fig.\,\ref{fig:morph} shows the flux-to-total ratios of the individual structural components. 3C-MCGs exhibit an approximately equal division of flux among bulge, disk, and envelope. In contrast, while 3C-CSGs have envelope-to-total ratios similar to those of 3C-MCGs, they show a larger disk-to-total ratio ($\sim 0.49$) and a correspondingly lower bulge-to-total ratio ($\sim 0.17$).

\subsection{Multi-component decomposition: Stellar Mass - Size Relations}

\begin{figure}
\centering
 \includegraphics[width=\columnwidth, trim=50 10 90 05,clip]{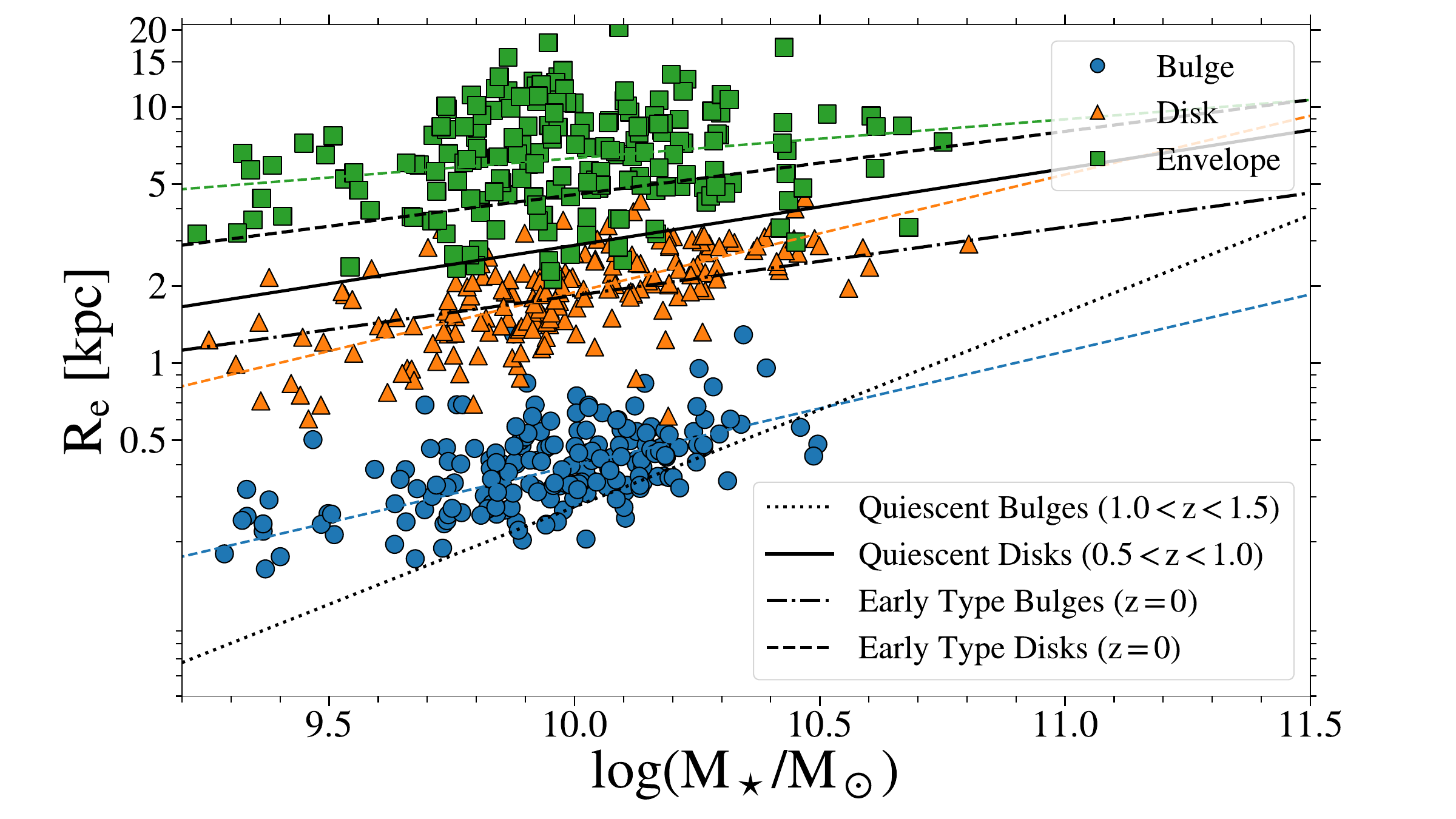}
 \caption{Mass-size relation of the the bulge (blue circles), disk (orange triangles) and envelope (green squares) of 3C-MCGs. Best fitting relations are shown as colored dashed lines. Bulge and disk mass--size relations for $z \sim 0$ early-type disks (dashed black line) and bulges (dot--dashed black line) from \citet{Lange.etal.2016}, and relations for $0.5 < z < 1.0$ quiescent disks (solid black line) and $1.0 < z < 1.5$ quiescent bulges (dotted black line) from \citet{Nedkova.etal.2024} are also shown.}
 \label{fig:mass_size_relation}
\end{figure}

In Fig.\,\ref{fig:mass_size_relation} we present a mass--size diagram for the bulge (blue circles), disk (orange triangles), and envelope (green squares) components of 3C-MCGs. The best-fitting relations for each component are shown as dashed lines. For comparison, we include bulge and disk mass--size relations for $z \sim 0$ early-type disks (dashed black line) and bulges (dot--dashed black line) from \citet{Lange.etal.2016}, as well as relations for $0.5 < z < 1.0$ quiescent disks (solid black line) and $1.0 < z < 1.5$ quiescent bulges (dotted black line) from \citet{Nedkova.etal.2024}. The stellar mass of each structural component was estimated by assuming a constant mass-to-light ratio across the galaxy and scaling the total stellar mass by the corresponding flux fraction of each component.

Both the bulge and disk components exhibit strong mass--size correlations, with Spearman rank coefficients of $r=0.52$ and $r=0.66$, respectively, and $p$-values $< 10^{-10}$. In contrast, the envelope component shows only a weak correlation, characterized by large scatter and marginal statistical significance (Spearman $r=0.20$, $p$-$\mathrm{value}=0.01$). A comparison with the mass--size relations of local early-type disks \citep{Lange.etal.2016} and quenched disks at $z \sim 0.75$ \citep{Nedkova.etal.2024} shows that 3C-MCG disks are systematically smaller than both populations. The bulges of 3C-MCGs are significantly more compact than those of $z \sim 0$ early-type galaxies \citep{Lange.etal.2016}, and are instead comparable to—though slightly larger than—quiescent bulges at $1.0 < z < 1.5$ \citep{Nedkova.etal.2024}.

\subsection{Environment and Morphology}

\begin{figure}
\centering

\parbox{0.9\linewidth}{
\centering
{\large\bfseries Centrals}\\[0.4em]
\includegraphics[width=\linewidth, trim=30 30 30 30, clip]{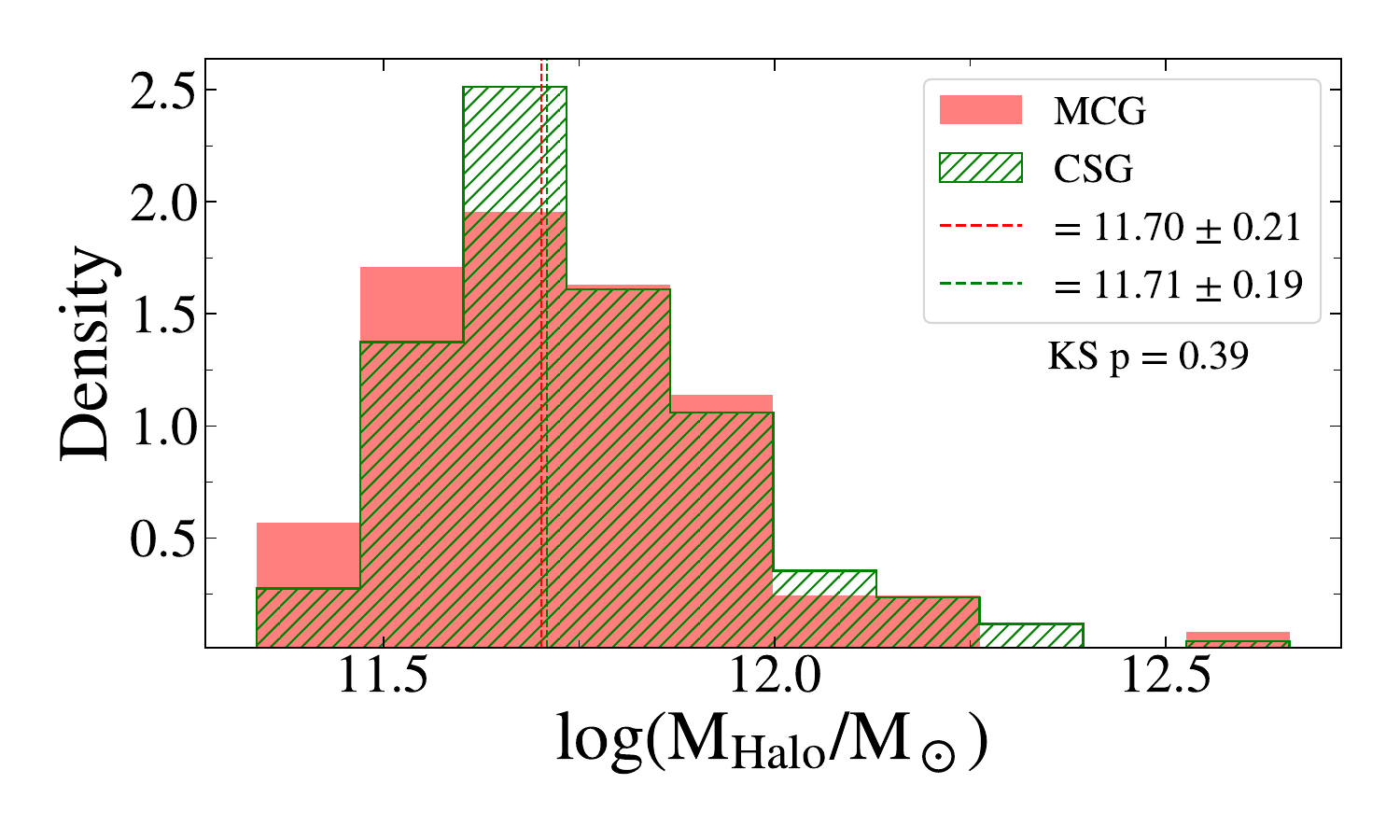}
}

\vspace{0.8em}

\parbox{0.9\linewidth}{
\centering
{\large\bfseries Satellites}\\[0.4em]
\includegraphics[width=\linewidth, trim=30 30 30 30, clip]{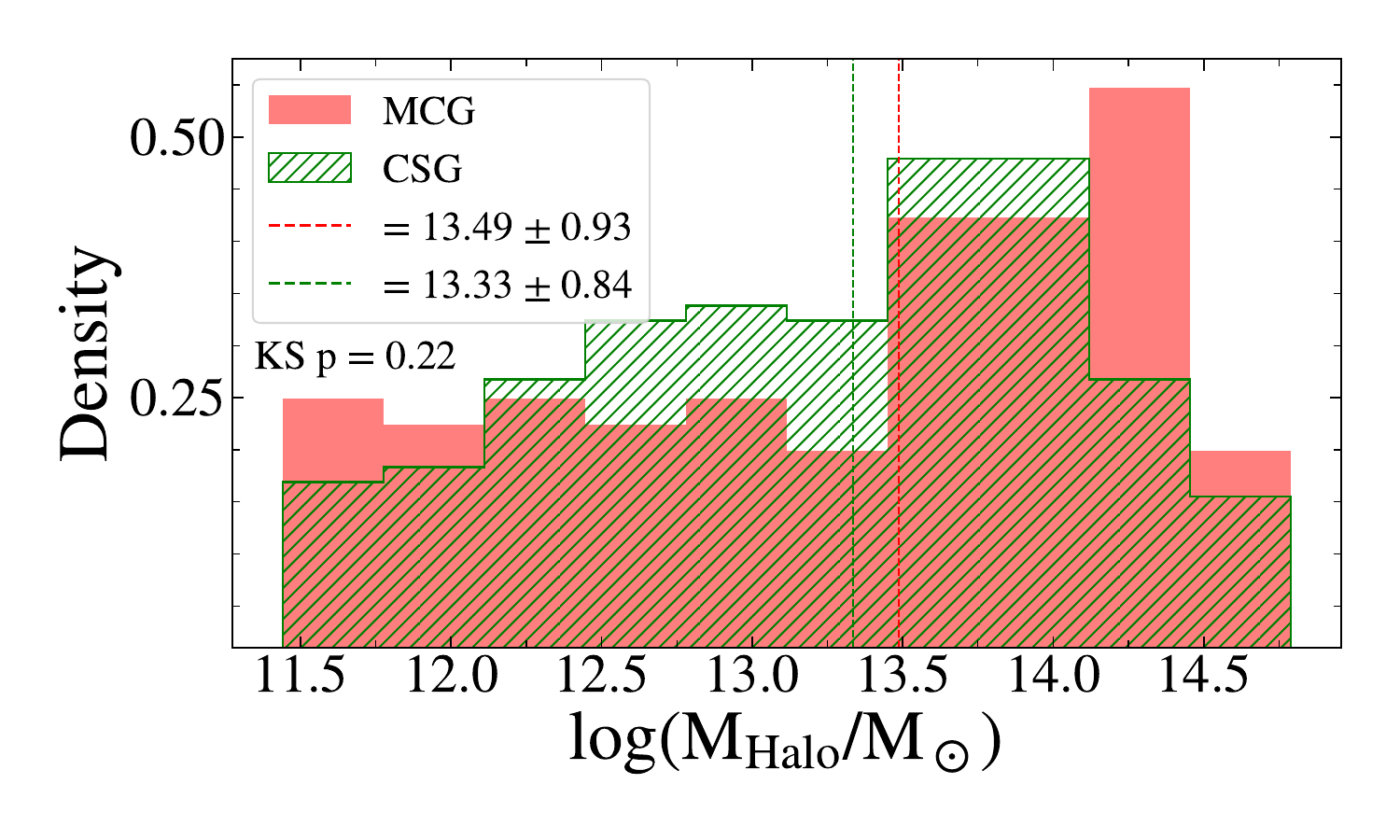}
}

\caption{Dark matter halo masses of MCGs and CSGs. Central galaxies are shown in the top panel, satellites on the bottom panel. Both MCGs and CSGs are found across a variety of environments. Most are central galaxies of low-mass halos or satellites in massive groups and galaxy clusters.}
\label{fig:env}
\end{figure}

\begin{figure*}
    \centering
    \begin{minipage}{0.32\linewidth}
        \centering
        \includegraphics[width=\linewidth, trim=30 30 30 30, clip]{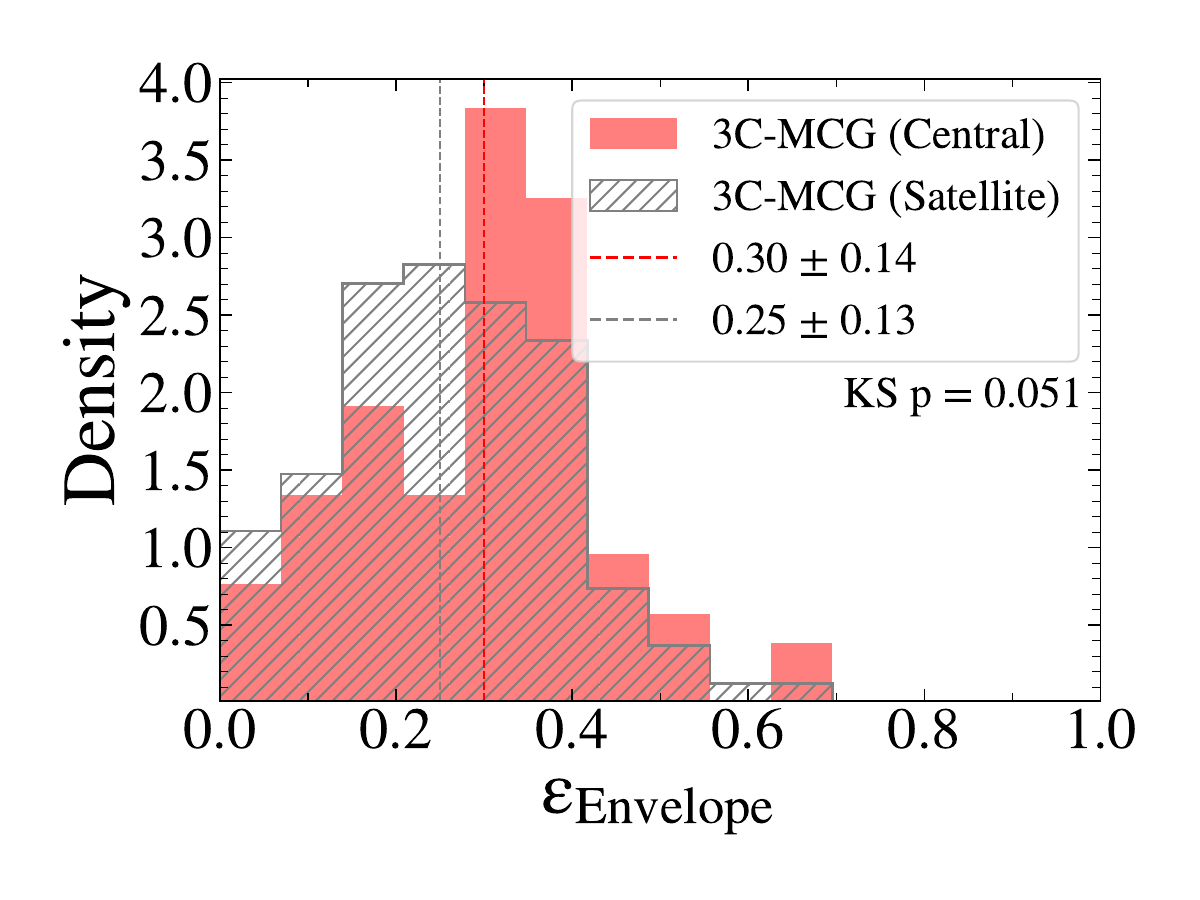}
    \end{minipage}
    \hfill
    \begin{minipage}{0.32\linewidth}
        \centering
        \includegraphics[width=\linewidth, trim=30 30 30 30, clip]{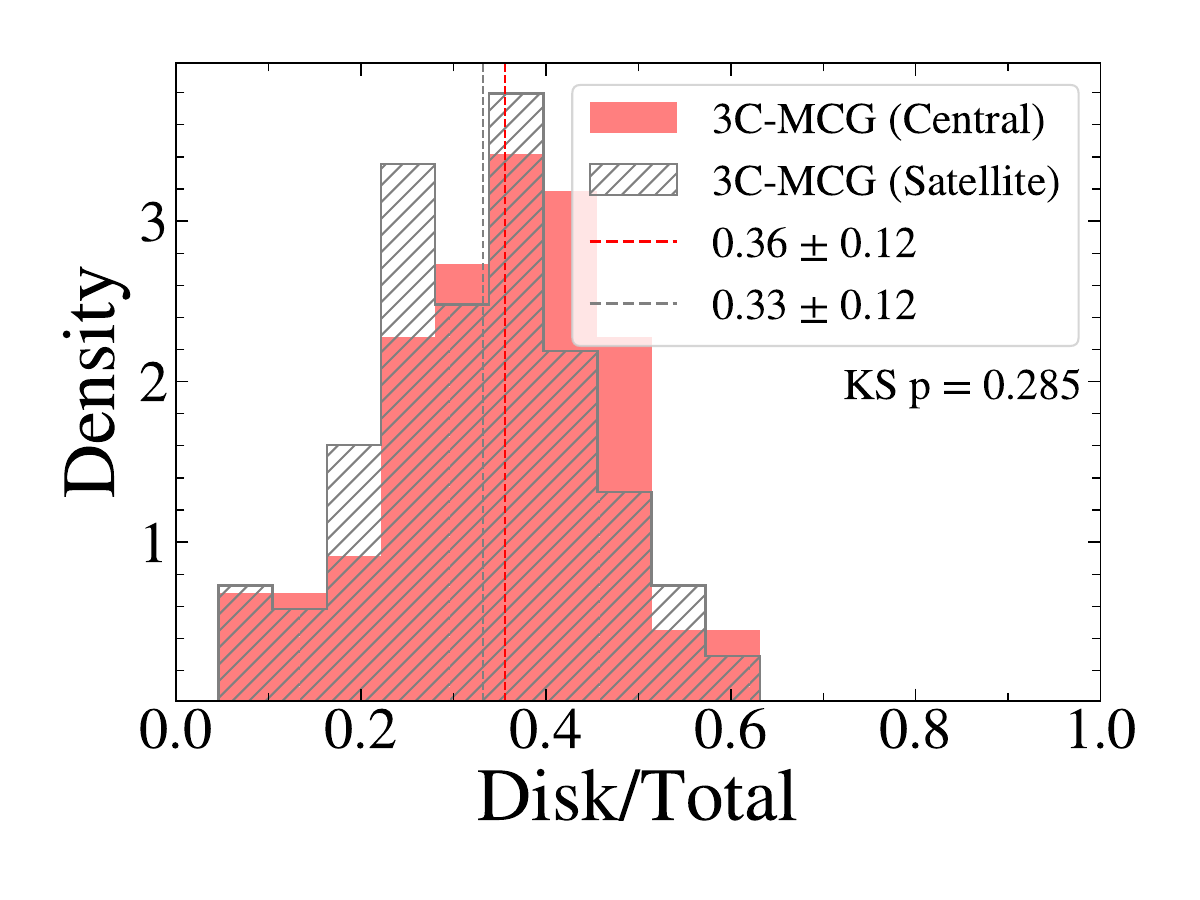}
    \end{minipage}
    \hfill
    \begin{minipage}{0.32\linewidth}
        \centering
        \includegraphics[width=\linewidth, trim=30 30 30 30, clip]{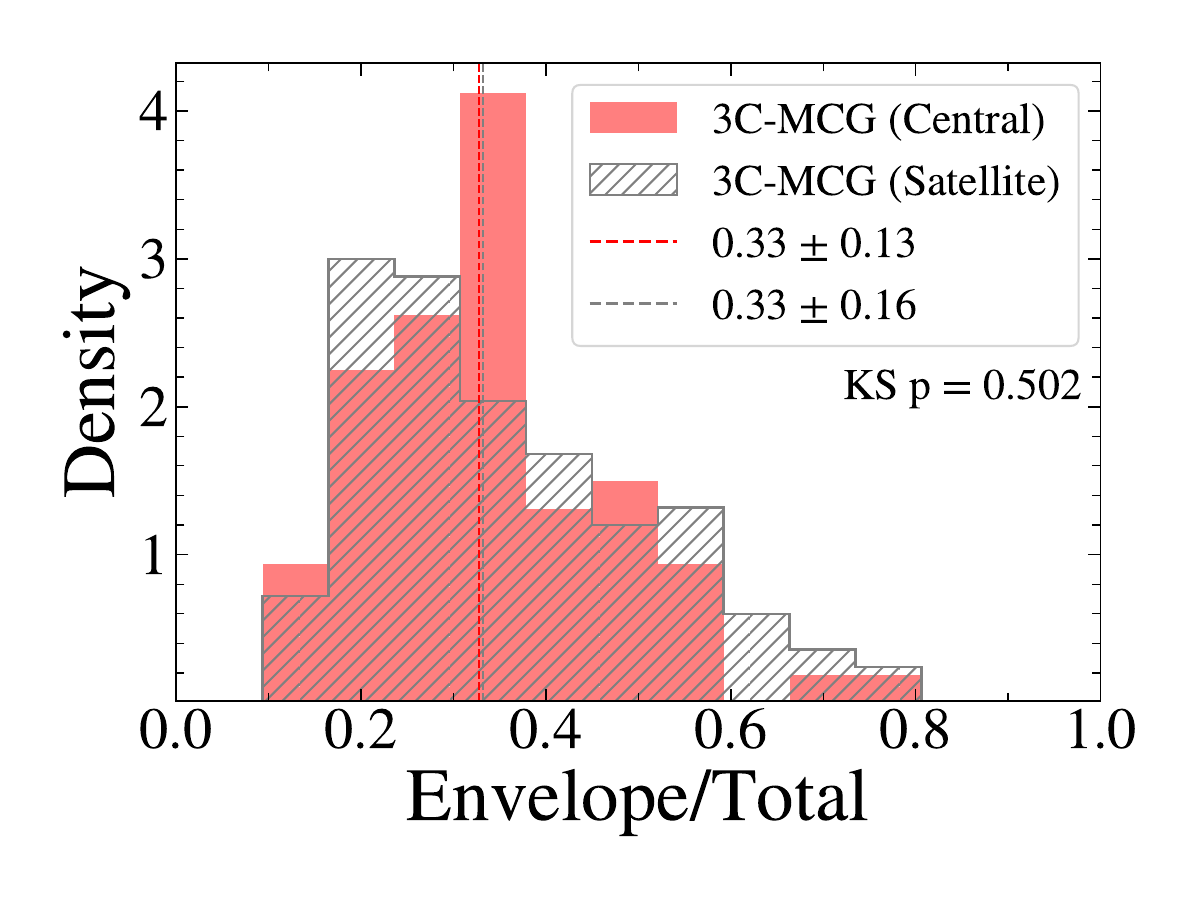}
    \end{minipage}
    \caption{Envelope ellipticity (left panel), disk-to-total (center panel), and envelope-to-total (right panel) flux ratios for central (75 galaxies) and satellite (117 galaxies) 3C-MCGs. Median values are indicated by dashed lines. No statistically significant differences are found between the samples, disfavoring a major role for disk heating by frequent tidal interactions in shaping the outer structure of 3C-MCGs.}
    \label{fig:env_comp3}
\end{figure*}

It is well established that galaxy environment and morphology are correlated, in the sense that galaxies inhabiting denser regions are more likely to be quiescent and to exhibit early-type morphologies \citep{dressler80,kauffmann04,blanton05}. Closely related to the present work is the possibility, raised by several studies \citep{valentinuzzi10,stringer15,peralta16}, that compact quiescent galaxies are preferentially found in galaxy clusters, where galaxy mergers are rare owing to high orbital velocities, thereby suppressing size growth through mergers. Also pertinent is the idea that the envelope component in MCGs could originate from disk heating driven by frequent tidal interactions, as suggested by \citet{kormendy12} in their discussion of S0 galaxies in the Virgo Cluster. While a detailed analysis of the environments of MCGs is beyond the scope of this work and will be presented in a future study, a brief comparison between the environments inhabited by MCGs and CSGs allows us to assess these possibilities.

Figure\,\ref{fig:env} shows the dark matter halo mass distributions of the full MCG and CSG samples, separated into central (top panel) and satellite (bottom panel) galaxies. The majority of both MCGs and CSGs are central galaxies residing in low-mass haloes ($\log (M_\mathrm{halo}/M_\odot) \lesssim 12.5$) accounting for $38\%$ of MCGs and $39\%$ of CSGs. Satellite galaxies span a wide range of halo masses, but are preferentially found in massive groups and galaxy clusters ($\log M_\mathrm{halo}/M_\odot \gtrsim 13.5$), where $23\%$ of MCGs and $19\%$ of CSGs are located. We find no statistically significant differences between the halo mass distributions of the two samples, either for centrals or satellites. This result is consistent with the findings of \citet{tortora20}, who showed that compact and normal-sized quiescent galaxies with $\log M_\star/M_\odot \geq 10.9$ inhabit similar environments. Our analysis extends this conclusion to lower stellar masses, down to $\log M_\star/M_\odot \sim 10.0$.

If frequent tidal interactions play an important role in shaping the morphology of MCGs, then one would expect systematic differences between central and satellite systems. In particular, since central MCGs are frequently isolated, satellite MCGs should exhibit more prominent outer envelopes and lower envelope ellipticities. To test this expectation, in Fig.\,\ref{fig:env_comp3} we compare the envelope ellipticity, as well as the disk-to-total and envelope-to-total flux ratios, between central and satellite 3C-MCGs. We find no statistically significant differences between these two populations, thereby disfavoring a scenario in which disk heating by frequent tidal interactions are the primary driver of envelope formation in 3C-MCGs.

\section{Discussion} \label{sec:discuss}

\subsection{Three-Component Structures in Massive Compact Galaxies and Non-Compact S0s}

Most of the evidence for outer components in S0 galaxies comes from analyses of one-dimensional surface brightness profiles. We briefly summarize the main results of these studies below, before discussing in more detail those that are most pertinent to the analyses presented here.

The surface brightness profiles of many disk galaxies exhibit an abrupt change in slope at large radii. Systems with a steeper outer slope are classified as Type~II (or truncated) disks, whereas those showing an excess of light relative to the outward extrapolation of the inner exponential profile are classified as Type~III (or anti-truncated) disks \citep{erwin05,pohlen06,gutierrez11,maltby15}. Disks that show no significant change in slope are referred to as Type~I. Type~III disks are further divided into two subclasses: Type~III-s disks, in which the outer excess light is associated with a decrease in isophotal ellipticity at large radii, and Type~III-d disks, in which no significant change in ellipticity is observed and the excess is instead interpreted as a disk-related phenomenon. \citet{erwin05} suggested that the surface brightness profiles of Type~III-s disks arise from an inclined disk embedded within a more spheroidal outer component, which may be associated with the bulge or with an additional structural component, such as a stellar halo. The combination of excess light and decreasing ellipticity at large radii observed in Type~III-s disks is qualitatively consistent with the three-component structures identified in the MCG and CSG samples, suggesting that these phenomena are likely related. We adopt this interpretation for the remainder of this discussion, while leaving a direct test of this connection to future work.

Using a sample of 280 S0 galaxies, \citet{maltby15} assessed the prevalence of different disk types and found that $\sim 50\%$ host Type~III disks, with an upper limit of about half of these classified as Type~III-s. Pure exponential (Type~I) disks comprise $\sim 25\%$ of the population, while Type~II disks account for less than $5\%$. The remaining $\sim 20\%$ of galaxies show no discernible exponential component. Thus, at most $\sim 25\%$ of S0 galaxies exhibit a three-component structure broadly comparable to that observed in 3C-MCGs. This fraction is substantially larger than the $7\%$ of three-component systems found in the CSG sample. We caution, however, that these results are not directly comparable: our methodology is tailored to MCGs, in which the envelope contributes a substantial fraction of the total flux, and may therefore miss galaxies with less prominent envelopes.

Nonetheless, even a $25\%$ fraction of three-component CSGs remains significantly lower than the $75\%$ of three-component systems among MCGs (which should be regarded as a lower limit, given that the detectability of the envelope component depends on the inclination, see Fig.\,\ref{fig:ell}). The high prevalence of three-component systems among MCGs therefore points to a remarkable degree of structural homogeneity, in contrast to the pronounced heterogeneity of the CSG sample. This is further supported by the markedly different bar fractions: bars are absent in MCGs, whereas $\sim 29\%$ of CSGs host stellar bars. One might wonder whether the bias of MCGs toward higher inclinations could affect our ability to detect bars. However, in this case we would expect to observe boxy/peanut-shaped bulges. Studies of local barred galaxies have shown that such structures are present in $\sim 79\%$ of galaxies with $\log M_{\star}/M_\odot \gtrsim 10.4$ \citep{erwin17}. Among the MCGs, however, there is only one ambiguous case: the residual map shows a structure resembling a peanut bulge, while the $r$- and $i$-band image shows no clear evidence for the presence of such structure.

In conclusion, the heterogeneity of S0 galaxies—the dominant morphological class among CSGs—is well established and is commonly interpreted as evidence for multiple evolutionary pathways. In contrast, the structural homogeneity of MCGs suggests that their formation is dominated by a single, or at most a small number of, evolutionary channels.

\subsection{The Nature of the Envelope Component: An extension of the Bulge, a Stellar Halo or a Thick Disk?}

\begin{figure}
    \centering
    \includegraphics[width=0.98\linewidth, trim=30 30 30 30, clip]{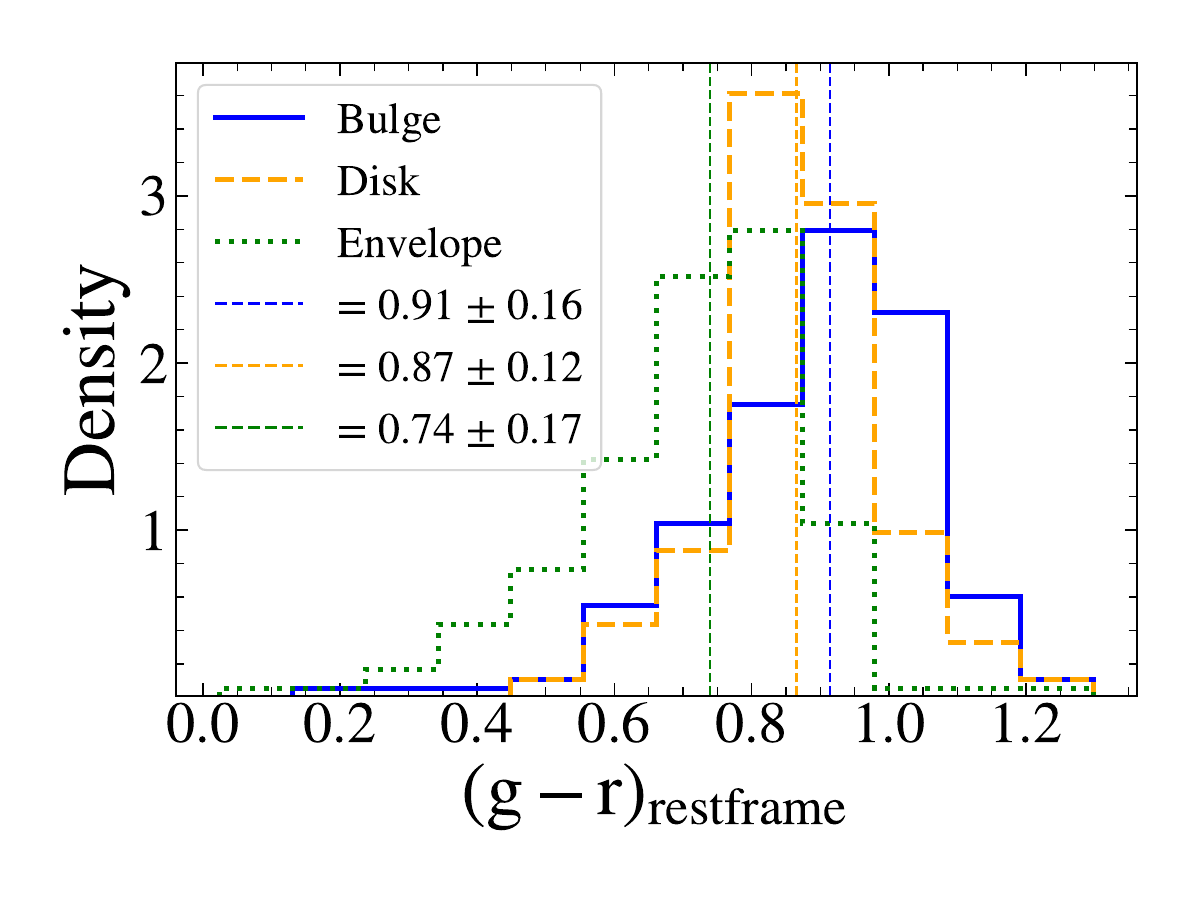}
    \vspace{0.1cm}
    \caption{$(g-r)$ restframe color distribution of the disk (orange), bulge (blue) and envelope (green) components of 3C-MCGs. Median values are indicated by dashed lines. While the bulge and disk exhibit similar distributions, the envelope show bluer colors.}
    \label{fig:mcg_colors}
\end{figure}

\begin{figure*}
    \centering

    \includegraphics[width=0.98\textwidth]{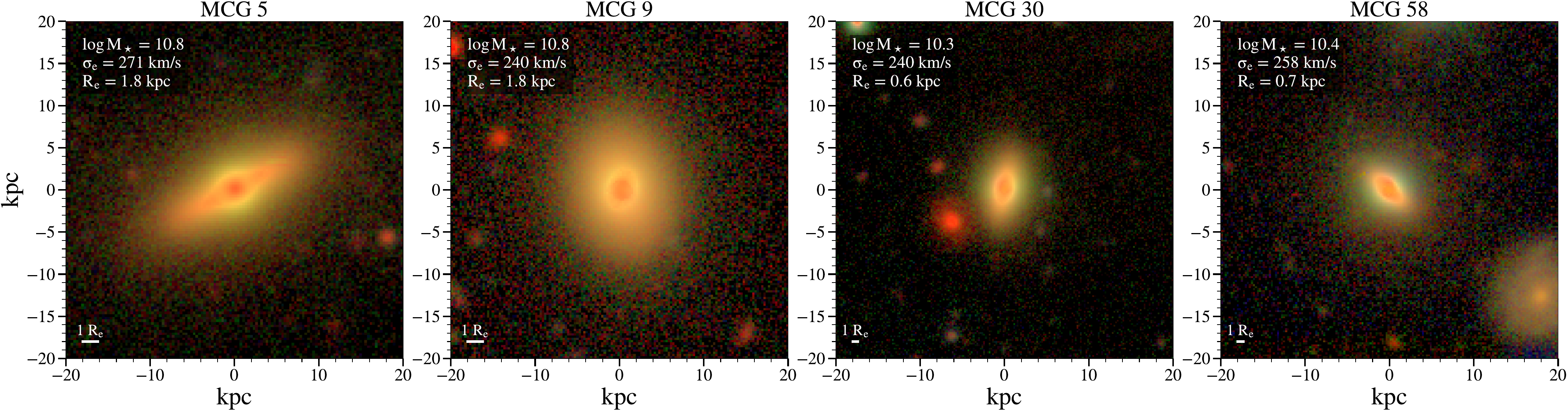}
    \vspace{0.1cm}

    \includegraphics[width=0.98\textwidth]{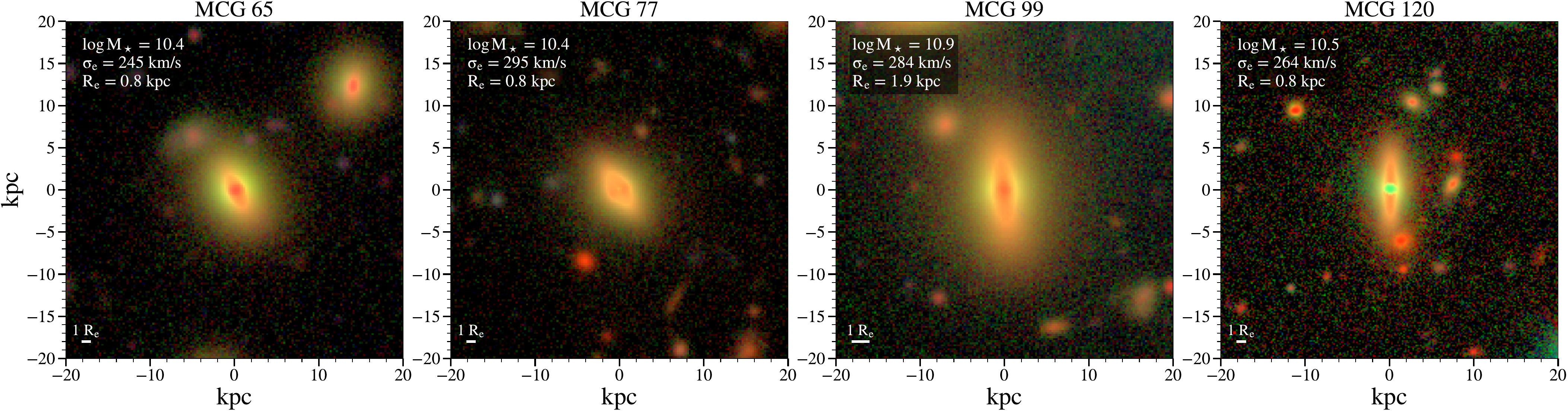}
    \vspace{0.1cm}

    \includegraphics[width=0.5\textwidth]{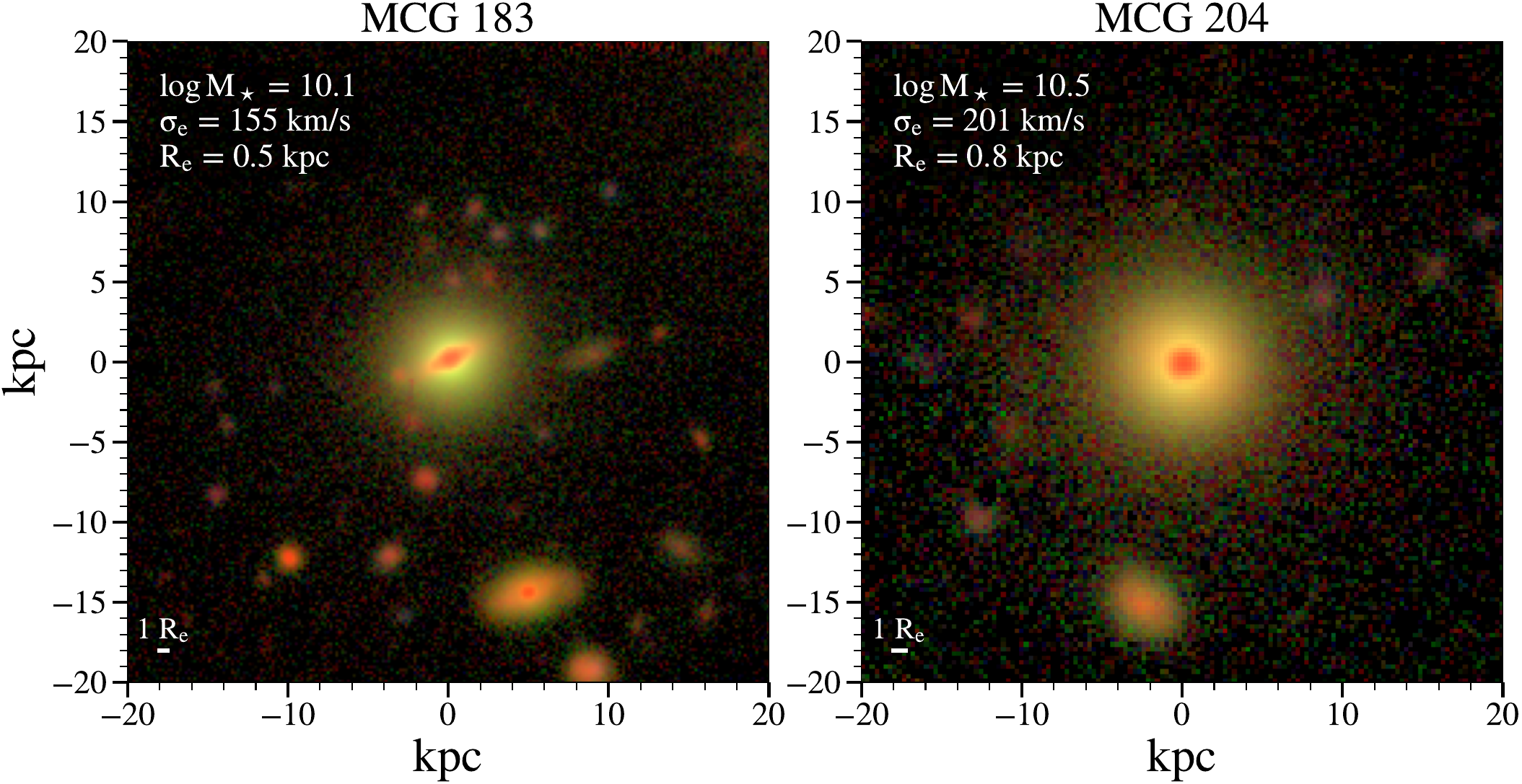}

    \caption{RGB composite images of relic candidates. Except for MCG 9, all are best-fitted by three-component models. These galaxies do not show any significant difference in relation to the parent MCG sample except for their extremely compact sizes.}
    \label{fig:mcg_relics}
\end{figure*}

\begin{table*}
\centering
\caption{Structural parameters of the relic candidate galaxies. Effective velocity dispersions are given in km/s and effective radii in kpc.}
\label{tab:relic_candidates}
\begin{tabular}{rccccccccccccc}
\hline
ID & $\log (M_\star/M_\odot)$ & $\sigma_e$ & $R_e$ & $R_{\mathrm{e,Bulge}}$ & $\epsilon_\mathrm{Disk}$ & $R_{\mathrm{e,Disk}}$ & $\epsilon_\mathrm{Env}$ & $n_\mathrm{Env}$ & $R_{\mathrm{e,Env}}$ & $B/T$ & $D/T$ & $Env/T$ \\
\hline
5   & 10.8 & 271 & 1.8 & 0.6 & 0.8 & 4.0 & 0.3 & 1.1 & 10.6 & 0.41 & 0.40 & 0.18 \\
9   & 10.8 & 240 & 1.8 & 0.6 & 0.3 & 4.3 & --  & --  & --   & 0.50 & 0.50 & -- \\
30  & 10.3 & 240 & 0.6 & 0.4 & 0.6 & 1.5 & 0.3 & 2.0 & 5.2  & 0.60 & 0.22 & 0.17 \\
58  & 10.4 & 258 & 0.7 & 0.2 & 0.7 & 0.9 & 0.3 & 2.5 & 5.9  & 0.23 & 0.57 & 0.20 \\
65  & 10.4 & 245 & 0.8 & 0.5 & 0.6 & 1.4 & 0.3 & 2.1 & 6.6  & 0.46 & 0.32 & 0.23 \\
77  & 10.4 & 295 & 0.8 & 0.3 & 0.7 & 1.3 & 0.2 & 1.4 & 4.7  & 0.43 & 0.42 & 0.15 \\
99  & 10.9 & 284 & 1.9 & 0.6 & 0.7 & 2.7 & 0.3 & 2.9 & 6.7  & 0.29 & 0.35 & 0.36 \\
120 & 10.5 & 264 & 0.9 & 0.3 & 0.8 & 1.9 & 0.4 & 2.0 & 12.6 & 0.37 & 0.39 & 0.24 \\
183 & 10.1 & 155 & 0.5 & 0.3 & 0.8 & 0.7 & 0.0 & 3.1 & 2.4  & 0.48 & 0.23 & 0.29 \\
204 & 10.5 & 201 & 0.8 & 0.3 & 0.0 & 1.4 & 0.0 & 8.0 & 5.6  & 0.38 & 0.22 & 0.39 \\
226 & 10.9 & 240 & 2.3 & 0.5 & 0.7 & 4.4 & 0.3 & 7.6 & 9.4  & 0.15 & 0.41 & 0.45 \\
\hline
\end{tabular}
\end{table*}

\citet{erwin05} suggested that the outer, rounder component observed in Type~III-s disks could correspond to an extended bulge or a stellar halo. By analogy, one may ask whether the envelope component identified in MCGs could likewise represent an extension of the bulge. A definitive answer to this question will likely require detailed dynamical studies; however, several arguments disfavor (but do not rule out) this interpretation.

First, in Fig.\,\ref{fig:mcg_colors} we compare the rest-frame $(g-r)$ colors of the bulge, disk, and envelope components. While the bulge and disk exhibit similar color distributions, the envelope is systematically bluer, with a median offset of $\sim 0.15$\,mag. Considering that color gradients in the outer regions of early type galaxies are interpreted as driven mainly by metallicity gradients \citep{labarbera12}, our results imply a more metal-poor envelope. This is not surprising as, when compared to the thin disk component, the stellar halos and thick disks of early-type galaxies show lower metallicities \citep{labarbera12,greene15,comeron16,pinna19a,pinna19b}.

Second, as discussed previously, two-component models fail to reproduce the observed radial variation in ellipticity. Interpreting the envelope as an extended bulge would therefore require a bulge whose ellipticity varies significantly with radius, which is a relatively contrived scenario. By contrast, interpreting the envelope as either a stellar halo or thick disk naturally accounts for the observed ellipticity gradients.

Further support for this interpretation comes from an analogy with NGC\,4594 (the ``Sombrero Galaxy''). \citet{Gadotti.etal.2012} reported a three-component structure in this system that closely resembles that observed in MCGs. After exploring several models, including a two-component configuration with an $n=4$ bulge of varying ellipticity, they concluded that NGC\,4594 is best described by a three-component model consisting of a flattened compact bulge, an inclined disk, and an extended outer halo.

Whether the envelope is predominantly a thick disk or a stellar halo cannot be assessed with our current data, beyond stating that both components are likely present given the broad distribution of $\epsilon_\mathrm{Envelope}$. Discriminating between these possibilities requires constraints on the intrinsic ellipticity, which can only be obtained for edge-on systems. We therefore defer a more detailed investigation of the nature of the envelope to the second paper in this series, where we carry out a thorough analysis of the subset of MCGs with edge-on inclinations.

\subsection{Parallels Between the Morphology and Kinematics of Compact Quiescent Galaxies at Low and High Redshifts}

We find that $96\%$ of MCGs require two or more structural components to achieve satisfactory fits, demonstrating the ubiquity of disks in these systems. This result is consistent with kinematic studies of smaller samples of massive compact galaxies at $z \sim 0$, which find them to be predominantly fast rotators \citep{yildirim17,Schnorr.et.al.2021}. Extending these works, we show that disks contribute a substantial fraction of the total flux, with a median disk-to-total ratio of $0.35$ and a standard deviation of $0.15$.

Are disks similarly ubiquitous in compact quiescent galaxies at high redshift? Morphological studies of quiescent galaxies at $z \gtrsim 1$ have reached apparently conflicting conclusions. Based on ellipticity distributions, \citet{vanderwel12} argued that the majority of quiescent galaxies at $z \sim 2$ are disk-dominated, a result later supported by \citet{Hill.etal.2019}. In contrast, \citet{lustig21} suggested that quiescent galaxies at $z \sim 3$ are predominantly bulge-dominated, based on the high Sérsic indices and low ellipticities (median values of $\sim 4.5$ and $0.27$, respectively) derived from single-Sérsic fits to a sample of ten galaxies.

While resolved kinematic studies could clarify the picture emerging from these morphological studies, the number of quiescent galaxies at $z \gtrsim 1$ with spatially resolved kinematics remains too small to draw firm conclusions about the prevalence of disks or the degree of rotational support. Nevertheless, most systems with available kinematic measurements are classified as fast rotators, albeit with lower rotational support than star-forming galaxies of comparable mass \citep{newman18,slob25}. If disks are indeed common in these systems, this implies that they are dynamically hotter than the disks of their star-forming counterparts.

Assuming that high-redshift quiescent galaxies are structurally similar to the MCGs in our sample -- a fair assumption given that in \citet{Clerici.etal.2024} we have shown that MCGs and $z \sim 1.5$ quiescent galaxies occupy similar regions of the $\log M_{\star}$--$\log \sigma_{\mathrm{e}}$ and $\log M_{\star}$--$\log R_{\mathrm{e}}$ planes (see ther Fig.\,15) -- the seemingly discrepant results described above can be naturally reconciled. First, the outer isophotes of MCGs predominantly trace the stellar envelope, which exhibits a broad range of projected ellipticities, typically $\epsilon_{\mathrm{Envelope}} \sim 0$--$0.4$ with a median of $0.28$, although a small number of galaxies reach $\epsilon_{\mathrm{Envelope}} \sim 0.5$--$0.6$. When combined with the small sample sizes of high-redshift studies, such a broad distribution can lead to divergent conclusions based on ellipticity alone. Second, the ubiquity of disks in MCGs naturally implies a large fraction of fast rotators among high-redshift compact quiescent galaxies. Finally, since the bulge and envelope components together dominate the flux budget in most galaxies and they are kinematically hotter than the disk, a reduced level of rotational support relative to disk-dominated star-forming galaxies is expected.

\subsection{Relic Galaxy Candidates}

Relic galaxies are compact quiescent systems at low redshifts ($z < 1$) that formed early in the history of the Universe ($z_{\mathrm{form}} \gtrsim 2$) and have since evolved passively. As a result, their morphological and structural properties are expected to have remained largely unchanged since the quenching of their star formation. There is no single, universally accepted definition of relic galaxies; however, it is generally agreed that they should (i) have formed the large majority of their mass before $z = 2$, (ii) are extremely compact, having sizes consistent with or smaller than the $z \sim 2$ size--mass relation (with some studies adopting even more stringent size thresholds), and (iii) exhibit morphological and kinematic properties indicative of a quiet accretion history since quenching \citep{ferre-mateu15,ferre-mateu17,spiniello21}.

The SDSS spectra of MCGs cover, on average, $1.4 \pm 0.5\,R_e$ (or $2.1 \pm 0.5$\,kpc), probing only the bulge and the inner parts of the disk and generally not reaching the envelope-dominated regions. This prevents us from assessing whether the stellar populations are uniformly very old throughout the entire galaxy. Nonetheless, by combining the structural properties presented in this work with the mass-weighted stellar population ages reported by \citet{Clerici.etal.2024}, we can construct a sample of extremely compact MCGs with undisturbed morphologies and very old central regions, relic galaxy candidates to be targeted for follow-up studies. 

We identify relic galaxy candidates within the MCG sample by requiring mass-weighted stellar population ages older than 10\,Gyr in the region probed by SDSS spectra, effective radii below the compactness criterion of \citet{vanderwel.etal.2014} ($R_\mathrm{e} /[M_\star/10^{11}\,M_\odot]^{0.75} < 2.5~\mathrm{kpc}$) and the absence of morphological disturbances, even at low surface brightness. A total of $90\%$ of MCGs satisfy the age criterion, of which only 13 have sizes smaller than the compactness threshold. Two of these systems show clear signs of interactions or morphological disturbances and are therefore excluded, leaving a final sample of 11 relic galaxy candidates. RGB colour images of these systems are shown in Fig.\,\ref{fig:mcg_relics}, with the exception of MCG~226, for which $g$-band imaging is unavailable. The main properties of the relic candidates are listed in Tab.\,\ref{tab:relic_candidates}.

Morphologically, the relic candidates show no systematic differences relative to the parent MCG sample, aside from their smaller effective radii. With the exception of MCG~9, all relic candidates are best described by three-component models. The relic candidates also span a wide range of environments, similar to the full MCG sample. Two are located in galaxy clusters (MCGs~65 and 99), two reside in groups (MCGs~9 and 30), one is part of a galaxy pair (MCG~5), and the remaining five are listed as having no companions in the \citet{lim17} catalog.

The lack of systematic differences in both morphological and environmental properties between relic candidates and the parent MCG sample complements the absence of differences in stellar population properties reported by \citet{Clerici.etal.2024} (see their Appendix~B). Together, these results suggest that a galaxy’s position in the $\log M_\star$--$\sigma_e$ plane is more tightly linked to its morphological and stellar population properties than its position in the $\log M_\star$--$R_\mathrm{e}$ plane. A more thorough comparison between MCGs and relic galaxies will be presented in the second paper of this series.

Finally, we note that eight of the relic candidates have stellar masses $\log M_\star/M_\odot \leq 10.5$. This mass regime is rarely explored in relic galaxy studies, and our results raise the possibility that relic galaxies also exist at these lower stellar masses.

%***https://ui.adsabs.harvard.edu/abs/2024ApJ...974..135J/abstract***
\section{Conclusions} \label{sec:conclusions}

In this work, we investigated the structural properties of massive compact quiescent galaxies (MCGs) derived from multi-component photometric decompositions, and compared them with those of a control sample of average-sized quiescent galaxies (CSGs) matched in stellar mass, star formation rate, redshift, and $g$--$i$ colour. Our main conclusions can be summarized as follows:

\begin{itemize}
    \item Galaxies in both the MCG and CSG samples are predominantly classified as S0s, accounting for $93\%$ and $71\%$ of the samples, respectively. Ellipticals represent $4\%$ of MCGs and $11\%$ of CSGs, while $18\%$ of CSGs are classified as late-type galaxies;
    \item The fraction of interacting or morphologically disturbed galaxies is low in both samples, amounting to $13\%$ for MCGs and $16\%$ for CSGs;
    \item Multi-component photometric decompositions reveal that MCGs are predominantly three-component systems. This structure is characterized by a decrease in ellipticity at large radii, resulting from an inclined disk embedded within a low--surface-brightness envelope of lower ellipticity. Specifically, $75\%$ of MCGs require a three-component model, while $21\%$ are well described by two components and $4\%$ by a single Sérsic profile. We show that two-component systems generally have lower inclinations, suggesting that the true fraction of three-component MCGs is likely higher;
    \item In contrast, only $7\%$ of CSGs exhibit a comparable three-component structure. This difference cannot be explained by inclination effects alone, as several highly inclined CSGs are nonetheless well fitted by two-component models. Another significant difference between the samples is the prevalence of bars: while $29\%$ of CSGs host stellar bars, none are detected in MCGs;
    \item Three-component MCGs and CSGs have similar bulge ($R_\mathrm{e} \simeq 0.39$\,kpc vs.\ $0.45$\,kpc) and envelope ($R_\mathrm{e} \simeq 6.4$\,kpc vs.\ $5.8$\,kpc) effective radii. In contrast, the disks of MCGs are significantly more compact, with median effective radii of $1.9$\,kpc compared to $3.3$\,kpc for CSGs;    
    \item The envelope-to-total flux ratios of MCGs and CSGs are similar ($0.33$ vs.\ $0.29$). However, MCGs exhibit higher bulge-to-total flux ratios ($0.33$ vs.\ $0.17$) and correspondingly lower disk-to-total ratios ($0.35$ vs.\ $0.49$);    
    \item The nature of the envelope component remains uncertain. Its broad ellipticity distribution ($\epsilon_\mathrm{Envelope} \sim 0.0$--$0.6$) suggests that it may correspond to a stellar halo in some systems and to a thick disk in others. Given that the envelope is systematically bluer than both the bulge and disk by $\sim 0.15$\,mag, we argue that it is unlikely to represent an extension of the bulge. Further insight will require assessing the intrinsic envelope ellipticity through the study of edge-on systems;
    \item MCGs are found in a wide range of environments, most commonly as central galaxies in low-mass haloes (38\%) or as satellites in massive groups and galaxy clusters (23\%). We find no statistically significant differences between the environments inhabited by MCGs and CSGs;
    \item Finally, we identify a sample of 11 MCGs that qualify as relic galaxy candidates, showing no signs of morphological disturbance and satisfying the strict compactness criterion of \citet{vanderwel.etal.2014}.
\end{itemize}

\section*{Acknowledgements}

KSC acknowledges the Coordination for the Improvement of Higher Education Personnel (CAPES) for the financial support
(88887.629089/2021-00). ASM acknowledges the financial support from the Brazilian National Council for Scientific and Technological Development (CNPq) and from the Fundação de Amparo à Pesquisa do Estado do Rio Grande do Sul (FAPERGS). ACSM acknowledges support from the European Southern Observatory (ESO) as an SCV visitor at the ESO Science Office in Vitacura, as well as financial support from CAPES (process no. 88887.001289/2024-00). RMD acknowledges the financial support from CNPq (132927/2025-0).

\bibliography{sample701}{}

@ARTICLE{erwin05,
       author = {{Erwin}, Peter and {Beckman}, John E. and {Pohlen}, Michael},
        title = "{Antitruncation of Disks in Early-Type Barred Galaxies}",
      journal = {\apjl},
         year = 2005,
        month = jun,
       volume = {626},
       number = {2},
        pages = {L81-L84},
          doi = {10.1086/431739},
archivePrefix = {arXiv},
       eprint = {astro-ph/0505216},
 primaryClass = {astro-ph},
       adsurl = {https://ui.adsabs.harvard.edu/abs/2005ApJ...626L..81E}
}

@ARTICLE{pinna19a,
       author = {{Pinna}, F. and {Falc{\'o}n-Barroso}, J. and {Martig}, M. and {Sarzi}, M. and {Coccato}, L. and {Iodice}, E. and {Corsini}, E.~M. and {de Zeeuw}, P.~T. and {Gadotti}, D.~A. and {Leaman}, R. and {Lyubenova}, M. and {McDermid}, R.~M. and {Minchev}, I. and {Morelli}, L. and {van de Ven}, G. and {Viaene}, S.},
        title = "{The Fornax 3D project: Unveiling the thick disk origin in FCC 170; possible signs of accretion}",
      journal = {\aap},
         year = 2019,
        month = mar,
       volume = {623},
          eid = {A19},
        pages = {A19},
          doi = {10.1051/0004-6361/201833193},
archivePrefix = {arXiv},
       eprint = {1901.04310},
 primaryClass = {astro-ph.GA},
       adsurl = {https://ui.adsabs.harvard.edu/abs/2019A&A...623A..19P}
}

@ARTICLE{Greene15,
       author = {{Greene}, Jenny E. and {Janish}, Ryan and {Ma}, Chung-Pei and {McConnell}, Nicholas J. and {Blakeslee}, John P. and {Thomas}, Jens and {Murphy}, Jeremy D.},
        title = "{The MASSIVE Survey. II. Stellar Population Trends Out to Large Radius in Massive Early-type Galaxies}",
      journal = {\apj},
         year = 2015,
        month = jul,
       volume = {807},
       number = {1},
          eid = {11},
        pages = {11},
          doi = {10.1088/0004-637X/807/1/11},
archivePrefix = {arXiv},
       eprint = {1504.02483},
 primaryClass = {astro-ph.GA},
       adsurl = {https://ui.adsabs.harvard.edu/abs/2015ApJ...807...11G}
}

@ARTICLE{buta95,
       author = {{Buta}, R.},
        title = "{The Catalog of Southern Ringed Galaxies}",
      journal = {\apjs},
         year = 1995,
        month = jan,
       volume = {96},
        pages = {39},
          doi = {10.1086/192113},
       adsurl = {https://ui.adsabs.harvard.edu/abs/1995ApJS...96...39B}
}

@ARTICLE{ferre-mateu15,
       author = {{Ferr{\'e}-Mateu}, Anna and {Mezcua}, Mar and {Trujillo}, Ignacio and {Balcells}, Marc and {van den Bosch}, Remco C.~E.},
        title = "{Massive Relic Galaxies Challenge the Co-evolution of Super-massive Black Holes and Their Host Galaxies}",
      journal = {\apj},
         year = 2015,
        month = jul,
       volume = {808},
       number = {1},
          eid = {79},
        pages = {79},
          doi = {10.1088/0004-637X/808/1/79},
archivePrefix = {arXiv},
       eprint = {1506.02663},
 primaryClass = {astro-ph.GA},
       adsurl = {https://ui.adsabs.harvard.edu/abs/2015ApJ...808...79F}
}

@BOOK{vaucouleurs91,
       author = {{de Vaucouleurs}, Gerard and {de Vaucouleurs}, Antoinette and {Corwin}, Jr., Herold G. and {Buta}, Ronald J. and {Paturel}, Georges and {Fouque}, Pascal},
        title = "{Third Reference Catalogue of Bright Galaxies}",
         year = 1991,
       adsurl = {https://ui.adsabs.harvard.edu/abs/1991rc3..book.....D}
}

@ARTICLE{deVaucouleurs59,
       author = {{de Vaucouleurs}, Gerard},
        title = "{Classification and Morphology of External Galaxies.}",
      journal = {Handbuch der Physik},
         year = 1959,
        month = jan,
       volume = {53},
        pages = {275},
          doi = {10.1007/978-3-642-45932-0_7},
       adsurl = {https://ui.adsabs.harvard.edu/abs/1959HDP....53..275D}
}

@ARTICLE{comeron16,
       author = {{Comer{\'o}n}, S. and {Salo}, H. and {Peletier}, R.~F. and {Mentz}, J.},
        title = "{A monolithic collapse origin for the thin and thick disc structure of the S0 galaxy <ASTROBJ>ESO 243-49</ASTROBJ>}",
      journal = {\aap},
         year = 2016,
        month = sep,
       volume = {593},
          eid = {L6},
        pages = {L6},
          doi = {10.1051/0004-6361/201629292},
archivePrefix = {arXiv},
       eprint = {1608.04238},
 primaryClass = {astro-ph.GA},
       adsurl = {https://ui.adsabs.harvard.edu/abs/2016A&A...593L...6C}
}

@ARTICLE{pinna19b,
       author = {{Pinna}, F. and {Falc{\'o}n-Barroso}, J. and {Martig}, M. and {Coccato}, L. and {Corsini}, E.~M. and {de Zeeuw}, P.~T. and {Gadotti}, D.~A. and {Iodice}, E. and {Leaman}, R. and {Lyubenova}, M. and {Mart{\'\i}n-Navarro}, I. and {Morelli}, L. and {Sarzi}, M. and {van de Ven}, G. and {Viaene}, S. and {McDermid}, R.~M.},
        title = "{The Fornax 3D project: Thick disks in a cluster environment}",
      journal = {\aap},
         year = 2019,
        month = may,
       volume = {625},
          eid = {A95},
        pages = {A95},
          doi = {10.1051/0004-6361/201935154},
archivePrefix = {arXiv},
       eprint = {1904.01260},
 primaryClass = {astro-ph.GA},
       adsurl = {https://ui.adsabs.harvard.edu/abs/2019A&A...625A..95P}
}

@ARTICLE{labarbera12,
       author = {{La Barbera}, F. and {Ferreras}, I. and {de Carvalho}, R.~R. and {Bruzual}, G. and {Charlot}, S. and {Pasquali}, A. and {Merlin}, E.},
        title = "{SPIDER - VII. Revealing the stellar population content of massive early-type galaxies out to 8R$_{e}$}",
      journal = {\mnras},
         year = 2012,
        month = nov,
       volume = {426},
       number = {3},
        pages = {2300-2317},
          doi = {10.1111/j.1365-2966.2012.21848.x},
archivePrefix = {arXiv},
       eprint = {1208.0587},
 primaryClass = {astro-ph.CO},
       adsurl = {https://ui.adsabs.harvard.edu/abs/2012MNRAS.426.2300L}
}

@ARTICLE{erwin17,
       author = {{Erwin}, Peter and {Debattista}, Victor P.},
        title = "{The frequency and stellar-mass dependence of boxy/peanut-shaped bulges in barred galaxies}",
      journal = {\mnras},
         year = 2017,
        month = jun,
       volume = {468},
       number = {2},
        pages = {2058-2080},
          doi = {10.1093/mnras/stx620},
archivePrefix = {arXiv},
       eprint = {1703.01602},
 primaryClass = {astro-ph.GA},
       adsurl = {https://ui.adsabs.harvard.edu/abs/2017MNRAS.468.2058E}
}

@ARTICLE{bruce12,
       author = {{Bruce}, V.~A. and {Dunlop}, J.~S. and {Cirasuolo}, M. and
         {McLure}, R.~J. and {Targett}, T.~A. and {Bell}, E.~F. and
         {Croton}, D.~J. and {Dekel}, A. and {Faber}, S.~M. and
         {Ferguson}, H.~C. and {Grogin}, N.~A. and {Kocevski}, D.~D. and
         {Koekemoer}, A.~M. and {Koo}, D.~C. and {Lai}, K. and {Lotz}, J.~M. and
         {McGrath}, E.~J. and {Newman}, J.~A. and {van der Wel}, A.},
        title = "{The morphologies of massive galaxies at 1 \&lt; z \&lt; 3 in the CANDELS-UDS field: compact bulges, and the rise and fall of massive discs}",
      journal = {\mnras},
         year = 2012,
        month = dec,
       volume = {427},
       number = {2},
        pages = {1666-1701},
          doi = {10.1111/j.1365-2966.2012.22087.x},
archivePrefix = {arXiv},
       eprint = {1206.4322},
 primaryClass = {astro-ph.CO},
       adsurl = {https://ui.adsabs.harvard.edu/abs/2012MNRAS.427.1666B}
}

@ARTICLE{conselice14,
       author = {{Conselice}, Christopher J.},
        title = "{The Evolution of Galaxy Structure Over Cosmic Time}",
      journal = {\araa},
         year = 2014,
        month = aug,
       volume = {52},
        pages = {291-337},
          doi = {10.1146/annurev-astro-081913-040037},
archivePrefix = {arXiv},
       eprint = {1403.2783},
 primaryClass = {astro-ph.GA},
       adsurl = {https://ui.adsabs.harvard.edu/abs/2014ARA&A..52..291C}
}

@ARTICLE{lim17,
       author = {{Lim}, S.~H. and {Mo}, H.~J. and {Lu}, Yi and {Wang}, Huiyuan and {Yang}, Xiaohu},
        title = "{Galaxy groups in the low-redshift Universe}",
      journal = {\mnras},
         year = 2017,
        month = sep,
       volume = {470},
       number = {3},
        pages = {2982-3005},
          doi = {10.1093/mnras/stx1462},
archivePrefix = {arXiv},
       eprint = {1706.02307},
 primaryClass = {astro-ph.GA},
       adsurl = {https://ui.adsabs.harvard.edu/abs/2017MNRAS.470.2982L}
}

@ARTICLE{gadotti26,
       author = {{Gadotti}, Dimitri A.},
        title = "{Robust galaxy image decompositions with differential evolution optimization and the problem of classical bulges in and beyond the nearby Universe}",
      journal = {\mnras},
         year = 2026,
        month = feb,
       volume = {545},
       number = {4},
          eid = {staf2072},
        pages = {staf2072},
          doi = {10.1093/mnras/staf2072},
archivePrefix = {arXiv},
       eprint = {2511.13823},
 primaryClass = {astro-ph.GA},
       adsurl = {https://ui.adsabs.harvard.edu/abs/2026MNRAS.545S2072G}
}

@ARTICLE{nair10,
       author = {{Nair}, Preethi B. and {Abraham}, Roberto G.},
        title = "{A Catalog of Detailed Visual Morphological Classifications for 14,034 Galaxies in the Sloan Digital Sky Survey}",
      journal = {\apjs},
         year = 2010,
        month = feb,
       volume = {186},
       number = {2},
        pages = {427-456},
          doi = {10.1088/0067-0049/186/2/427},
archivePrefix = {arXiv},
       eprint = {1001.2401},
 primaryClass = {astro-ph.CO},
       adsurl = {https://ui.adsabs.harvard.edu/abs/2010ApJS..186..427N}
}

@ARTICLE{aihara18,
       author = {{Aihara}, Hiroaki and {Armstrong}, Robert and {Bickerton}, Steven and {Bosch}, James and {Coupon}, Jean and {Furusawa}, Hisanori and {Hayashi}, Yusuke and {Ikeda}, Hiroyuki and {Kamata}, Yukiko and {Karoji}, Hiroshi and {Kawanomoto}, Satoshi and {Koike}, Michitaro and {Komiyama}, Yutaka and {Lang}, Dustin and {Lupton}, Robert H. and {Mineo}, Sogo and {Miyatake}, Hironao and {Miyazaki}, Satoshi and {Morokuma}, Tomoki and {Obuchi}, Yoshiyuki and {Oishi}, Yukie and {Okura}, Yuki and {Price}, Paul A. and {Takata}, Tadafumi and {Tanaka}, Manobu M. and {Tanaka}, Masayuki and {Tanaka}, Yoko and {Uchida}, Tomohisa and {Uraguchi}, Fumihiro and {Utsumi}, Yousuke and {Wang}, Shiang-Yu and {Yamada}, Yoshihiko and {Yamanoi}, Hitomi and {Yasuda}, Naoki and {Arimoto}, Nobuo and {Chiba}, Masashi and {Finet}, Francois and {Fujimori}, Hiroki and {Fujimoto}, Seiji and {Furusawa}, Junko and {Goto}, Tomotsugu and {Goulding}, Andy and {Gunn}, James E. and {Harikane}, Yuichi and {Hattori}, Takashi and {Hayashi}, Masao and {He{\l}miniak}, Krzysztof G. and {Higuchi}, Ryo and {Hikage}, Chiaki and {Ho}, Paul T.~P. and {Hsieh}, Bau-Ching and {Huang}, Kuiyun and {Huang}, Song and {Imanishi}, Masatoshi and {Iwata}, Ikuru and {Jaelani}, Anton T. and {Jian}, Hung-Yu and {Kashikawa}, Nobunari and {Katayama}, Nobuhiko and {Kojima}, Takashi and {Konno}, Akira and {Koshida}, Shintaro and {Kusakabe}, Haruka and {Leauthaud}, Alexie and {Lee}, Chien-Hsiu and {Lin}, Lihwai and {Lin}, Yen-Ting and {Mandelbaum}, Rachel and {Matsuoka}, Yoshiki and {Medezinski}, Elinor and {Miyama}, Shoken and {Momose}, Rieko and {More}, Anupreeta and {More}, Surhud and {Mukae}, Shiro and {Murata}, Ryoma and {Murayama}, Hitoshi and {Nagao}, Tohru and {Nakata}, Fumiaki and {Niida}, Mana and {Niikura}, Hiroko and {Nishizawa}, Atsushi J. and {Oguri}, Masamune and {Okabe}, Nobuhiro and {Ono}, Yoshiaki and {Onodera}, Masato and {Onoue}, Masafusa and {Ouchi}, Masami and {Pyo}, Tae-Soo and {Shibuya}, Takatoshi and {Shimasaku}, Kazuhiro and {Simet}, Melanie and {Speagle}, Joshua and {Spergel}, David N. and {Strauss}, Michael A. and {Sugahara}, Yuma and {Sugiyama}, Naoshi and {Suto}, Yasushi and {Suzuki}, Nao and {Tait}, Philip J. and {Takada}, Masahiro and {Terai}, Tsuyoshi and {Toba}, Yoshiki and {Turner}, Edwin L. and {Uchiyama}, Hisakazu and {Umetsu}, Keiichi and {Urata}, Yuji and {Usuda}, Tomonori and {Yeh}, Sherry and {Yuma}, Suraphong},
        title = "{First data release of the Hyper Suprime-Cam Subaru Strategic Program}",
      journal = {\pasj},
         year = 2018,
        month = jan,
       volume = {70},
          eid = {S8},
        pages = {S8},
          doi = {10.1093/pasj/psx081},
archivePrefix = {arXiv},
       eprint = {1702.08449},
 primaryClass = {astro-ph.IM},
       adsurl = {https://ui.adsabs.harvard.edu/abs/2018PASJ...70S...8A}
}

@ARTICLE{simard11,
       author = {{Simard}, Luc and {Mendel}, J. Trevor and {Patton}, David R. and {Ellison}, Sara L. and {McConnachie}, Alan W.},
        title = "{A Catalog of Bulge+disk Decompositions and Updated Photometry for 1.12 Million Galaxies in the Sloan Digital Sky Survey}",
      journal = {\apjs},
         year = 2011,
        month = sep,
       volume = {196},
       number = {1},
          eid = {11},
        pages = {11},
          doi = {10.1088/0067-0049/196/1/11},
archivePrefix = {arXiv},
       eprint = {1107.1518},
 primaryClass = {astro-ph.CO},
       adsurl = {https://ui.adsabs.harvard.edu/abs/2011ApJS..196...11S}
}

@ARTICLE{cappellari06,
       author = {{Cappellari}, Michele and {Bacon}, R. and {Bureau}, M. and {Damen}, M.~C. and {Davies}, Roger L. and {de Zeeuw}, P.~T. and {Emsellem}, Eric and {Falc{\'o}n-Barroso}, Jes{\'u}s and {Krajnovi{\'c}}, Davor and {Kuntschner}, Harald and {McDermid}, Richard M. and {Peletier}, Reynier F. and {Sarzi}, Marc and {van den Bosch}, Remco C.~E. and {van de Ven}, Glenn},
        title = "{The SAURON project - IV. The mass-to-light ratio, the virial mass estimator and the Fundamental Plane of elliptical and lenticular galaxies}",
      journal = {\mnras},
         year = 2006,
        month = mar,
       volume = {366},
       number = {4},
        pages = {1126-1150},
          doi = {10.1111/j.1365-2966.2005.09981.x},
archivePrefix = {arXiv},
       eprint = {astro-ph/0505042},
 primaryClass = {astro-ph},
       adsurl = {https://ui.adsabs.harvard.edu/abs/2006MNRAS.366.1126C}
}

@ARTICLE{salim18,
       author = {{Salim}, Samir and {Boquien}, M{\'e}d{\'e}ric and {Lee}, Janice C.},
        title = "{Dust Attenuation Curves in the Local Universe: Demographics and New Laws for Star-forming Galaxies and High-redshift Analogs}",
      journal = {\apj},
         year = 2018,
        month = may,
       volume = {859},
       number = {1},
          eid = {11},
        pages = {11},
          doi = {10.3847/1538-4357/aabf3c},
archivePrefix = {arXiv},
       eprint = {1804.05850},
 primaryClass = {astro-ph.GA},
       adsurl = {https://ui.adsabs.harvard.edu/abs/2018ApJ...859...11S}
}

@ARTICLE{tortora20,
       author = {{Tortora}, C. and {Napolitano}, N.~R. and {Radovich}, M. and {Spiniello}, C. and {Hunt}, L. and {Roy}, N. and {Moscardini}, L. and {Scognamiglio}, D. and {Spavone}, M. and {Brescia}, M. and {Cavuoti}, S. and {D`Ago}, G. and {Longo}, G. and {Bellagamba}, F. and {Maturi}, M. and {Roncarelli}, M.},
        title = "{Nature versus nurture: relic nature and environment of the most massive passive galaxies at z < 0.5}",
      journal = {\aap},
         year = 2020,
        month = jun,
       volume = {638},
          eid = {L11},
        pages = {L11},
          doi = {10.1051/0004-6361/202038373},
archivePrefix = {arXiv},
       eprint = {2006.13235},
 primaryClass = {astro-ph.GA},
       adsurl = {https://ui.adsabs.harvard.edu/abs/2020A&A...638L..11T}
}

@ARTICLE{pohlen06,
       author = {{Pohlen}, M. and {Trujillo}, I.},
        title = "{The structure of galactic disks. Studying late-type spiral galaxies using SDSS}",
      journal = {\aap},
         year = 2006,
        month = aug,
       volume = {454},
       number = {3},
        pages = {759-772},
          doi = {10.1051/0004-6361:20064883},
archivePrefix = {arXiv},
       eprint = {astro-ph/0603682},
 primaryClass = {astro-ph},
       adsurl = {https://ui.adsabs.harvard.edu/abs/2006A&A...454..759P}
}

@ARTICLE{stringer15,
       author = {{Stringer}, Martin and {Trujillo}, Ignacio and {Dalla Vecchia}, Claudio and {Martinez-Valpuesta}, Inma},
        title = "{A cosmological context for compact massive galaxies}",
      journal = {\mnras},
         year = 2015,
        month = may,
       volume = {449},
       number = {3},
        pages = {2396-2404},
          doi = {10.1093/mnras/stv455},
archivePrefix = {arXiv},
       eprint = {1503.03078},
 primaryClass = {astro-ph.GA},
       adsurl = {https://ui.adsabs.harvard.edu/abs/2015MNRAS.449.2396S}
}

@ARTICLE{valentinuzzi10,
       author = {{Valentinuzzi}, T. and {Fritz}, J. and {Poggianti}, B.~M. and {Cava}, A. and {Bettoni}, D. and {Fasano}, G. and {D'Onofrio}, M. and {Couch}, W.~J. and {Dressler}, A. and {Moles}, M. and {Moretti}, A. and {Omizzolo}, A. and {Kj{\ae}rgaard}, P. and {Vanzella}, E. and {Varela}, J.},
        title = "{Superdense Massive Galaxies in Wings Local Clusters}",
      journal = {\apj},
         year = 2010,
        month = mar,
       volume = {712},
       number = {1},
        pages = {226-237},
          doi = {10.1088/0004-637X/712/1/226},
archivePrefix = {arXiv},
       eprint = {0907.2392},
 primaryClass = {astro-ph.CO},
       adsurl = {https://ui.adsabs.harvard.edu/abs/2010ApJ...712..226V}
}

@ARTICLE{peralta16,
       author = {{Peralta de Arriba}, Luis and {Quilis}, Vicent and {Trujillo}, Ignacio and {Cebri{\'a}n}, Mar{\'\i}a and {Balcells}, Marc},
        title = "{Massive relic galaxies prefer dense environments}",
      journal = {\mnras},
         year = 2016,
        month = sep,
       volume = {461},
       number = {1},
        pages = {156-163},
          doi = {10.1093/mnras/stw1240},
archivePrefix = {arXiv},
       eprint = {1605.06503},
 primaryClass = {astro-ph.GA},
       adsurl = {https://ui.adsabs.harvard.edu/abs/2016MNRAS.461..156P}
}

@ARTICLE{kauffmann04,
       author = {{Kauffmann}, Guinevere and {White}, Simon D.~M. and {Heckman}, Timothy M. and {M{\'e}nard}, Brice and {Brinchmann}, Jarle and {Charlot}, St{\'e}phane and {Tremonti}, Christy and {Brinkmann}, Jon},
        title = "{The environmental dependence of the relations between stellar mass, structure, star formation and nuclear activity in galaxies}",
      journal = {\mnras},
         year = 2004,
        month = sep,
       volume = {353},
       number = {3},
        pages = {713-731},
          doi = {10.1111/j.1365-2966.2004.08117.x},
archivePrefix = {arXiv},
       eprint = {astro-ph/0402030},
 primaryClass = {astro-ph},
       adsurl = {https://ui.adsabs.harvard.edu/abs/2004MNRAS.353..713K}
}

@ARTICLE{dressler80,
       author = {{Dressler}, A.},
        title = "{Galaxy morphology in rich clusters: implications for the formation and evolution of galaxies.}",
      journal = {\apj},
         year = 1980,
        month = mar,
       volume = {236},
        pages = {351-365},
          doi = {10.1086/157753},
       adsurl = {https://ui.adsabs.harvard.edu/abs/1980ApJ...236..351D}
}

@ARTICLE{blanton05,
       author = {{Blanton}, Michael R. and {Eisenstein}, Daniel and {Hogg}, David W. and {Schlegel}, David J. and {Brinkmann}, J.},
        title = "{Relationship between Environment and the Broadband Optical Properties of Galaxies in the Sloan Digital Sky Survey}",
      journal = {\apj},
         year = 2005,
        month = aug,
       volume = {629},
       number = {1},
        pages = {143-157},
          doi = {10.1086/422897},
archivePrefix = {arXiv},
       eprint = {astro-ph/0310453},
 primaryClass = {astro-ph},
       adsurl = {https://ui.adsabs.harvard.edu/abs/2005ApJ...629..143B}
}

@ARTICLE{laurikainen09,
       author = {{Laurikainen}, E. and {Salo}, H. and {Buta}, R. and {Knapen}, J.~H.},
        title = "{Bars, Ovals, and Lenses in Early-Type Disk Galaxies: Probes of Galaxy Evolution}",
      journal = {\apjl},
         year = 2009,
        month = feb,
       volume = {692},
       number = {1},
        pages = {L34-L39},
          doi = {10.1088/0004-637X/692/1/L34},
archivePrefix = {arXiv},
       eprint = {0901.0641},
 primaryClass = {astro-ph.GA},
       adsurl = {https://ui.adsabs.harvard.edu/abs/2009ApJ...692L..34L}
}

@ARTICLE{dominguez-sanchez18,
       author = {{Dom{\'\i}nguez S{\'a}nchez}, H. and {Huertas-Company}, M. and {Bernardi}, M. and {Tuccillo}, D. and {Fischer}, J.~L.},
        title = "{Improving galaxy morphologies for SDSS with Deep Learning}",
      journal = {\mnras},
         year = 2018,
        month = feb,
       volume = {476},
       number = {3},
        pages = {3661-3676},
          doi = {10.1093/mnras/sty338},
archivePrefix = {arXiv},
       eprint = {1711.05744},
 primaryClass = {astro-ph.GA},
       adsurl = {https://ui.adsabs.harvard.edu/abs/2018MNRAS.476.3661D}
}

@ARTICLE{laurikainen11,
       author = {{Laurikainen}, E. and {Salo}, H. and {Buta}, R. and {Knapen}, J.~H.},
        title = "{Near-infrared atlas of S0-Sa galaxies (NIRS0S)}",
      journal = {\mnras},
         year = 2011,
        month = dec,
       volume = {418},
       number = {3},
        pages = {1452-1490},
          doi = {10.1111/j.1365-2966.2011.19283.x},
archivePrefix = {arXiv},
       eprint = {1110.1996},
 primaryClass = {astro-ph.CO},
       adsurl = {https://ui.adsabs.harvard.edu/abs/2011MNRAS.418.1452L}
}

@ARTICLE{burstein05,
       author = {{Burstein}, David and {Ho}, Luis C. and {Huchra}, John P. and {Macri}, Lucas M.},
        title = "{The K-Band Luminosities of Galaxies: Do S0s Come from Spiral Galaxies?}",
      journal = {\apj},
         year = 2005,
        month = mar,
       volume = {621},
       number = {1},
        pages = {246-255},
          doi = {10.1086/427408},
       adsurl = {https://ui.adsabs.harvard.edu/abs/2005ApJ...621..246B}
}

@ARTICLE{gao18,
       author = {{Gao}, Hua and {Ho}, Luis C. and {Barth}, Aaron J. and {Li}, Zhao-Yu},
        title = "{The Carnegie-Irvine Galaxy Survey. VII. Constraints on the Origin of S0 Galaxies from Their Photometric Structure}",
      journal = {\apj},
         year = 2018,
        month = aug,
       volume = {862},
       number = {2},
          eid = {100},
        pages = {100},
          doi = {10.3847/1538-4357/aacdac},
archivePrefix = {arXiv},
       eprint = {1806.06350},
 primaryClass = {astro-ph.GA},
       adsurl = {https://ui.adsabs.harvard.edu/abs/2018ApJ...862..100G}
}

@ARTICLE{huang13b,
       author = {{Huang}, Song and {Ho}, Luis C. and {Peng}, Chien Y. and {Li}, Zhao-Yu and {Barth}, Aaron J.},
        title = "{Fossil Evidence for the Two-phase Formation of Elliptical Galaxies}",
      journal = {\apjl},
         year = 2013,
        month = may,
       volume = {768},
       number = {2},
          eid = {L28},
        pages = {L28},
          doi = {10.1088/2041-8205/768/2/L28},
archivePrefix = {arXiv},
       eprint = {1304.2299},
 primaryClass = {astro-ph.CO},
       adsurl = {https://ui.adsabs.harvard.edu/abs/2013ApJ...768L..28H}
}

@ARTICLE{vanderberg76,
       author = {{van den Bergh}, S.},
        title = "{A new classification system for galaxies.}",
      journal = {\apj},
         year = 1976,
        month = jun,
       volume = {206},
        pages = {883-887},
          doi = {10.1086/154452},
       adsurl = {https://ui.adsabs.harvard.edu/abs/1976ApJ...206..883V}
}

@ARTICLE{huertas16,
       author = {{Huertas-Company}, M. and {Bernardi}, M. and
         {P{\'e}rez-Gonz{\'a}lez}, P.~G. and {Ashby}, M.~L.~N. and {Barro}, G. and
         {Conselice}, C. and {Daddi}, E. and {Dekel}, A. and {Dimauro}, P. and
         {Faber}, S.~M. and {Grogin}, N.~A. and {Kartaltepe}, J.~S. and
         {Kocevski}, D.~D. and {Koekemoer}, A.~M. and {Koo}, D.~C. and
         {Mei}, S. and {Shankar}, F.},
        title = "{Mass assembly and morphological transformations since z {\ensuremath{\sim}} 3 from CANDELS}",
      journal = {\mnras},
         year = 2016,
        month = nov,
       volume = {462},
       number = {4},
        pages = {4495-4516},
          doi = {10.1093/mnras/stw1866},
archivePrefix = {arXiv},
       eprint = {1606.04952},
 primaryClass = {astro-ph.GA},
       adsurl = {https://ui.adsabs.harvard.edu/abs/2016MNRAS.462.4495H}
}

@ARTICLE{davari17,
       author = {{Davari}, Roozbeh H. and {Ho}, Luis C. and {Mobasher}, Bahram and
         {Canalizo}, Gabriela},
        title = "{Detection of Prominent Stellar Disks in the Progenitors of Present-day Massive Elliptical Galaxies}",
      journal = {\apj},
         year = 2017,
        month = feb,
       volume = {836},
       number = {1},
          eid = {75},
        pages = {75},
          doi = {10.3847/1538-4357/836/1/75},
archivePrefix = {arXiv},
       eprint = {1606.07571},
 primaryClass = {astro-ph.GA},
       adsurl = {https://ui.adsabs.harvard.edu/abs/2017ApJ...836...75D}
}

@ARTICLE{spiniello21,
       author = {{Spiniello}, C. and {Tortora}, C. and {D'Ago}, G. and {Coccato}, L. and {La Barbera}, F. and {Ferr{\'e}-Mateu}, A. and {Napolitano}, N.~R. and {Spavone}, M. and {Scognamiglio}, D. and {Arnaboldi}, M. and {Gallazzi}, A. and {Hunt}, L. and {Moehler}, S. and {Radovich}, M. and {Zibetti}, S.},
        title = "{INSPIRE: INvestigating Stellar Population In RElics. I. Survey presentation and pilot study}",
      journal = {\aap},
         year = 2021,
        month = feb,
       volume = {646},
          eid = {A28},
        pages = {A28},
          doi = {10.1051/0004-6361/202038936},
archivePrefix = {arXiv},
       eprint = {2011.05347},
 primaryClass = {astro-ph.GA},
       adsurl = {https://ui.adsabs.harvard.edu/abs/2021A&A...646A..28S}
}

@ARTICLE{trujillo14,
       author = {{Trujillo}, Ignacio and {Ferr{\'e}-Mateu}, Anna and {Balcells}, Marc and {Vazdekis}, Alexandre and {S{\'a}nchez-Bl{\'a}zquez}, Patricia},
        title = "{NGC 1277: A Massive Compact Relic Galaxy in the Nearby Universe}",
      journal = {\apjl},
         year = 2014,
        month = jan,
       volume = {780},
       number = {2},
          eid = {L20},
        pages = {L20},
          doi = {10.1088/2041-8205/780/2/L20},
archivePrefix = {arXiv},
       eprint = {1310.6367},
 primaryClass = {astro-ph.CO},
       adsurl = {https://ui.adsabs.harvard.edu/abs/2014ApJ...780L..20T}
}

@ARTICLE{ferre-mateu12,
       author = {{Ferr{\'e}-Mateu}, A. and {Vazdekis}, A. and {Trujillo}, I. and {S{\'a}nchez-Bl{\'a}zquez}, P. and {Ricciardelli}, E. and {de la Rosa}, I.~G.},
        title = "{Young ages and other intriguing properties of massive compact galaxies in the local Universe}",
      journal = {\mnras},
         year = 2012,
        month = jun,
       volume = {423},
       number = {1},
        pages = {632-646},
          doi = {10.1111/j.1365-2966.2012.20897.x},
archivePrefix = {arXiv},
       eprint = {1203.2623},
 primaryClass = {astro-ph.CO},
       adsurl = {https://ui.adsabs.harvard.edu/abs/2012MNRAS.423..632F}
}

@ARTICLE{ferreira22,
       author = {{Ferreira}, Leonardo and {Adams}, Nathan and {Conselice}, Christopher J. and {Sazonova}, Elizaveta and {Austin}, Duncan and {Caruana}, Joseph and {Ferrari}, Fabricio and {Verma}, Aprajita and {Trussler}, James and {Broadhurst}, Tom and {Diego}, Jose and {Frye}, Brenda L. and {Pascale}, Massimo and {Wilkins}, Stephen M. and {Windhorst}, Rogier A. and {Zitrin}, Adi},
        title = "{Panic! at the Disks: First Rest-frame Optical Observations of Galaxy Structure at z > 3 with JWST in the SMACS 0723 Field}",
      journal = {\apjl},
         year = 2022,
        month = oct,
       volume = {938},
       number = {1},
          eid = {L2},
        pages = {L2},
          doi = {10.3847/2041-8213/ac947c},
archivePrefix = {arXiv},
       eprint = {2207.09428},
 primaryClass = {astro-ph.GA},
       adsurl = {https://ui.adsabs.harvard.edu/abs/2022ApJ...938L...2F}
}

@ARTICLE{lee24,
       author = {{Lee}, Jeong Hwan and {Park}, Changbom and {Hwang}, Ho Seong and {Kwon}, Minseong},
        title = "{Morphology of Galaxies in JWST Fields: Initial Distribution and Evolution of Galaxy Morphology}",
      journal = {\apj},
         year = 2024,
        month = may,
       volume = {966},
       number = {1},
          eid = {113},
        pages = {113},
          doi = {10.3847/1538-4357/ad3448},
archivePrefix = {arXiv},
       eprint = {2312.04899},
 primaryClass = {astro-ph.GA},
       adsurl = {https://ui.adsabs.harvard.edu/abs/2024ApJ...966..113L}
}

@ARTICLE{huertas24,
       author = {{Huertas-Company}, M. and {Iyer}, K.~G. and {Angeloudi}, E. and {Bagley}, M.~B. and {Finkelstein}, S.~L. and {Kartaltepe}, J. and {McGrath}, E.~J. and {Sarmiento}, R. and {Vega-Ferrero}, J. and {Arrabal Haro}, P. and {Behroozi}, P. and {Buitrago}, F. and {Cheng}, Y. and {Costantin}, L. and {Dekel}, A. and {Dickinson}, M. and {Elbaz}, D. and {Grogin}, N.~A. and {Hathi}, N.~P. and {Holwerda}, B.~W. and {Koekemoer}, A.~M. and {Lucas}, R.~A. and {Papovich}, C. and {P{\'e}rez-Gonz{\'a}lez}, P.~G. and {Pirzkal}, N. and {Seill{\'e}}, L.-M. and {de la Vega}, A. and {Wuyts}, S. and {Yang}, G. and {Yung}, L.~Y.~A.},
        title = "{Galaxy morphology from z {\ensuremath{\sim}} 6 through the lens of JWST}",
      journal = {\aap},
         year = 2024,
        month = may,
       volume = {685},
          eid = {A48},
        pages = {A48},
          doi = {10.1051/0004-6361/202346800},
archivePrefix = {arXiv},
       eprint = {2305.02478},
 primaryClass = {astro-ph.GA},
       adsurl = {https://ui.adsabs.harvard.edu/abs/2024A&A...685A..48H}
}

@ARTICLE{hubble26,
       author = {{Hubble}, E.~P.},
        title = "{Extragalactic nebulae.}",
      journal = {\apj},
         year = 1926,
        month = dec,
       volume = {64},
        pages = {321-369},
          doi = {10.1086/143018},
       adsurl = {https://ui.adsabs.harvard.edu/abs/1926ApJ....64..321H}
}

@ARTICLE{newman18,
       author = {{Newman}, Andrew B. and {Belli}, Sirio and {Ellis}, Richard S. and {Patel}, Shannon G.},
        title = "{Resolving Quiescent Galaxies at z {\ensuremath{\gtrsim}} 2. II. Direct Measures of Rotational Support}",
      journal = {\apj},
         year = 2018,
        month = aug,
       volume = {862},
       number = {2},
          eid = {126},
        pages = {126},
          doi = {10.3847/1538-4357/aacd4f},
archivePrefix = {arXiv},
       eprint = {1806.06815},
 primaryClass = {astro-ph.GA},
       adsurl = {https://ui.adsabs.harvard.edu/abs/2018ApJ...862..126N}
}

@ARTICLE{slob25,
       author = {{Slob}, Martje and {Kriek}, Mariska and {de Graaff}, Anna and {Cheng}, Chloe M. and {Beverage}, Aliza G. and {Bezanson}, Rachel and {F{\"o}rster Schreiber}, Natascha M. and {Lorenz}, Brian and {Mancera Pi{\~n}a}, Pavel E. and {Marchesini}, Danilo and {Muzzin}, Adam and {Newman}, Andrew B. and {Price}, Sedona H. and {Suess}, Katherine A. and {van de Sande}, Jesse and {van Dokkum}, Pieter and {Weisz}, Daniel R.},
        title = "{Fast rotators at cosmic noon: Stellar kinematics for 15 quiescent galaxies from JWST-SUSPENSE}",
      journal = {\aap},
         year = 2025,
        month = oct,
       volume = {702},
          eid = {A110},
        pages = {A110},
          doi = {10.1051/0004-6361/202555812},
archivePrefix = {arXiv},
       eprint = {2506.04310},
 primaryClass = {astro-ph.GA},
       adsurl = {https://ui.adsabs.harvard.edu/abs/2025A&A...702A.110S}
}

@ARTICLE{yildirim17,
       author = {{Y{\i}ld{\i}r{\i}m}, Ak{\i}n and {van den Bosch}, Remco C.~E. and {van de Ven}, Glenn and {Mart{\'\i}n-Navarro}, Ignacio and {Walsh}, Jonelle L. and {Husemann}, Bernd and {G{\"u}ltekin}, Kayhan and {Gebhardt}, Karl},
        title = "{The structural and dynamical properties of compact elliptical galaxies}",
      journal = {\mnras},
         year = 2017,
        month = jul,
       volume = {468},
       number = {4},
        pages = {4216-4245},
          doi = {10.1093/mnras/stx732},
archivePrefix = {arXiv},
       eprint = {1701.05898},
 primaryClass = {astro-ph.GA},
       adsurl = {https://ui.adsabs.harvard.edu/abs/2017MNRAS.468.4216Y}
}

@ARTICLE{lustig21,
       author = {{Lustig}, Peter and {Strazzullo}, Veronica and {D'Eugenio}, Chiara and {Daddi}, Emanuele and {Pannella}, Maurilio and {Renzini}, Alvio and {Cimatti}, Andrea and {Gobat}, Raphael and {Jin}, Shuowen and {Mohr}, Joseph J. and {Onodera}, Masato},
        title = "{Compact, bulge-dominated structures of spectroscopically confirmed quiescent galaxies at z {\ensuremath{\approx}} 3}",
      journal = {\mnras},
         year = 2021,
        month = feb,
       volume = {501},
       number = {2},
        pages = {2659-2676},
          doi = {10.1093/mnras/staa3766},
archivePrefix = {arXiv},
       eprint = {2012.02766},
 primaryClass = {astro-ph.GA},
       adsurl = {https://ui.adsabs.harvard.edu/abs/2021MNRAS.501.2659L}
}

@ARTICLE{vanderwel12,
       author = {{van der Wel}, Arjen and {Rix}, Hans-Walter and {Wuyts}, Stijn and {McGrath}, Elizabeth J. and {Koekemoer}, Anton M. and {Bell}, Eric F. and {Holden}, Bradford P. and {Robaina}, Aday R. and {McIntosh}, Daniel H.},
        title = "{The Majority of Compact Massive Galaxies at z \raisebox{-0.5ex}\textasciitilde 2 are Disk Dominated}",
      journal = {\apj},
         year = 2011,
        month = mar,
       volume = {730},
       number = {1},
          eid = {38},
        pages = {38},
          doi = {10.1088/0004-637X/730/1/38},
archivePrefix = {arXiv},
       eprint = {1101.2423},
 primaryClass = {astro-ph.CO},
       adsurl = {https://ui.adsabs.harvard.edu/abs/2011ApJ...730...38V}
}

@ARTICLE{gutierrez11,
       author = {{Guti{\'e}rrez}, Leonel and {Erwin}, Peter and {Aladro}, Rebeca and {Beckman}, John E.},
        title = "{The Outer Disks of Early-type Galaxies. II. Surface-brightness Profiles of Unbarred Galaxies and Trends with Hubble Type}",
      journal = {\aj},
         year = 2011,
        month = nov,
       volume = {142},
       number = {5},
          eid = {145},
        pages = {145},
          doi = {10.1088/0004-6256/142/5/145},
archivePrefix = {arXiv},
       eprint = {1108.3662},
 primaryClass = {astro-ph.CO},
       adsurl = {https://ui.adsabs.harvard.edu/abs/2011AJ....142..145G}
}

@ARTICLE{maltby15,
       author = {{Maltby}, David T. and {Arag{\'o}n-Salamanca}, Alfonso and {Gray}, Meghan E. and {Hoyos}, Carlos and {Wolf}, Christian and {Jogee}, Shardha and {B{\"o}hm}, Asmus},
        title = "{The environmental dependence of the structure of galactic discs in STAGES S0 galaxies: implications for S0 formation}",
      journal = {\mnras},
         year = 2015,
        month = feb,
       volume = {447},
       number = {2},
        pages = {1506-1530},
          doi = {10.1093/mnras/stu2536},
archivePrefix = {arXiv},
       eprint = {1412.3167},
 primaryClass = {astro-ph.GA},
       adsurl = {https://ui.adsabs.harvard.edu/abs/2015MNRAS.447.1506M}
}

@ARTICLE{kormendy12,
       author = {{Kormendy}, John and {Bender}, Ralf},
        title = "{A Revised Parallel-sequence Morphological Classification of Galaxies: Structure and Formation of S0 and Spheroidal Galaxies}",
      journal = {\apjs},
         year = 2012,
        month = jan,
       volume = {198},
       number = {1},
          eid = {2},
        pages = {2},
          doi = {10.1088/0067-0049/198/1/2},
archivePrefix = {arXiv},
       eprint = {1110.4384},
 primaryClass = {astro-ph.CO},
       adsurl = {https://ui.adsabs.harvard.edu/abs/2012ApJS..198....2K}
}

@article{Cappellari.etal.2011,
       author = {{Cappellari}, Michele and {Emsellem}, Eric and {Krajnovi{\'c}}, Davor and {McDermid}, Richard M. and {Serra}, Paolo and {Alatalo}, Katherine and {Blitz}, Leo and {Bois}, Maxime and {Bournaud}, Fr{\'e}d{\'e}ric and {Bureau}, M. and {Davies}, Roger L. and {Davis}, Timothy A. and {de Zeeuw}, P.~T. and {Khochfar}, Sadegh and {Kuntschner}, Harald and {Lablanche}, Pierre-Yves and {Morganti}, Raffaella and {Naab}, Thorsten and {Oosterloo}, Tom and {Sarzi}, Marc and {Scott}, Nicholas and {Weijmans}, Anne-Marie and {Young}, Lisa M.},
        title = "{The ATLAS$^{3D}$ project - VII. A new look at the morphology of nearby galaxies: the kinematic morphology-density relation}",
      journal = {\mnras},
         year = 2011,
        month = sep,
       volume = {416},
       number = {3},
        pages = {1680-1696},
          doi = {10.1111/j.1365-2966.2011.18600.x},
archivePrefix = {arXiv},
       eprint = {1104.3545},
 primaryClass = {astro-ph.CO},
       adsurl = {https://ui.adsabs.harvard.edu/abs/2011MNRAS.416.1680C}
}

@ARTICLE{Lange.etal.2016,
       author = {{Lange}, Rebecca and {Moffett}, Amanda J. and {Driver}, Simon P. and {Robotham}, Aaron S.~G. and {Lagos}, Claudia del P. and {Kelvin}, Lee S. and {Conselice}, Christopher and {Margalef-Bentabol}, Berta and {Alpaslan}, Mehmet and {Baldry}, Ivan and {Bland-Hawthorn}, Joss and {Bremer}, Malcolm and {Brough}, Sarah and {Cluver}, Michelle and {Colless}, Matthew and {Davies}, Luke J.~M. and {H{\"a}u{\ss}ler}, Boris and {Holwerda}, Benne W. and {Hopkins}, Andrew M. and {Kafle}, Prajwal R. and {Kennedy}, Rebecca and {Liske}, Jochen and {Phillipps}, Steven and {Popescu}, Cristina C. and {Taylor}, Edward N. and {Tuffs}, Richard and {van Kampen}, Eelco and {Wright}, Angus H.},
        title = "{Galaxy And Mass Assembly (GAMA): M\_star - R\_e relations of z = 0 bulges, discs and spheroids}",
      journal = {\mnras},
         year = 2016,
        month = oct,
       volume = {462},
       number = {2},
        pages = {1470-1500},
          doi = {10.1093/mnras/stw1495},
archivePrefix = {arXiv},
       eprint = {1607.01096},
 primaryClass = {astro-ph.GA},
       adsurl = {https://ui.adsabs.harvard.edu/abs/2016MNRAS.462.1470L}
}

@ARTICLE{Nedkova.etal.2024,
       author = {{Nedkova}, Kalina V. and {H{\"a}u{\ss}ler}, Boris and {Marchesini}, Danilo and {Brammer}, Gabriel B. and {Feinstein}, Adina D. and {Johnston}, Evelyn J. and {Kartaltepe}, Jeyhan S. and {Koekemoer}, Anton M. and {Martis}, Nicholas S. and {Muzzin}, Adam and {Rafelski}, Marc and {Shipley}, Heath V. and {Skelton}, Rosalind E. and {Stefanon}, Mauro and {van der Wel}, Arjen and {Whitaker}, Katherine E.},
        title = "{Bulge+disc decomposition of HFF and CANDELS galaxies: UVJ diagrams and stellar mass-size relations of galaxy components at 0.2 {\ensuremath{\leq}} z {\ensuremath{\leq}} 1.5}",
      journal = {\mnras},
         year = 2024,
        month = aug,
       volume = {532},
       number = {4},
        pages = {3747-3777},
          doi = {10.1093/mnras/stae1702},
archivePrefix = {arXiv},
       eprint = {2406.14613},
 primaryClass = {astro-ph.GA},
       adsurl = {https://ui.adsabs.harvard.edu/abs/2024MNRAS.532.3747N}
}

@ARTICLE{Clerici.etal.2024,
       author = {{Clerici}, K. Slodkowski and {Schnorr-M{\"u}ller}, A. and {Trevisan}, M. and {Ricci}, T.~V.},
        title = "{Massive compact quiescent galaxies in the M$_{{\ensuremath{\star}}}$ versus {\ensuremath{\sigma}}$_{e}$ plane: insights from stellar population properties}",
      journal = {\mnras},
         year = 2024,
        month = jun,
       volume = {531},
       number = {1},
        pages = {1034-1055},
          doi = {10.1093/mnras/stae1213},
archivePrefix = {arXiv},
       eprint = {2405.02348},
 primaryClass = {astro-ph.GA},
       adsurl = {https://ui.adsabs.harvard.edu/abs/2024MNRAS.531.1034C}
}

@ARTICLE{Bertin.and.Arnouts.1996,
       author = {{Bertin}, E. and {Arnouts}, S.},
        title = "{SExtractor: Software for source extraction.}",
      journal = {\aaps},
         year = 1996,
        month = jun,
       volume = {117},
        pages = {393-404},
          doi = {10.1051/aas:1996164},
       adsurl = {https://ui.adsabs.harvard.edu/abs/1996A&AS..117..393B}
}

@ARTICLE{Erwin.2015,
       author = {{Erwin}, Peter},
        title = "{IMFIT: A Fast, Flexible New Program for Astronomical Image Fitting}",
      journal = {\apj},
         year = 2015,
        month = feb,
       volume = {799},
       number = {2},
          eid = {226},
        pages = {226},
          doi = {10.1088/0004-637X/799/2/226},
archivePrefix = {arXiv},
       eprint = {1408.1097},
 primaryClass = {astro-ph.IM},
       adsurl = {https://ui.adsabs.harvard.edu/abs/2015ApJ...799..226E}
}

@ARTICLE{Huang.etal.2013,
       author = {{Huang}, Song and {Ho}, Luis C. and {Peng}, Chien Y. and {Li}, Zhao-Yu and {Barth}, Aaron J.},
        title = "{The Carnegie-Irvine Galaxy Survey. III. The Three-component Structure of Nearby Elliptical Galaxies}",
      journal = {\apj},
         year = 2013,
        month = mar,
       volume = {766},
       number = {1},
          eid = {47},
        pages = {47},
          doi = {10.1088/0004-637X/766/1/47},
archivePrefix = {arXiv},
       eprint = {1212.2639},
 primaryClass = {astro-ph.CO},
       adsurl = {https://ui.adsabs.harvard.edu/abs/2013ApJ...766...47H}
}

@Article{Ho.et.al.2011,
  title = {{MatchIt}: Nonparametric Preprocessing for Parametric Causal Inference},
  author = {Daniel E. Ho and Kosuke Imai and Gary King and Elizabeth A. Stuart},
  journal = {Journal of Statistical Software},
  year = {2011},
  volume = {42},
  number = {8},
  pages = {1--28},
  url = {http://www.jstatsoft.org/v42/i08/},
}

@Article{Rosenbaum.and.Rubin.1983,
	author = "P. R. {Rosenbaum} and D. B. {Rubin}",
	title = "{The Central Role of the Propensity Score in Observational Studies for Causal Effects}",
	journal = "Biometrika",
	year = 1983,
	volume = 70,
	pages = "41--55",
	doi = "10.1093/biomet/70.1.41"
}

@ARTICLE{Oh.etal.2017,
       author = {{Oh}, Semyeong and {Greene}, Jenny E. and {Lackner}, Claire N.},
        title = "{Testing the Presence of Multiple Photometric Components in Nearby Early-type Galaxies using SDSS}",
      journal = {\apj},
         year = 2017,
        month = feb,
       volume = {836},
       number = {1},
          eid = {115},
        pages = {115},
          doi = {10.3847/1538-4357/836/1/115},
archivePrefix = {arXiv},
       eprint = {1612.06495},
 primaryClass = {astro-ph.GA},
       adsurl = {https://ui.adsabs.harvard.edu/abs/2017ApJ...836..115O}
}

@ARTICLE{Gadotti.etal.2012,
       author = {{Gadotti}, Dimitri A. and {S{\'a}nchez-Janssen}, Rub{\'e}n.},
        title = "{Surprises in image decomposition of edge-on galaxies: does Sombrero have a (classical) bulge?}",
      journal = {\mnras},
         year = 2012,
        month = jun,
       volume = {423},
       number = {1},
        pages = {877-888},
          doi = {10.1111/j.1365-2966.2012.20925.x},
archivePrefix = {arXiv},
       eprint = {1101.2900},
 primaryClass = {astro-ph.CO},
       adsurl = {https://ui.adsabs.harvard.edu/abs/2012MNRAS.423..877G}
}

@ARTICLE{Yildirm.etal.2015,
       author = {{Y{\i}ld{\i}r{\i}m}, Ak{\i}n and {van den Bosch}, Remco C.~E. and {van de Ven}, Glenn and {Husemann}, Bernd and {Lyubenova}, Mariya and {Walsh}, Jonelle L. and {Gebhardt}, Karl and {G{\"u}ltekin}, Kayhan},
        title = "{MRK 1216 and NGC 1277 - an orbit-based dynamical analysis of compact, high-velocity dispersion galaxies}",
      journal = {\mnras},
         year = 2015,
        month = sep,
       volume = {452},
       number = {2},
        pages = {1792-1816},
          doi = {10.1093/mnras/stv1381},
archivePrefix = {arXiv},
       eprint = {1506.06762},
 primaryClass = {astro-ph.GA},
       adsurl = {https://ui.adsabs.harvard.edu/abs/2015MNRAS.452.1792Y}
}

@ARTICLE{Aihara.etal.2022,
       author = {{Aihara}, Hiroaki and {AlSayyad}, Yusra and {Ando}, Makoto and {Armstrong}, Robert and {Bosch}, James and {Egami}, Eiichi and {Furusawa}, Hisanori and {Furusawa}, Junko and {Harasawa}, Sumiko and {Harikane}, Yuichi and {Hsieh}, Bau-Ching and {Ikeda}, Hiroyuki and {Ito}, Kei and {Iwata}, Ikuru and {Kodama}, Tadayuki and {Koike}, Michitaro and {Kokubo}, Mitsuru and {Komiyama}, Yutaka and {Li}, Xiangchong and {Liang}, Yongming and {Lin}, Yen-Ting and {Lupton}, Robert H. and {Lust}, Nate B. and {MacArthur}, Lauren A. and {Mawatari}, Ken and {Mineo}, Sogo and {Miyatake}, Hironao and {Miyazaki}, Satoshi and {More}, Surhud and {Morishima}, Takahiro and {Murayama}, Hitoshi and {Nakajima}, Kimihiko and {Nakata}, Fumiaki and {Nishizawa}, Atsushi J. and {Oguri}, Masamune and {Okabe}, Nobuhiro and {Okura}, Yuki and {Ono}, Yoshiaki and {Osato}, Ken and {Ouchi}, Masami and {Pan}, Yen-Chen and {Plazas Malag{\'o}n}, Andr{\'e}s A. and {Price}, Paul A. and {Reed}, Sophie L. and {Rykoff}, Eli S. and {Shibuya}, Takatoshi and {Simunovic}, Mirko and {Strauss}, Michael A. and {Sugimori}, Kanako and {Suto}, Yasushi and {Suzuki}, Nao and {Takada}, Masahiro and {Takagi}, Yuhei and {Takata}, Tadafumi and {Takita}, Satoshi and {Tanaka}, Masayuki and {Tang}, Shenli and {Taranu}, Dan S. and {Terai}, Tsuyoshi and {Toba}, Yoshiki and {Turner}, Edwin L. and {Uchiyama}, Hisakazu and {Vijarnwannaluk}, Bovornpratch and {Waters}, Christopher Z. and {Yamada}, Yoshihiko and {Yamamoto}, Naoaki and {Yamashita}, Takuji},
        title = "{Third data release of the Hyper Suprime-Cam Subaru Strategic Program}",
      journal = {\pasj},
         year = 2022,
        month = apr,
       volume = {74},
       number = {2},
        pages = {247-272},
          doi = {10.1093/pasj/psab122},
archivePrefix = {arXiv},
       eprint = {2108.13045},
 primaryClass = {astro-ph.IM},
       adsurl = {https://ui.adsabs.harvard.edu/abs/2022PASJ...74..247A}
}

@BOOK{Sersic.1968,
       author = {{Sersic}, Jose Luis},
        title = "{Atlas de Galaxias Australes}",
         year = 1968,
       adsurl = {https://ui.adsabs.harvard.edu/abs/1968adga.book.....S}
}

@ARTICLE{ferre-mateu17,
       author = {{Ferr{\'e}-Mateu}, Anna and {Trujillo}, Ignacio and {Mart{\'\i}n-Navarro}, Ignacio and {Vazdekis}, Alexandre and {Mezcua}, Mar and {Balcells}, Marc and {Dom{\'\i}nguez}, Lilian},
        title = "{Two new confirmed massive relic galaxies: red nuggets in the present-day Universe}",
      journal = {\mnras},
         year = 2017,
        month = may,
       volume = {467},
       number = {2},
        pages = {1929-1939},
          doi = {10.1093/mnras/stx171},
archivePrefix = {arXiv},
       eprint = {1701.05197},
 primaryClass = {astro-ph.GA},
       adsurl = {https://ui.adsabs.harvard.edu/abs/2017MNRAS.467.1929F}
}

@ARTICLE{vanderwel.etal.2014,
       author = {{van der Wel}, A. and {Franx}, M. and {van Dokkum}, P.~G. and {Skelton}, R.~E. and {Momcheva}, I.~G. and {Whitaker}, K.~E. and {Brammer}, G.~B. and {Bell}, E.~F. and {Rix}, H. -W. and {Wuyts}, S. and {Ferguson}, H.~C. and {Holden}, B.~P. and {Barro}, G. and {Koekemoer}, A.~M. and {Chang}, Yu-Yen and {McGrath}, E.~J. and {H{\"a}ussler}, B. and {Dekel}, A. and {Behroozi}, P. and {Fumagalli}, M. and {Leja}, J. and {Lundgren}, B.~F. and {Maseda}, M.~V. and {Nelson}, E.~J. and {Wake}, D.~A. and {Patel}, S.~G. and {Labb{\'e}}, I. and {Faber}, S.~M. and {Grogin}, N.~A. and {Kocevski}, D.~D.},
        title = "{3D-HST+CANDELS: The Evolution of the Galaxy Size-Mass Distribution since z = 3}",
      journal = {\apj},
         year = 2014,
        month = jun,
       volume = {788},
       number = {1},
          eid = {28},
        pages = {28},
          doi = {10.1088/0004-637X/788/1/28},
archivePrefix = {arXiv},
       eprint = {1404.2844},
 primaryClass = {astro-ph.GA},
       adsurl = {https://ui.adsabs.harvard.edu/abs/2014ApJ...788...28V}
}

@ARTICLE{Schnorr.et.al.2021,
       author = {{Schnorr-M{\"u}ller}, A. and {Trevisan}, M. and {Riffel}, R. and {Chies-Santos}, A.~L. and {Furlanetto}, C. and {Ricci}, T.~V. and {Lohmann}, F.~S. and {Flores-Freitas}, R. and {Mallmann}, N.~D. and {Alamo-Mart{\'\i}nez}, K.~A.},
        title = "{The puzzling origin of massive compact galaxies in MaNGA}",
      journal = {\mnras},
         year = 2021,
        month = oct,
       volume = {507},
       number = {1},
        pages = {300-317},
          doi = {10.1093/mnras/stab2116},
archivePrefix = {arXiv},
       eprint = {2104.12737},
 primaryClass = {astro-ph.GA},
       adsurl = {https://ui.adsabs.harvard.edu/abs/2021MNRAS.507..300S}
}

@ARTICLE{Hill.etal.2019,
       author = {{Hill}, Allison R. and {van der Wel}, Arjen and {Franx}, Marijn and {Muzzin}, Adam and {Skelton}, Rosalind E. and {et al.}},
        title = "{High-redshift Massive Quiescent Galaxies Are as Flat as Star-forming Galaxies: The Flattening of Galaxies and the Correlation with Structural Properties in CANDELS/3D-HST}",
      journal = {\apj},
         year = 2019,
        month = jan,
       volume = {871},
       number = {1},
          eid = {76},
        pages = {76},
          doi = {10.3847/1538-4357/aaf50a},
archivePrefix = {arXiv},
       eprint = {1901.02009},
 primaryClass = {astro-ph.GA},
       adsurl = {https://ui.adsabs.harvard.edu/abs/2019ApJ...871...76H}
}
\bibliographystyle{aasjournal}



\end{document}